\documentclass[%
 reprint,
 amsmath,amssymb,amsthm,
 aps,
 prx
floatfix,
]{revtex4-2}

\usepackage{graphicx}
\usepackage{dcolumn,amsthm}
\usepackage{bm}

\usepackage{mathtools}
\usepackage{txfonts}
\usepackage{dsfont}
\usepackage[english]{babel}

\usepackage[colorlinks=true,linkcolor=blue,citecolor=blue,urlcolor=blue]{hyperref}

\newcommand{\<}{\langle}
\renewcommand{\>}{\rangle}
\newcommand{\abs}[1]{\left\lvert#1\right\rvert}
\newcommand{\norm}[1]{\lVert#1\rVert}

\providecommand{\tr}{{\rm tr}}
\renewcommand{\phi}{\varphi}
\newcommand{\bra}[1]{\left\langle #1\right\rvert}
\newcommand{\ket}[1]{\left\lvert #1\right\rangle}

\begin{document}

\title{Non-Gaussian Quantum States and Where to Find Them}

\author{Mattia Walschaers}
\email{mattia.walschaers@lkb.upmc.fr}
\affiliation{Laboratoire Kastler Brossel, Sorbonne Universit\'{e}, CNRS, ENS-Universit\'e PSL, Coll\`{e}ge de France, 4 place Jussieu, F-75252 Paris, France}

\date{\today}

\begin{abstract}
Gaussian states have played on important role in the physics of continuous-variable quantum systems. They are appealing for the experimental ease with which they can be produced, and for their compact and elegant mathematical description. Nevertheless, many proposed quantum technologies require us to go beyond the realm of Gaussian states and introduce non-Gaussian elements. In this Tutorial, we provide a roadmap for the physics of non-Gaussian quantum states. We introduce the phase-space representations as a framework to describe the different properties of quantum states in continuous-variable systems. We then use this framework in various ways to explore the structure of the state space. We explain how non-Gaussian states can be characterised not only through the negative values of their Wigner function, but also via other properties such as quantum non-Gaussianity and the related stellar rank. For multimode systems, we are naturally confronted with the question of how non-Gaussian properties behave with respect to quantum correlations. To answer this question, we first show how non-Gaussian states can be created by performing measurements on a subset of modes in a Gaussian state. Then, we highlight that these measured modes must be correlated via specific quantum correlations to the remainder of the system to create quantum non-Gaussian or Wigner-negative states. On the other hand, non-Gaussian operations are also shown to enhance or even create quantum correlations. Finally, we will demonstrate that Wigner negativity is a requirement to violate Bell inequalities and to achieve a quantum computational advantage. At the end of the Tutorial, we also provide an overview of several experimental realisations of non-Gaussian quantum states in quantum optics and beyond.
\end{abstract}

\maketitle

\tableofcontents

\section{Introduction}


Gaussian states have a long history in quantum physics, which dates back to Schr\"odinger's introduction of the coherent state as a means to study the harmonic oscillator \cite{Schrodinger-Coherent}. In later times, Gaussian states rose to prominence due to their importance in the description of Bose gases \cite{doi:10.1063/1.1704002,robinson_ground_1965,verbeure_many-body_2011} and in the theory of optical coherence \cite{PhysRev.131.2766,PhysRevLett.10.277}. With the advent of quantum information theory, the elegant mathematical structure of Gaussian states made them important objects in the study of continuous-variable (CV) quantum information theory \cite{RevModPhys.77.513,RevModPhys.84.621,doi:10.1142/S1230161214400010}. In this Tutorial, we focus on bosonic systems which means that the continuous variables of interest are field quadratures. Gaussian quantum states are then defined as the states for which measurement statistics of these field quadratures is Gaussian.


Gaussian states can be fully described by their mean field and covariance matrix, and, due to Williamson's decomposition \cite{Williamson}, the latter can be studied using a range of tools from symplectic vector spaces. As such, one can directly relate quadrature squeezing to Gaussian entanglement via the Bloch-Messiah decomposition \cite{PhysRevA.71.055801}. In the full state space of CV systems, Gaussian states are furthermore known to play a specific role: of all possible states with the same covariance matrix, the Gaussian state will always have the weakest entanglement \cite{PhysRevLett.96.080502} and the highest entropy \cite{PhysRevA.59.1820}. From a theoretical point of view, Gaussian quantum states provide, thus, an elegant and highly relevant framework for quantum information theory. On an experimental level, CV quantum information has long been motivated by advances in quantum optics, due to the capability of on-demand generation of ever larger entangled states using either spatial modes \cite{PhysRevA.78.012301,Su:12, PhysRevApplied.14.044025} or time-frequency modes \cite{roslund_wavelength-multiplexed_2014,PhysRevLett.112.120505,cai-2017,Asavanant:2019aa,Larsen:2019aa,yang2021squeezed}. Furthermore, Gaussian states also play a key role in the recent demonstration of a quantum advantage with Gaussian Boson Sampling \cite{Zhong1460}. These developments have made the CV quantum optics an important platform for quantum computation \cite{Bourassa2021blueprintscalable}.\\

Regardless of all the experimental and theoretical successes of Gaussian states, they have a major shortcoming in the context of quantum technologies: all Gaussian measurements of such states can be efficiently simulated \cite{PhysRevLett.88.097904}. In pioneering work on CV quantum computation, it is already argued that a non-Gaussian operation is necessary to implement a universal quantum computer in CV \cite{PhysRevLett.82.1784}. Later works that laid the groundwork for CV measurement-based quantum computing have left the question of this non-Gaussian operation somewhat in the open \cite{PhysRevLett.97.110501,gu_quantum_2009,PhysRevLett.112.120504}. Common schemes, based on the cubic phase gate, turn out to be particularly hard to implement in realistic setups \cite{arzani_polynomial_2017}. Furthermore, these protocols require highly non-Gaussian states, such as Gottesman-Kitaev-Preskill (GKP) states \cite{PhysRevA.64.012310}, to encode information. Even though such states could also serve as a non-Gaussian resource for implementing non-Gaussian gates \cite{PhysRevLett.123.200502}, these states remain notoriously challenging to produce. In spite of the practical problems involved with non-Gaussian states, one is obliged to venture into non-Gaussian territory to reach a quantum computational advantage in the CV regime \cite{mari_positive_2012}. This emphasises the importance of a general understanding of non-Gaussian states and their properties. In this Tutorial, we attempt to provide a roadmap to navigate within this quickly developing field.\\

In Section \ref{sec:CVSystems}, we take an unusual start to introduce CV systems. We first present some elements many-boson physics, by treating Fock space. This mathematical environment is probably familiar to most readers to describe photons. We then explain how such a Fock space can also be described in phase space, which is the more natural framework from CV quantum optics. We will introduce phase-space representations of states and observables in CV systems such as the Wigner function, and to familiarise the reader with the language of multimode systems. By first reviewing the basics of Fock space, we can make interesting connections between what is known as the discrete-variable (DV) approach and the CV approach to quantum optics. We will see that there is often a shady region between these two frameworks, where techniques that are typically associated with one framework can be applied in the other.We will finally argue that the main distinction lies in whether one measures photons (DV) or field quadratures (CV).

In Section \ref{sec:NonGaussianStates}, we provide the reader with an introduction to some of the different structures that can be identified in the space of CV quantum states. When pure states are considered, all non-Gaussian states are known to have a non-positive Wigner function \cite{HUDSON1974249,Soto1983}, but this no longer holds when mixed states enter the game \cite{PhysRevA.79.062302}. In the entirety of the state space, non-Gaussian states occupy such a vast territory that it is impossible to describe all of them within one single formalism. Nevertheless, there has recently been considerable progress in the classification of non-Gaussian states \cite{PhysRevLett.123.043601,PhysRevLett.124.063605}. We will introduce some key ideas behind quantum non-Gaussianity, the stellar rank, and Wigner negativity as tools to characterise non-Gaussian states.

Section \ref{sec:CreationNonGauss} introduces two main families of techniques to create non-Gaussian states starting from Gaussian inputs. The first approach concentrates on deterministic methods, which rely on the implementation of non-Gaussian unitary transformations. We will show how such transformations can be built by using a specific non-Gaussian gate. We then introduce the second class of techniques which are probabilisitic and rely on performing non-Gaussian measurements on a Gaussian state and conditioning on a certain measurement result. We introduce our recently developed approach to describe these systems \cite{PRXQuantum.1.020305} and present mode-selective photon subtraction as a case study.

Then all the pieces are set to discuss the interplay between non-Gaussian effects and quantum correlations in Section \ref{sec:quantumCorr}. First, we will consider the resources that are required to conditionally prepare certain non-Gaussian states. The conditional scheme relies on performing a non-Gaussian measurement on one part of a bipartite Gaussian state, and will show that the nature of the quantum correlations in this bipartite state is essential. We will show that we can only generate quantum non-Gaussian states if the initial bipartite state is entangled. Furthermore, to conditionally generate Wigner negativity we even require quantum steering. In the second part of Section \ref{sec:quantumCorr}, we show how non-Gaussian operations can in return enhance or create quantum correlations. Finally, we will show that Wigner negativity (in either the state or the measurement) is necessary to violate Bell inequalities in CV systems.

In a similar fashion, we will spend most of Section \ref{sec:Advantages} to explain the result of Ref.~\cite{mari_positive_2012}, which shows that Wigner negativity is also necessary to reach a quantum advantage. To show this, we explicitly construct a protocol to efficiently simulate the measurement outcomes of a setup with states, operations, and detectors that are described by positive Wigner functions. In the remainder of the section, we provide comments on the quantum computational advantage reached with Gaussian Boson Sampling.

Finally, in Section \ref{sec:Exp}, we provide a quick overview of non-Gaussian states in CV experiments. Due to the author's background, the first half of this overview will focus on quantum optics.In the second part, we also discuss some key developments in other branches of experimental quantum physics. Readers should be warned that this is by no means an extensive review of all the relevant experimental progress. A more general conclusion and outlook on what the future may have in store is presented in Section \ref{sec:conclusions}.

\section{Continuous-Variable Quantum States}\label{sec:CVSystems}

Before we can start our endeavour to classify non-Gaussian states of CV systems and study their properties, we must develop some basic formalism for dealing with multimode bosonic systems. At the root of bosonic systems lies the canonical commutation relation, $[\hat x, \hat p] \sim i \mathds{1}$, which can be traced back to the early foundations of quantum mechanics. The study of the algebra of such non-commuting observables has given birth to rich branches of mathematics and mathematical physics that ponder on the subtleties of these observables and their associated states. In this Tutorial, we will keep a safe distance from the representation theory of the associated $C^*$-algebras that describes bosonic field theories in their most general sense. We do refer interested readers to a rich but technical literature \cite{verbeure_many-body_2011,petz_invitation_1990,bratteli_operator_1987,bratteli_operator_1997}. 

In this Tutorial, we will exclusively work within the Fock representation, which implies that we consider systems with a finite expectation value for the number of particles. In quantum optics, this assumption translates to the logical requirement that energies remain finite. There are many approaches to mathematically construct such systems (luckily for us they are all equivalent \cite{stone_linear_1930,von_neumann_eindeutigkeit_1931,stone_one-parameter_1932,von_neumann_uber_1932}). Here, we briefly present two such approaches that nicely capture one of the key dualities on quantum physics. First we will take the particle approach by introducing the Fock space that describes identical bosonic particles in Subsection \ref{sec:Fock}. Subsequently, in Subsection \ref{sec:Phase}, we take the approach that starts out from a wave-picture, by concentrating on the phase-space representation of the electromagnetic field. Here we will also introduce the phase-space representations of CV quantum states that will prove to be crucial tools in the remainder of this Tutorial. We will show how these approaches are quite naturally two sides of the same coin. In Subsection \ref{sec:Modes}, we briefly discuss the concept of modes and the role they play in CV quantum systems. This subsection is both intended to provide some clarification about common jargon and to eliminate common misconceptions. We finish this section by presenting a brief case study of Gaussian states in Subsection \ref{sec:Gaussian}, reviewing some key results. After all, it is difficult to appreciate the subtleties of non-Gaussian states without having a flavour from their Gaussian counterparts.

\subsection{Fock Space}\label{sec:Fock}

In typical quantum mechanics text books, the story of identical particles usually starts by considering a set of $n$ particles, which are each described by a quantum state vector in a single-particle Hilbert space $\cal H$, thus for the $i$th particle we ascribe a state vector $\ket{\psi_i} \in \cal H$. The joint state of these $n$ particles is then given by the tensor product of the state vectors $\ket{\psi_1}, \dots, \ket{\psi_n}$. However, if the particles are identical in all their internal degrees of freedom, we should be free to permute them without changing the observed physics. Formally, such permutation is implemented by a unitary operator $U_{\sigma}$, for the permutation $\sigma \in S_n$, which acts as
\begin{equation}
U_{\sigma} \ket{\psi_1}\otimes \dots \otimes \ket{\psi_n} = \ket{\psi_{\sigma(1)}}\otimes \dots \otimes\ket{\psi_{\sigma(n)}}.
\end{equation}
Invariance of physical observables under such permutations can be achieved by either imposing the $n$-particle states vector to be fully symmetric (bosons) or fully antisymmetric (fermions) under these permutations of particles. In this Tutorial, we focus exclusively on bosons, and thus the condition that must be imposed to obtain a bosonic $n$-particle state is
\begin{equation}\label{eq:symmetryCond}
U_{\sigma} \ket{\Psi^{(n)}} = \ket{\Psi^{(n)}}.
\end{equation}
Because these are the only states that are permitted to described the bosonic system, we commonly use the Hilbert space ${\cal H}^{(n)}_s$ which is a subspace of ${\cal H}^{\otimes n}$ that contains only those states that fulfil (\ref{eq:symmetryCond}). It is usually convenient to generate these spaces with a set of elementary tensors, known as Fock states, which we define as
\begin{equation}
\ket{\psi_1}\vee \dots \vee \ket{\psi_n} \coloneqq \sum_{\sigma \in S_n}  \ket{\psi_{\sigma(1)}}\otimes \dots \ket{\psi_{\sigma(n)}},
\end{equation}
such that
\begin{equation}\label{eq:nPhotonSpace}
{\cal H}^{(n)}_s = \overline{{\rm span} \big\{\ket{\psi_1}\vee \dots \vee \ket{\psi_n} \mid \ket{\psi_i} \in {\cal H}\big\}},
\end{equation}
where we refer to Appendix \ref{sec:Span} for some further details on the span. This fully describes a system of $n$ bosonic particles in what is often referred to as first quantisation. It is interesting to note that these identical particles appear to be entangled with respect to the tensor product structure of ${\cal H}^{\otimes n}$. There is still debate on whether this is a mathematical artefact of our description or rather a genuine physical feature of identical particles \cite{PhysRevX.10.041012}. Nevertheless, it is undeniable that this structure leads to physical interference phenomena that do not exist for distinguishable particles \cite{tichy_interference_2014}.\\

The name ``first quantisation'' suggests the existence of a second quantisation, which turns out to be more appropriate for this Tutorial. Second quantisation finds its origins in models where particle numbers are not fixed or conserved. This formalism is largely based on creation and annihilation operators, denoted $\hat a^{\dag}$ and $\hat a$, respectively, that add or remove particles. To accommodate these operators in our mathematical framework, we must equip our Hilbert space to describe a varying number of particles. Therefore, we introduce the Fock space
\begin{equation}\label{eq:FockH}
\Gamma({\cal H}) \coloneqq {\cal H}^{(0)}_s\oplus {\cal H}^{(1)}_s \oplus {\cal H}^{(2)}_s \oplus \dots,
\end{equation}
where the single-particle Hilbert space is given by ${\cal H}^{(1)}_s = {\cal H}$. Furthermore, we retrieve a peculiar component $ {\cal H}^{(0)}_s$ which describes the fraction of the system that contains no particles at all. On its own, ${\cal H}^{(0)}_s$ is thus populated by a single state $\ket{0}$ that we refer to as the vacuum. This implies that technically ${\cal H}^{(0)}_s \cong \mathbb{C}$ the zero-particle Hilbert space is just described by a complex number that corresponds to the overlap of the state with the vacuum. A general pure state in Fock space $\ket{\Psi} \in \Gamma({\cal H})$ can then be described using the structure \eqref{eq:FockH} as
\begin{equation}
\ket{\Psi} = \Psi^{(0)}\oplus \Psi^{(1)}\oplus \Psi^{(2)}\oplus \dots,
\end{equation}
where $\Psi^{(i)} \in {\cal H}^{(i)}_s$ are non-normalised vectors (and therefore we omit the $\ket{.}$) in the $i$-particle Hilbert space. Because $\ket{\Psi}$ is a state, we must impose the normalisation condition $\norm{\Psi}^2 = \sum_{i=0}^{\infty} \norm{\Psi^{(i)}}^2 = 1$ 

We can now define a creation operator $\hat a^{\dag}(\phi)$ for every $\phi \in {\cal H}$ \footnote{Note that the notation is not a coincidence, we can indeed construct a formal linear map from the single-particle Hilbert space into the operator algebra of linear operators on the Fock space. This mapping takes single-particle state vectors $\psi \in {\cal H}$ and maps them to a creation operators. The mapping is linear in the sense that $\hat a^{\dag}(x \psi +y \phi) = x \hat a^{\dag}(\psi) + y \hat a^{\dag}(\phi)$, for all $x,y \in \mathbb{C}$}, which acts as
\begin{equation}\label{eq:CreationOp}
\hat a^{\dag}(\phi) \ket{\Psi} = 0 \oplus \left(\Psi^{(0)}\ket{\phi}\right)\oplus \left( \ket{\phi}\vee\Psi^{(1)}\right)\oplus \left(\ket{\phi}\vee \Psi^{(2)}\right)\oplus \dots
\end{equation}
In the same spirit, it is possible to provide an explicit construction of the annihilation operators $\hat a(\phi)$, but here we will content ourselves by just introducing the annihilation operator as the hermitian conjugate of the creation operator. Just as the creation operator that literally adds a particle to the system, the annihilation operator literally removes one. One additional property of the annihilation operators is that they destroy the vacuum state:
\begin{equation}
\hat a(\phi) \ket{0} = 0.
\end{equation}
We can now use creation and annihilation operators to build an arbitrary Fock state by creating particles on the vacuum state
\begin{equation}
\ket{\psi_1}\vee \dots \vee \ket{\psi_n} = \hat a^{\dag} (\psi_1)\hat a^{\dag} (\psi_2) \dots \hat a^{\dag} (\psi_n) \ket{0}
\end{equation}
and by considering superpositions of such Fock states, we can ultimately generate the entire Fock space. By considering any basis of the single-particle Hilbert space $\cal H$ and constructing all possible Fock states of all possible lengths that can be formed by generating particles in these basis vectors we construct a basis of the Fock space $\Gamma(\cal H)$. We refer to this basis as the Fock basis.

The beauty of second quantisation lies in the natural appearance of states which have no fixed particle number. The most important example is the coherent state
\begin{equation}\label{eq:10}
\ket{\alpha} \coloneqq e^{-\frac{\norm{\alpha}^2}{8}} \sum_{j=0}^\infty \frac{[\hat a^{\dag} (\alpha) ]^j}{2^j j!} \ket{0},
\end{equation}
where $\alpha \in {\cal H}$ is a non-normalised vector in the single particle Hilbert space. One can, indeed, simply generalise \eqref{eq:CreationOp} to non-normalised vectors in ${\cal H}$ which we use explicitly in \eqref{eq:10}. Second, we note that an unusual factor $2$ was included to make the definition consistent with (\ref{eq:CoherentStatePhase}). 

In quantum optics, these ciherent states are crucial objects as they describe perfectly coherent light \cite{PhysRev.131.2766}. It is important to remark that a coherent state is always generated by a single vector in the single-particle Hilbert space. Coherent states often provides a good approximation for the state that is produced by a single-mode laser far above threshold \cite{mandel_wolf_1995}. More generally, the study of laser light is a whole field in its own right and often the light deviates from the fully coherent approximation.\\

The creation and annihilation operators are not only important objects because they populate the Fock space; they are also of key importance for describing observables in a many-boson system. These operators are the  generators of the algebra of observables that represents the canonical commutation relations on Fock space. This implies that any observable can ultimately be approximated by a polynomial of creation and annihilation operators. At the heart of this mathematical formalism lies the canonical commutation relation (CCR):
\begin{equation}
[\hat a(\phi) , \hat a^{\dag}(\psi)] = \langle \phi \mid \psi \rangle,
\end{equation}
which describes the algebra of observables. Note that this relation holds for any vectors $\ket{\phi}$ and $\ket{\psi}$ in the single-particle Hilbert space ${\cal H}$. These vectors should not form a basis, nor should they be orthogonal. When $\ket{\phi} = \ket{\psi}$, we find that $[\hat a(\psi) , \hat a^{\dag}(\psi)] = 1$. On the other hand, when the single-particle states $\ket{\phi}$ and  $\ket{\psi}$ are fully orthogonal, we find that $[\hat a(\phi) , \hat a^{\dag}(\psi)] = 0$. In these cases we recover the typical creation and annihilation operators for harmonic oscillators. However, by introducing the creations and annihilation operators through \eqref{eq:CreationOp}, we can also deal with more general cases. Furthermore, all definitions and the form of the CCR are still valid when $\phi$ and $\psi$ are unnormalised vectors in ${\cal H}$. A more detailed discussion can be found in \cite{Walschaers_2020}.

When we leave the realm of pure states, the description of quantum states becomes tedious. Commonly, one uses a density operator $\hat \rho$ with $\tr \,\hat \rho =1$ to formally describe a state. However, we can generally think of these density operators as infinite-dimensional matrices with an infinite number of components in the Fock basis. In other words, this is not necessarily a convenient description. In an operational sense, any state is considered to be characterised when we know all the moments of all the possible observables. Because the creation and annihilation operators generate the algebra, one knows all the moments of all the observables if one knows all the correlation functions $\tr[\hat \rho\, \hat a^{\dag}(\psi_1) \dots \hat a^{\dag} (\psi_n) \hat a(\phi_1) \dots \hat a (\phi_m) ]$, for all possible lengths $n$ and $m$. Even though this might seem like an equally challenging endeavour, much of quantum statistical mechanics boils down to finding expressions of the correlation functions for relevant classes of states.\\

\subsection{Phase Space}\label{sec:Phase}
In the previous subsection, we started our analysis by extending a system of one quantum particle to a system of many quantum particles. Here we follow a different route, where we start by considering the classical electric field. With some effort, we can apply such an analysis to any bosonic field, but in this Tutorial we focus on quantum optics as our main field of application. For a more extensive introduction from a quantum optics perspective we recommend Ref.~\cite{mandel_wolf_1995,BookSchleich,RevModPhys.92.035005}, whereas a general introduction to quantum physics in phase space can be found in Ref.~\cite{de_almeida_weyl_1998}.

A travelling electromagnetic wave is described by a solution of Maxwell's equations. As is commonly the case in optics, we focus on the complex representation of the electric field which is generally given by ${\bf E}^{(+)}({\bf r},t)$. It is related to the real-valued electric field ${\bf E}({\bf r},t)$ that is encountered in standard electrodynamics textbooks by ${\bf E}({\bf r},t) = {\bf E}^{(+)}({\bf r},t)+\left({\bf E}^{(+)}({\bf r},t)\right)^*$. To express the electric field, it is useful to introduce an orthonormal mode basis $\{{\bf u}_i({\bf r},t)\}$. These modes are solutions to Maxwell's equations
\begin{align}
&\nabla \cdot {\bf u}_i({\bf r},t) = 0,\\
&\left(\Delta - \frac{1}{c^2} \frac{\partial^2}{\partial t^2}\right){\bf u}_i({\bf r},t) = 0.
\end{align}
The orthogonalisation property is implemented by the following condition
\begin{equation}
\frac{1}{V}\int_V {\rm d}^3{\bf r} \left({\bf u}_i({\bf r},t)\right)^* {\bf u}_j({\bf r},t) = \delta_{i,j},
\end{equation}
where $V$ is some large volume that contains the entire physical system. This assumption serves the practical purpose of allowing us to consider a discrete mode basis and on top it makes physical sense. Note that we do not integrate over $t$, which implies that at every instant of time $t$ we consider a mode basis that is normalised with respect to the spatial degrees of freedom. It is practical to assume that all relevant physics can be described by a (possibly large) finite number of modes $m$. These modes now form a basis in which we can expand any solution to Maxwell's equations and thus we may write 
\begin{equation}
{\bf E}^{(+)}({\bf r},t) = \sum_{j=1}^m {\cal E}_j {\bf u}_j({\bf r},t),
\end{equation}
where ${\cal E}_i$ are a set of complex numbers which can be written in terms of the real and imaginary parts 
\begin{equation}\label{eq:quadratures}
{\cal E}_j = E^{(x)}_j + i E^{(p)}_j.
\end{equation}
These real and imaginary parts of the field are known as the amplitude and phase quadrature, respectively. We can interpret these quantities $\vec{E} \coloneqq (E^{(x)}_1, E^{(p)}_1, \dots,E^{(x)}_m, E^{(p)}_m) \in \mathbb{R}^{2m}$ as the coordinate in optical phase space that describes the light field.

The space of solutions of Maxwell's equations forms a Hilbert space which we will call the mode space ${\cal M}$ and the mode basis chosen to describe this space is far from unique. As with all Hilbert spaces, we can define unitary transformations and use them to change from one basis to another. As such, let us introduce the unitary operator $U$ to change between bases
\begin{align}
&{\bf u}_i({\bf r},t) = \sum_{j=1}^m U_{ji}{\bf v}_j({\bf r},t),\\
&{\bf v}_i({\bf r},t) = \sum_{j=1}^m U^{\dag}_{ji}{\bf u}_j({\bf r},t),\label{eq:vUu}
\end{align}
where we can in principle obtain $U$ as an infinite-dimensional matrix with
\begin{equation}
U_{ji} = \frac{1}{V}\int_V {\rm d}^3{\bf r} \left({\bf v}_j({\bf r},t)\right)^* {\bf u}_i({\bf r},t),
\end{equation}
which remarkably does not depend on time due to the normalisation properties of the mode bases. We can analogously expand the electric field in the new mode basis 
\begin{equation}\label{eq:FieldConst}
{\bf E}^{(+)}({\bf r},t) = \sum_i {\cal E}'_i {\bf v}_i({\bf r},t),
\end{equation}
where ${\cal E}'_i = \sum_j U_{ij} {\cal E}_j$. This observation is of great importance when we quantise the electric field. The change of mode basis also imposes a change of coordinates in the optical phase space. Like in (\ref{eq:quadratures}) the new components can also be divided in real and imaginary parts, which leads to a new coordinate $\vec E '$. Because the coordinate vectors in optical phase space are real $2m$-dimensional vectors, we obtain
\begin{equation}\label{eq:quadratureTranspform}
\vec E ' = O \vec E,
\end{equation}
where $O$ is an orthonormal transformation. However, the orthogonal transformation $O$ on the phase space must correspond to the unitary transformation $U$ on the modes, which imposes the constraint
\begin{align}
&O_{2i-1,2j-1} = \frac{1}{2} (U_{ij}+U_{ij}^*), \\
&O_{2i-1,2j} = -\frac{1}{2i} (U_{ij}-U_{ij}^*),\\
&O_{2i,2j-1} = \frac{1}{2i} (U_{ij}-U_{ij}^*),\\
&O_{2i,2j}  = \frac{1}{2} (U_{ij}+U_{ij}^*).
\end{align} 
This imposes a symplectic structure to the transformation $O$ such that the optical phase space, just like the phase space of analytical mechanics, can be treated as a symplectic space. The conserved symplectic structure associated with this space is given by
\begin{equation}
\Omega = \bigoplus_{j=1}^m \omega, \text{ with } \omega = \begin{pmatrix} 0 & -1 \\ 1 & 0\end{pmatrix},
\end{equation}
such that $O^T\Omega O = \Omega$. Note that $\Omega$ can be interpreted as a matrix representation of the imaginary $i$, in the sense that it has the properties $\Omega^T = - \Omega$ and $\Omega^2 = - \mathds{1}$ \footnote{In literature, one will encounter various different choices for the symplectic structure which correspond to different forms of ordering the amplitude and phase quadratures. In general, any matrix $J$ that satisfies the conditions $J^T = - J$ and $J^2 = - \mathds{1}$ defines a symplectic structure. A popular alternative to the ordering $(E^{(x)}_1, E^{(p)}_1, \dots,E^{(x)}_m, E^{(p)}_m)$ that we follow in this Tutorial is the ordering $(E^{(x)}_1, \dots,E^{(x)}_m, E^{(p)}_1, \dots , E^{(p)}_m)$. The choice in this Tutorial is motivated by the study of entanglement, where it is more convenient to group the quadratures that correspond to the same modes together.}. \\

In quantum optics, the electric field of light is treated as a quantum observable $\hat {\bf E}^{(+)}({\bf r},t)$. In this quantisation, the modes, i.e., the normalised solutions to Maxwell's equations, remain classical objects and all the quantum features are absorbed in the coefficients. We can thus write
\begin{equation}
\hat {\bf E}^{(+)}({\bf r},t) = \sum_i {\cal E}^{(1)}_i \frac{\hat x_i + i \hat p_i}{2} {\bf u}_i({\bf r},t),
\end{equation}
where ${\cal E}^{(1)}_i$ is a constant that carries the dimensions of the field, which can be interpreted as the electric field of a single photon. Glossing over many subtleties of the quantisation of the electromagnetic field, we recall the reader that any system that is described on phase space can be quantised through canonical quantisation. The quadrature operators $\hat x_j$ and $\hat p_k$ therefore follow the canonical commutation relations $[\hat x_j, \hat p_k] = 2i\delta_{j,k}$, such that they satisfy the Heisenberg relation $\Delta \hat x \Delta \hat p \geqslant 1$. As they are introduced above, the quadrature operators are specifically related to the specific mode basis. Indeed, $\hat x_j$ and $\hat p_j$ are the quadrature operators that describe the field in mode ${\bf u}_j({\bf r},t)$. Thus, when we change the basis of modes, we should change the quadrature operators accordingly in line with (\ref{eq:quadratureTranspform}). To overcome these difficulties, it is often convenient to introduce a basis-independent expression for the quadrature operators, which can be done by mapping any point in the optical phase space $\vec f \in \mathbb{R}^{2m}$ to an observable $\hat q (\vec{f})$, given by
\begin{equation}\label{eq:GeneralQ}
\hat q (\vec{f}) \coloneqq \sum_{j=1}^m f_{2j-1} \hat x_j + f_{2j} \hat p_j. 
\end{equation}
These quadrature operators follow a generalised version of the canonical commutation relation (CCR), given by
\begin{equation}\label{eq:CCRHere}
[\hat q (\vec{f}_1) , \hat q (\vec{f}_2) ] = - 2i \vec{f}_1^T \Omega\vec f_2 , \text{ for all } \vec f_1, \vec f_2 \in \mathbb{R}^{2m} ,
\end{equation}
We highlight the particular case where $[\hat q (\vec{f}) , \hat q (\Omega\vec{f})] = 2i \norm{\vec f}^2$, such that we recover the typical form of the CCR for $\norm{\vec f} = 1$. This highlights that $\Omega$ maps an amplitude quadrature to its associated phase quadrature. From a mathematical point of view, everything is perfectly well-defined for arbitrary $\vec f \in \mathbb{R}^{2m}$ and no normalisation conditions have to be imposed. From \eqref{eq:GeneralQ} we can see that the norm of $\vec f$ can be factored out, such that it serves as general rescaling factor of the quadrature operator. In a physical context, when a quadrature is measured, it is common to renormalise measurements to units of vacuum noise, which practically means that we set $\norm{\vec f} = 1$. Unless explicitly stated otherwise, we assume that $\norm{\vec f} = 1$ throughout this Tutorial. We can use these general quadrature operators to express electric field operator as
\begin{equation}\label{eq:fieldGenQuad}
\hat {\bf E}^{(+)}({\bf r},t) = \sum_{j=1}^m {\cal E}^{(1)}_j \frac{ \hat q (\vec{e}_j) + i \hat q (\Omega\vec{e}_j)}{2} {\bf u}_j({\bf r},t),
\end{equation}
and we can use the basis transformation (\ref{eq:quadratureTranspform}) to equivalently express the electric field operator in a different mode basis as
\begin{equation}
\hat {\bf E}^{(+)}({\bf r},t) = \sum_{j=1}^m {\cal E}^{(1)}_j \frac{ \hat q (O\vec{e}_j) + i \hat q (\Omega O \vec{e}_j)}{2} {\bf v}_j({\bf r},t).
\end{equation}
This procedure shows us that optical elements, that change the mode basis, change the associated quadrature operators accordingly. 

Equation (\ref{eq:fieldGenQuad}) shows us explicitly that quadrature operators $\hat q (\vec{f})$ and $\hat q (\Omega\vec{f})$ correspond to the same mode, regardless of the mode basis. This reflects the fact that $\vec f$ generates one axis in the optical phase space and $\Omega \vec f$ generates the second axis that corresponds to the same mode. As such, any arbitrary mode comes with an associated two-dimensional phase space that mathematically can be denoted as ${\rm span}(\vec f, \Omega \vec f)$. Because this phase space is uniquely associated with a specific mode, we introduce the notation
\begin{equation}
{\bf f} = {\rm span} (\vec f, \Omega \vec f),
\end{equation}
and we refer to this as ``mode ${\bf f}$''. This allows us to concentrate on the multimode quantum states within this Tutorial, while the specifications of the modes can be left ambiguous. The modes can be seen as the physical implementations of the quantum system and are of major importance in the experimental setting as multimode quantum optics experiments rely on the manipulation of these modes.\\ 

Multimode quantum states define expectation values of the field, and when we consider CV quantum optics, we primarily focus on the expectation values of the quadrature operators $\hat q (\vec{f})$. These operators are unbounded and have a continuous spectrum. The measurement of a field quadrature thus leads to a continuum of possible outcomes and the continuous variable approach to quantum optics implies that this characterises quantum properties of light through the measurement of such quadrature operators.

Formally, we can again describe a quantum state on such a system by a density operator $\hat \rho$ but this description is rather inconvenient. It turns out that the quadrature operators $\hat q (\vec{f})$ generate the algebra of observables for the quantum system that is comprised within our multimode light. In other words, any observable can be approximated by a polynomial of quadrature operators. This generally implies that we can fully characterise the quantum state $\hat \rho$ by correlation functions of the type $\tr [\hat \rho \hat q (\vec{f}_1) \dots \hat q (\vec{f}_n)]$. When we know these correlation functions for all lengths $n$ and normalised vectors in phase space, we have fully characterised the state. 

To go beyond the information that is contained in correlation functions, it is often convenient to consider probability distributions as a whole. For a single quadrature $\hat q (\vec f)$ we can introduce the characteristic function for any $\lambda \in \mathbb{R}$ as
\begin{equation}
\chi(\lambda) = \tr [\hat \rho e^{i \lambda \hat q (\vec f)}] = \sum_{n=0}^{\infty} \frac{(i\lambda)^n}{n!}\tr[ \hat \rho { \hat q (\vec f)}^n],
\end{equation}
which is clearly related to the moments $\tr[ \hat \rho { \hat q (\vec f)}^n]$. The characteristic function is the Fourier transform of the probability distribution of the outcomes of observable $\hat q (\vec f)$. We can thus obtain the probability distribution as
\begin{equation}\label{eq:px}
p(x) = \frac{1}{2\pi} \int_{\mathbb{R}}{\rm d}\lambda \, \chi(\lambda) e^{-i x\lambda}.
\end{equation}
This approach can be readily generalised to the joint probability distribution for a set of commuting quadrature operators. We thus consider $\vec f_1, \dots \vec f_n$ with $[\hat q (\vec{f}_j) , \hat q (\vec{f}_k) ] = 0$ for all $j,k$, and we define for all $\vec \lambda = \lambda_1\vec f_1+ \lambda_2\vec f_2 + \dots + \lambda_n \vec f_n$ (note that $\vec \lambda$ is not normalised). We can then use the properties of the quadrature operators to construct $\hat q(\vec\lambda) = \sum_{k=1}^n \lambda_k \hat q (\vec f_k)$ and define the function
\begin{equation}
\chi(\vec \lambda) = \tr [\hat \rho e^{i \hat q (\vec \lambda)}].
\end{equation}
This function generates all the correlations between observables $\hat q (\vec{f}_1), \dots , \hat q (\vec{f}_n)$ and it can be used to obtain the multivariate probability distribution
\begin{equation}\label{eq:pxmulti}
p(\vec x) = \frac{1}{(2\pi)^{n}}\int_{\mathbb{R}^n} {\rm d}\vec \lambda\, \chi(\vec \lambda) e^{-i \vec\lambda^T \vec x},
\end{equation}
where ${\rm d}\vec \lambda = {\rm d}\lambda_1\dots {\rm d}\lambda_n$. The function $p(\vec x)$ describes the probability density to jointly obtain $x_1, \dots, x_n$ as measurement outcomes for the measurements of $\hat q (\vec{f}_1), \dots , \hat q (\vec{f}_n)$, respectively. This approach relies on the fact that commuting observables can be jointly measured and what we presented to derive (\ref{eq:px}) and (\ref{eq:pxmulti}) is ultimately just classical probability theory. However, not all quadrature operators commute such that joint measurements are not always possible. This implies that a quantum state cannot be straightforwardly defined by a probability distribution of the optical phase space.

Intriguingly, we can carry out the same procedure for a full multimode system over a set of $m$ modes. To this goal, let us define the quantum characteristic function 
\begin{equation}
\chi: \mathbb{R}^{2m} \rightarrow \mathbb{C} : \vec \lambda \mapsto \chi(\vec \lambda) \coloneqq \tr [\hat \rho e^{i \hat q (\vec \lambda)}].
\end{equation}
This function, defined on the full optical phase space, can be used to generate all correlation functions between all quadrature operators. As such, it does characterise the full quantum state, but it is common practice to rather study its inverse Fourier transform which is known as the Wigner function \cite{PhysRev.40.749,HILLERY1984121,doi:https://doi.org/10.1002/3527602976.ch3}
\begin{equation}\label{eq:wigDef}
W(\vec x) \coloneqq \frac{1}{(2\pi)^{2m}}\int_{\mathbb{R}^{2m}} {\rm d}\vec \lambda\, \chi(\vec \lambda) e^{-i \vec\lambda^T \vec x}.
\end{equation}
This function has many appealing properties even though it is not a probability distribution but rather a quasiprobability distribution. First of all, the Wigner function is normalised, i.e., $\int_{\mathbb{R}^{2m}}{\rm d}\vec x \, W(\vec x) = 1$. Furthermore, its marginals consistently describe all the joint probability distributions for sets of commuting quadratures in the system. Formally, this implies that $p(\vec x)$ of (\ref{eq:pxmulti}) can be obtained by integrating over all the phase space axes that are not contained within ${\rm span} (\vec f_1, \dots \vec f_n)$. To do so, let us introduce the $n$-dimensional vector $\vec x_M$ that is associated with the measured quadratures, and the $2m-n$ dimensional vectors $\vec x_c$ which describe all other axes in phase space. An arbitrary point in phase space can thus be written as $\vec x = \vec x_M \oplus \vec x_c$. Then we find that 
\begin{equation}
P(\vec x_M) = \int_{\mathbb{R}^{2m - n}} {\rm d}\vec x_c \, W(\vec x_M \oplus \vec x_c).
 \end{equation}
 Finally, the Wigner function also produces the correct expectation values
\begin{equation}\label{eq:momentsFromWigner}
\int_{\mathbb{R}^{2m}} {\rm d}\vec x \, \vec f_1^T \vec x\dots \vec f_n^T \vec x  W(\vec x) = {\rm Re}\{\tr[\hat \rho \hat q (\vec{f}_1) \dots \hat q (\vec{f}_n)]\}.
 \end{equation} 
Note that considering the real part of $\tr[\hat \rho \hat q (\vec{f}_1) \dots \hat q (\vec{f}_n)$ is essentially equivalent to considering symmetric ordering of the operators. Regardless of these nice properties the Wigner function is by itself not a well-defined probability distribution. Due to complementarity, the function can reach negative values for some states. This Wigner negativity is consistent with the impossibility to jointly describe the measurement statistics of all quadratures while also complying with the laws of quantum physics (notably the Heisenberg relation). The profound relation between negativity of the Wigner functions and joint measurability is perhaps most strikingly illustrated by its connection to contextuality \cite{PhysRevLett.101.020401}.
.
The formalism of Wigner functions can be used to construct phase-space representations of arbitrary observables by introducing
\begin{equation}
\chi_A(\vec \lambda) = \tr[\hat A e^{i \hat q (\vec \lambda)}],
\end{equation}
such that the Wigner representation is given by
\begin{equation}\label{eq:WigA}
W_{A}(\vec x) = \frac{1}{(2\pi)^{2m}}\int_{\mathbb{R}^{2m}}{\rm d}\vec \lambda \, \chi_A(\vec \lambda) e^{-i \vec\lambda^T \vec x}.
\end{equation}
These Wigner representations have the appealing property that
\begin{equation}\label{eq:wignerProduct}
\tr [\hat A \hat \rho]= (4\pi)^m \int_{\mathbb{R}^{2m}} {\rm d}\vec x \,W^*_{A}(\vec x)W(\vec x).
\end{equation}
When $\hat A$ is an observable and thus has $\hat A = \hat A^{\dag}$, its Wigner function will be real such that $W^*_{A}(\vec x)=W_{A}(\vec x)$. However, it may sometimes be useful to extend the formalism to more general operators. As such the entire theory of continuous-variable quantum systems can be developed using Wigner functions. \\

Several aspects of the phase-space representations in this section are reminiscent of earlier results in Section \ref{sec:Fock} which was fully developed in a language of particles (also known as a discrete-variable approach). Indeed, the algebra of operators that is generated by the creation and annihilation operators is actually the same as the algebra generated by the quadrature operators. To formalise this, we must first stress that the optical phase space is isomorphic to an $m$-dimensional complex Hilbert space which can equally be interpreted as the single-particle Hilbert space of a photon. Formally, this equivalence is constructed through the bijection (see also Appendix \ref{sec:TopologicalVectorSpace})
\begin{equation}\label{eq:PhaseHilb}
\vec f \in \mathbb{R}^{2m} \mapsto \sum_{j} (f_{2j-1} +i f_{2j})\ket{\phi_j} \in \mathcal{H},
\end{equation}
where $\{\ket{\phi_j}\}$ is an arbitrary basis of $\mathcal{H}$.
We can introduce the operators 
\begin{equation}\label{creaAnn}\begin{split}
&\hat a (\vec f) = \frac{1}{2}[\hat q (\vec f) + i\hat q (\Omega\vec f)], \\
&\hat a^{\dag} (\vec f) = \frac{1}{2}[\hat q (\vec f) - i\hat q (\Omega\vec f)].
\end{split}
\end{equation}
By using (\ref{eq:PhaseHilb}), we can naturally associate these operators to creation and annihilation operators on the single particle Hilbert space. We retrieve the canonical commutation relation
\begin{equation}
[\hat a (\vec f_1) , \hat a^{\dag} (\vec f_ 2)] = \vec f_1^T \vec f_2 - i \vec f_1^T \Omega \vec f_2,
\end{equation}
which can be connected to the inner product on the Hilbert space ${\cal H}$ via (\ref{eq:PhaseHilb}).

The definition of creation and annihilation operators allows us to make sense of the vacuum state in our phase space picture. The vacuum state is completely characterised by the property 
\begin{equation}
\hat a (\vec f) \ket{0} = 0, \text{ for all } \vec f \in \mathbb{R}^{2m}.
\end{equation}
This simple fact can be used to evaluate the quantum characteristic function
\begin{align}
\chi_0(\vec \lambda) &= \tr (\ket{0}\bra{0} e^{i [\hat a^{\dag}(\vec \lambda) + \hat a (\vec \lambda) ]}) \\
&= \sum_{n=0}^\infty \frac{i^n}{n!}  \bra{0}[\hat a^{\dag}(\vec \lambda) + \hat a (\vec \lambda) ]^n\ket{0}\\
&=\sum_{n=0}^\infty -\frac{\norm{\vec \lambda}^{2n}}{2^nn!}\label{eq:combinatorics}\\
&=\exp\left[-\frac{\norm{\vec \lambda}^2}{2}\right].\label{eq:characVacuum}
\end{align}
To obtain equation (\ref{eq:combinatorics}) we need a considerable amount of combinatorics to evaluate $\bra{0}[\hat a^{\dag}(\vec \lambda) + \hat a (\vec \lambda) ]^n\ket{0}$. In general, it can be shown that $\bra{0}\hat a^{\dag}(\vec \lambda_1)\dots \hat a^{\dag}(\vec \lambda_{k})\hat a(\vec \lambda_{k+1}) \dots \hat a(\vec \lambda_{k+l}) \ket{0} = 0$. Thus it suffices to cast $[\hat a^{\dag}(\vec \lambda) + \hat a (\vec \lambda) ]^n$ in normal ordering and extract the term proportional to identity. Even though straightforward, this calculation is quite cumbersome and thus we do not present the details.

From (\ref{eq:characVacuum}) the Wigner function can be obtained via an inverse Fourier transformation that leads to
\begin{equation}\label{eq:WigVac}
W_0(\vec x) = \frac{e^{-\frac{1}{2} \norm{\vec x}^2}}{2\pi}. 
\end{equation}
This Wigner function  describes a Gaussian distribution on the phase space with unit variance along every axis. We can thus use (\ref{eq:momentsFromWigner}) to see that the vacuum state saturates Heisenberg's inequality, i.e., $\Delta \hat q(\vec f) \Delta \hat q(\Omega \vec f) = 1$.\\ 

The quadrature operators thus generate the same algebra of observables as the creation and annihilation operators. However, both sets of observables tend to cause mathematical problems because they are unbounded operators \cite{vonNeumannBook,WeylBook,Conway1985,bratteli_operator_1997}. The unboundedness means that, when $\ket{\Psi}$ is contained in the Fock space, there is no guarantee that $\hat q(\vec f) \ket{\Psi}$ will also be contain in the Fock space. One way of solving this problem explicitly is by only considering states for which $\bra{\Psi} \hat q(\vec f)^2 \ket{\Psi} < \infty$, such that $\hat q(\vec f) \ket{\Psi}$ is a well-defined state. Physically this assumption makes sense, as it ultimately implies that we only consider states with finite energies. However, the unboundedness of quadrature operators also disqualifies them as well-defined generators of the $C^*$-algebra of observables (since elements of such algebras must be bounded). $C^*$-algebras are essential tools as they allow to reconstruct the whole framework of Hilbert spaces based on representation theory of abstract algebras (which is essentially the idea of canonical quantisation). A highly formal and detailed treatment that considers all these subtleties for bosonic systems is found in \cite{bratteli_operator_1987}. The key idea is to rather consider a set of bounded operators that describes the same algebra of observables \cite{vonNeumannBook,WeylBook} and are known as the displacement operators:
\begin{equation}\label{eq:DispOp}
\hat D(\vec \alpha) = e^{- i \hat q(\Omega\vec \alpha)/2},
\end{equation}
where $\vec \alpha \in \mathbb{R}^{2m}$ need not be normalised.
Again, we can use the isomorphism (\ref{eq:PhaseHilb}) to identify the displacement operator on the quantised phase space to a displacement operator on Fock space. These operators can be seen as generators of the quadrature operators and they act in a very natural way on them:
\begin{equation}\label{eq:dispQuad}
\hat D^{\dag}(\vec{\alpha}) \hat q (\vec f) \hat D (\vec \alpha) = \hat q (\vec f) + \vec \alpha^T \vec f,
\end{equation}
which means that the value $ \vec \alpha^T \vec f$ is added to the measurement outcomes of $\hat q (\vec f)$. The displacement operator can be combined according to the rule
\begin{equation}\label{eq:CombDis}
\hat{D}(\vec \alpha_1)\hat{D}(\vec \alpha_2) = \hat{D}(\vec \alpha_1+\vec \alpha_2)e^{\frac{i}{4} \vec \alpha_1^T \Omega \vec \alpha_2}.
\end{equation}
This rule is yet another representation of the canonical commutation relation and it generates the same algebra of observables. This implies that any observable $\hat A$ can be written as a linear combination of displacement operators. We use the Hilbert-Schmidt inner-product $\<\hat A, \hat B\>_{\rm HS} = \tr [\hat A^{\dag} \hat B]$ to make this explicit
\begin{equation}\begin{split}
\hat A &= \int_{\mathbb{R}^{2m}}{\rm d}\vec \lambda \, \< \hat D(2\Omega\vec \lambda), \hat A\>_{\rm HS} \hat D(2\Omega\vec \lambda)\\
&= \int_{\mathbb{R}^{2m}}{\rm d}\vec \lambda \, \tr[\hat A \hat D(-2\Omega\vec \lambda)] \hat D(2\Omega\vec \lambda),
\end{split}
\end{equation}
and we can readily identify that
\begin{equation}
\tr[\hat A \hat D(-2\Omega\vec \lambda)] = \chi^*_A(\vec \lambda).
\end{equation}
It can then directly be seen that
\begin{align}
\tr[\hat A \hat \rho] &= \int_{\mathbb{R}^{2m}}{\rm d}\vec \lambda \, \chi^*_A(\vec \lambda) \tr[\hat D(2\Omega\vec \lambda) \hat \rho]
\\&=\int_{\mathbb{R}^{2m}}{\rm d}\vec \lambda \, \chi^*_A(\vec \lambda) \chi(\vec \lambda).
\end{align}
And we immediately obtain (\ref{eq:wignerProduct}) via Plancherel's theorem \cite{Plancherel,Conway1985}.

The displacement operators also implement a unitary operation on a quantum state. This unitary operation has a remarkably simple effect when it is expressed on the level of the Wigner function. Via the property (\ref{eq:CombDis}), we can calculate that
\begin{equation}
\hat \rho \mapsto \hat D(\vec \alpha)\hat \rho \hat D^{\dag}(\vec \alpha) \implies \chi(\vec \lambda) \mapsto \chi(\vec \lambda)e^{-i \vec \alpha^T \vec \lambda}.
\end{equation}
Performing the inverse Fourier transform of these quantum characteristic functions leads to
\begin{equation}\label{eq:WigDisp}
W(\vec x) \overset{D(\vec \alpha)}{\mapsto} W(\vec x - \vec{\alpha}).
\end{equation}
The displacement operator thus literally implements a displacement of the Wigner function by a vector $\vec \alpha \in \mathbb{R}^{2m}$ in phase space.\\

Displacement operators are also well-known as the generators of the coherent states that were introduced in (\ref{eq:10}). We can combine the bijection between phase space and Hilbert space (\ref{eq:PhaseHilb}), the expression of creation and annihilation operators in terms of quadratures (\ref{creaAnn}), and the definition of the displacement operator (\ref{eq:DispOp}) to derive that
\begin{equation}\label{eq:CoherentStatePhase}
\ket{\vec\alpha} = \hat D(\vec\alpha)\ket{0}.
\end{equation}
By combining (\ref{eq:WigVac}) and (\ref{eq:WigDisp}), we immediately see that the Wigner function for such a coherent state is given by
\begin{equation}
W_{\alpha}(\vec x) = W_{0}(\vec x - \vec \alpha) = \frac{e^{-\frac{1}{2} \norm{\vec x - \vec{\alpha}}^2}}{2\pi}.
\end{equation}
We emphasise that there is a slight difference between the coherent states as defined here, and coherent states as sometimes introduced in literature. The difference is a factor two, which appears because we normalised the shot noise to $1$ rather than to $1/2$. As such, our coherent states have an energy in mode ${\bf f}$ which is given by
\begin{equation}
\bra{\vec \alpha} \hat a^{\dag} (\vec f)a(\vec f) \ket{\vec \alpha} = \frac{1}{4} [(\vec \alpha^T \vec f)^2 + (\vec \alpha^T \Omega \vec f)^2].
\end{equation} 
The coherent states lead us to two other representations of quantum states and observables: the Q-function and P-function. The definition of the P-function is related to the idea that coherent states form an overcomplete basis of Fock space. This implies, notably, that for a $m$-dimensional single particle Hilbert space
\begin{equation}
 \frac{1}{(4\pi)^m} \int_{\mathbb{R}^{2m}} {\rm d}\vec\alpha \ket{\vec\alpha}\bra{\vec\alpha} = \mathds{1},
\end{equation}
with $\mathds{1}$ the identity operator. We can then show that any observable can be written as \cite{PhysRev.131.2766,PhysRevLett.10.277}
\begin{equation}
\hat A =  \frac{1}{(4\pi)^m} \int_{\mathbb{R}^{2m}} {\rm d}\vec\alpha P_A(\vec \alpha) \ket{\vec\alpha}\bra{\vec\alpha},
\end{equation}
where we refer to $P_{A}(\vec\alpha)$ as the P-function of the observable $\hat A$. Similarly, we can represent a density of operator $\hat \rho$ by its P-function $P(\vec\alpha)$. The reader should be warned that P-functions often have rather unpleasant mathematical properties. In particular, they often are not actual functions and can be highly singular. 

The P-function naturally comes with a dual representation that is known as the Q-function. As often in quantum physics, what actually counts is the expectation value of an observable in a specific state. We can use the P-function to write
\begin{align}
\tr [\hat A \hat \rho] &=  \frac{1}{(4\pi)^m} \int_{\mathbb{R}^{2m}} {\rm d}\vec \alpha  P_A(\vec \alpha ) \bra{\vec\alpha}\hat \rho \ket{\vec\alpha}\\
&=  \frac{1}{(4\pi)^m}\int_{\mathbb{R}^{2m}} {\rm d}\vec \alpha  P(\vec \alpha ) \bra{\vec\alpha}\hat A \ket{\vec\alpha}.
\end{align}
This naturally introduces the Q-function, given by
\begin{equation}
Q_{A}(\vec \alpha)= \frac{1}{(4\pi)^m}\bra{\vec\alpha}\hat A \ket{\vec\alpha},
\end{equation}
and in particular for the quantum state $\hat \rho$ we find that
\begin{equation}\label{eq:Q1}
Q(\vec \alpha) = \frac{1}{(4\pi)^m}\bra{\vec\alpha}\hat \rho \ket{\vec\alpha}.
\end{equation}
The latter is of particular interest because it represents the quantum state $\hat \rho$ as an actual probability distribution. This leads us to the general identity that
\begin{equation}\label{eq:ABQP}
\tr [\hat A \hat B] = \int_{\mathbb{R}^{2m}} {\rm d}\vec \alpha  P_A(\vec\alpha ) Q_{B}(\vec\alpha) = \int_{\mathbb{R}^{2m}} {\rm d}\vec \alpha  Q_A(\vec \alpha ) P_{B}(\vec \alpha).
\end{equation}
Thus finishing our introduction to the various descriptions of the quantum states and observables of bosonic many-particle systems.\\

The Q-function has a clear physical interpretation. It is directly proportional to the fidelity of the state $\hat \rho$ with respect to a target coherent state $\ket{\vec\alpha}$. Furthermore, it is always positive, which implies that it is a well-defined probability distribution. Because we can write $\bra{\vec\alpha}\hat \rho \ket{\vec\alpha} = \tr [ \hat \rho \ket{\vec\alpha}\bra{\vec\alpha}]$, we can use (\ref{eq:wignerProduct}) and (\ref{eq:WigVac}) to express the Q-function in terms of the Wigner function as
\begin{equation}\label{eq:Q2}
Q(\vec \alpha) = \int_{\mathbb{R}^{2m}} {\rm d}\vec x \, W(\vec x) W_0(\vec x - \vec \alpha).
\end{equation}
To satisfy both (\ref{eq:wignerProduct}) and (\ref{eq:ABQP}), we find that
\begin{align}
W(\vec x) &=\frac{1}{(4\pi)^m} \int_{\mathbb{R}^{2m}} {\rm d}\vec \alpha \, P(\vec \alpha) W_0(\vec x - \vec \alpha)\\
 &=  \frac{1}{ (4\pi)^m}\int_{\mathbb{R}^{2m}} {\rm d}\vec \alpha \, P(\vec \alpha) \frac{e^{-\frac{1}{2}\norm{\vec x - \vec \alpha}^2}}{(2\pi)^{m}}.
\end{align}
In turn, this implies that
\begin{equation}
Q(\vec \alpha) = \frac{1}{(8\pi)^m}\int_{\mathbb{R}^{2m}} {\rm d}\vec \beta \, P(\vec \beta) \frac{e^{-\frac{1}{4}\norm{\vec \beta - \vec \alpha}^2}}{(2\pi)^{m}}.
\end{equation}
These results thus show that all these phase-space representations are ultimately related to one another through convolution or deconvolution with a Gaussian (recall that the Wigner function of the vacuum (\ref{eq:WigVac}) is a Gaussian distribution on phase space). One can now follow Ref.~\cite{PhysRev.177.1882} to define a continuous family of phase-space representations ${\cal W}_{\sigma}$ for $\sigma \in [-1,1]$
\begin{equation}
{\cal W}_{\sigma}(\vec \alpha) = \left(\frac{1}{4\pi[1-\sigma]}\right)^{m} \int_{\mathbb{R}^{2m}} {\rm d}\vec \beta \, P(\vec \beta) \frac{e^{-\frac{1}{2[1-\sigma]}\norm{\vec \beta - \vec \alpha}^2}}{(2\pi)^{m}},
\end{equation}
where we convolute the P-function with an ever increasing Gaussian, smoothening its features.
We can then see that
\begin{equation}
\tr[\hat A \hat \rho] = (4\pi)^m \int_{\mathbb{R}^{2m}}{\rm d}\vec{\alpha}\, {\cal W}_{A, -\sigma}(\vec \alpha) {\cal W}_{\sigma}(\vec \alpha).
\end{equation}
We find, notably, that $W(\vec x) = {\cal W}_{\sigma=0}(\vec x)$, $Q(\vec \alpha) = {\cal W}_{\sigma=-1}(\vec \alpha)$, and $P(\vec \alpha) = (4\pi)^m {\cal W}_{\sigma=1}(\vec \alpha)$. This shows that the phase-space representation becomes more regular when decreasing $\sigma$. Other generalised probability distributions have been considered in literature \cite{RevModPhys.17.195,Sperling_2020}, often to circumvent the unappealing properties of the P-function. In this Tutorial, we mainly use the Wigner function and (to a lesser extent) the Q-function, as they are suitable tools to classify non-Gaussian quantum states. The P-function is often used in literature to characterise the non-classicality of a state, where the intuition is that classical light is a mixture of coherent states (and thus its P-function is a probability distribution) \cite{PhysRev.131.2766,PhysRevLett.10.277}.\\

Before we close this introductory section on the phase space description of CV quantum systems, we introduce one final tool that often comes in handy. The Wigner function can itself be obtained as the expectation value of an operator \cite{PhysRevA.15.449}. Formally, we write
\begin{equation}\label{eq:wigexpvalueparity}
W(\vec x) = \frac{1}{(2\pi)^m} \tr [\hat \rho \hat \Delta (\vec x)].
\end{equation}
Using linearity and (\ref{eq:wigDef}), we obtain the special case
\begin{equation}
\hat \Delta (\vec 0) = \frac{1}{(2\pi)^{m}}\int_{\mathbb{R}^{2m}} {\rm d}\vec \lambda\,  e^{i\hat q (\vec \lambda)}.
\end{equation}
By using techniques based on (\ref{eq:dispQuad}) and (\ref{eq:CombDis}), we can show that
\begin{equation}
\hat \Delta (\vec 0)\hat q (\vec f) \hat \Delta (\vec 0) = - \hat q (\vec f).
\end{equation}
This means that $\hat \Delta (\vec 0)$ is the parity operator. Its eigenstates are the Fock states, since 
\begin{equation}
\hat \Delta (\vec 0) a^{\dag}(\vec f_1) \dots a^{\dag}(\vec f_n) \ket{0} =  (-1)^n a^{\dag}(\vec f_1) \dots a^{\dag}(\vec f_n) \ket{0}.
\end{equation}
Thus we can formally identify 
\begin{equation}
\hat \Delta (\vec 0) = (-\mathds{1})^{\hat N},
\end{equation}
where $\hat N$ is the number operator. We define this operator by introducing a mode basis $\{\vec e_1, \Omega \vec e_1, \dots , \vec e_m, \Omega \vec e_m\}$ of the optical phase space, such that $\hat N \coloneqq \sum_{j=1}^m a^\dag(\vec e_j)a(\vec e_j)$. This definition can be combined with the properties of the displacement operator to obtain that
\begin{equation}\label{eq:Parity}
\hat \Delta (\vec x) = \hat D(-\vec x) (-\mathds{1})^{\hat N} \hat D(\vec x).
\end{equation}
Note that what we just obtained is the operator equivalent of a $\delta$-function, which becomes even more explicit when we explicitly write down its Wigner representation
\begin{equation}
W_{\Delta (\vec x')}(\vec{x}) = \frac{1}{(4\pi)^m} \delta (\vec x - \vec x '),
\end{equation}
which follows directly from (\ref{eq:wignerProduct}) and (\ref{eq:wigexpvalueparity}).

This result may seem somewhat artificial, but it turns out to be extremely useful. The observable $\hat \Delta (\vec x)$ can be measured experimentally by counting photons, which means that the combination of photon counting and displacements directly allows us to reconstruct the Wigner function of the quantum state \cite{PhysRevLett.76.4344}. Up to recently, the lack of good photon-number-resolving detectors in the optical frequency range has long made this method unfeasible for most states. Even though there was an early demonstration of the method for coherent states \cite{PhysRevA.60.674}, it is only due to recent developments in detector technologies that the method the method can be applied to more general states \cite{PhysRevLett.105.253603,Nehra:19}. The idea was also other settings \cite{PhysRevLett.78.2547}, and was used in pioneering CV experiments with trapped ions, such as \cite{PhysRevLett.77.4281}, and in cavity-QED \cite{PhysRevA.62.054101,PhysRevLett.89.200402}.

\subsection{Discrete and continuous variables}\label{sec:Modes}

In Subsection \ref{sec:Fock}, we have introduced a many-boson system, regardless of the physical realisation of these bosons. Such a many-boson system and its Fock space are built upon the structure that is determined by the single-particle Hilbert space ${\cal H}$. The Fock space that is constructed accordingly has a rich structure that is further explored in the Tutorial \cite{Walschaers_2020}. In optics, the bosons that we consider are photons, and quantum optics can thus be seen as the theory of a many-boson system in the context of \ref{sec:Fock}. This approach to quantum optics is referred to as the DV approach.

In Subsection \ref{sec:Phase}, we contrast this with the CV approach to quantum optics. This approach relies on the measurement of the field quadratures, and can thus been seen as a bosonic quantum field theory. Therefore we started this approach by introducing the classical electric field and its modes, which we subsequently quantised through canonical quantisation. We introduced the notion of optical phase space as a general way of describing CV quantum systems. The phase space is directly related to the modes of the field and manipulations of the modes also cause changes in the optical phase space. Nevertheless, any system with a phase space can be described by these techniques. 

Both of these approaches are ultimately equivalent. Bosonic creation and annihilation operators describe the same algebra of observables as bosonic quadrature operators, which means that on the level of mathematical structure, both approaches can be interchanged and even mixed. This is strikingly clear when the Wigner function, i.e, the phase-space representation of quantum states and observables that is most naturally associated with field quadratures, turns out to be directly measurable by counting photons. Notably, this implies that when it comes to mathematical structures, bosonic particles such as atoms can also be described on phase space. 

The real difference between CV and DV approaches is of an experimental nature. What is important is not the observables that are technically present in the quantum system, but the observables that are practically measured in the lab. For the CV approach we typically use homodyne detection to measure quadratures \cite{RevModPhys.77.513,lvovsky2020production}, whereas in DV approaches we count photons \cite{Flamini_2018}.\\

A common source of misunderstanding between the DV and CV community stems from the role they attribute to the single-particle Hilbert space and optical phase space, respectively. As we argued, both spaces are (at least for a finite-dimensional number of modes) isomorphic, see Appendix \ref{sec:TopologicalVectorSpace} for some additional mathematical intuition. However, the Hilbert space of a photon, which is inherently a quantum particle, is often interpreted as a quantum object. At the same time, the optical phase space represents the field quadratures of optical modes and is thus rather considered to be a classical object. The origin of this confusion lies in the fact that the optical modes, i.e., normalised solutions of Maxwell's equations, also form a Hilbert space that has its origins entirely in classical physics. 

The optical modes are the vessels that contain photons much in the same way as a set of electrons contains spins. The crucial difference is that optical modes are not uniquely defined, we can manipulate them, transform them from one mode basis to another with an interferometer and thus consider new superpositions of modes. In typical experimental settings, one would not consider a superposition of two electrons a new well-defined electron. 

Because creation and displacement operators always act in one specific mode (i.e., they are generated by a single vector on the single-photon Hilbert space), single-photon states and coherent states are always single-mode states. We may expand this single mode in a different mode basis, which can even be done physically by sending the state through a beamsplitter, to create some form of entanglement in the quantum states. However, this entanglement is just a manifestation of the fact that we are not considering the optimal mode basis. In the CV approach, this has led to the notion of ``intrinsic'' properties \cite{RevModPhys.92.035005}, which are those properties of quantum states that are independent of the chosen mode basis. The purity and entropy of a state are notable examples, but one can also introduce a notion of ``intrinsic entanglement'' to refer to a state that is entangled in any possible mode basis. In the next subsection, we introduce Gaussian states which will later be shown to never be intrinsically entangled.

\subsection{Gaussian States}\label{sec:Gaussian}

Now that we have introduced phase-space representations for states and observables of CV quantum systems, we still need one building block before we can tackle multimode non-Gaussian states: a good understanding of Gaussian states. It is not the goal of this subsection to delve deep into decades worth of research on Gaussian states. We rather highlight a few key results that set apart Gaussian quantum states from the rest of the vast states space. For more extended reviews, we refer the reader to \cite{RevModPhys.84.621,doi:10.1142/S1230161214400010}. These states are also extensively studied in the mathematical physics literature under the name ``quasi-free states of the CCR algebra''.

Gaussian states are by definition states that have a Wigner function which is a Gaussian:
\begin{equation}\label{eq:GaussStateWig}
W_G(\vec x) = \frac{e^{-\frac{1}{2}(\vec x - \vec \xi)^TV^{-1}(\vec x - \vec \xi)}}{(2\pi)^m \sqrt{\det V}},
\end{equation}
where $\vec \xi$ is referred to as the mean field (or displacement) and $V$ is known as the covariance matrix. With (\ref{eq:momentsFromWigner}), we can verify that the mean field indeed corresponds to the expectation value of the field quadrature
\begin{equation}
\tr [\hat \rho \hat q (\vec f)] = \vec \xi^T \vec f, 
\end{equation}
similarly, we find for the covariance matrix
\begin{equation}\begin{split}\label{eq:covQuad}
&\tr [\hat \rho \hat q (\vec f_1)\hat q (\vec f_2)] - \tr [\hat \rho \hat q (\vec f_1)]\tr [\hat \rho \hat q (\vec f_2)] \\
&= \vec f_1^T V \vec f_2 - i \vec f_1^T \Omega \vec f_2.
\end{split}
\end{equation} 
These quantities can thus be obtained for arbitrary quantum states, but for Gaussian states the covariance matrix and the mean field also determine all higher order expectation values. The most elegant way to see this is via the multivariate cumulants (also known as truncated correlation functions), which vanish beyond order two \cite{verbeure_many-body_2011}. This fact implies that all properties of Gaussian states can ultimately be deduced from their mean field and -often more importantly- from their covariance matrix.

Hitherto, we have encountered the vacuum state $\ket{0}$ and the coherent states $\ket{\vec \alpha}$ as examples of Gaussian states. Both of the examples have a covariance matrix $V = \mathds{1}$. However, there is much larger range of possible covariance matrices available and they have to satisfy certain constraints \cite{PhysRevA.49.1567}. At first instance, we note that a covariance matrix must be positive. An additional constraint is obtained by imposing that the variance $\Delta^2 \hat q (\vec f) \geqslant 0$ for all $\vec f$ in phase space. Eq.~(\ref{eq:covQuad}) then directly yields that $\vec f^T (V - i \Omega) \vec f \geqslant 0$, which implies that $V \geqslant 0$ and suggests that $(V - i \Omega) \geqslant 0$. However, the latter is not obvious, since $\vec f$ are real vectors, whereas $(V - i \Omega)$ is a complex matrix. We thus need an additional ingredient: the Heisenberg inequality. Formally, this inequality can be obtained through Robertson's more general inequality \cite{PhysRev.34.163}, such that we find:
\begin{equation}
\Delta^2 \hat q (\vec f_1)\Delta^2 \hat q (\vec f_2)  \geqslant  \frac{1}{4}\abs{\tr\{\hat \rho [\hat q (\vec f_1), \hat q (\vec f_2) ]\}}^2.
\end{equation}
We can now apply the CCR \eqref{eq:CCRHere} to obtain the general form
\begin{equation}
\Delta^2 \hat q (\vec f_1)\Delta^2 \hat q (\vec f_2)  \geqslant \abs{\vec f_1^T \Omega \vec f_2}^2.
\end{equation}
On the other hand, the definition \eqref{eq:covQuad} of the covariance matrix can be used to translate this result to
\begin{equation}
\vec f_1^T V \vec f_1 \, \vec f_2^T V \vec f_2  \geqslant \abs{\vec f_1^T \Omega \vec f_2}^2.
\end{equation}
This identity can then be used to prove $(\vec f_1^T - i \vec f_2^T) (V - i \Omega) (\vec f_1 + i \vec f_2) \geqslant 0$ for all $\vec f_1, \vec f_2 \in \mathbb{R}^{2m}$. As a consequence, we find that
\begin{equation}
V - i \Omega \geqslant 0,
\end{equation}
an important constraint on the covariance matrix $V$, which can be understood as combining the positivity conditions and the Heisenberg inequality. 

To further understand the structure of covariance matrices and the Gaussian states that they describe, we highlight some important results on symplectic matrices. The first of these results is Williamson's decomposition \cite{Williamson}, which states that any positive-definite real matrix $V$ can be diagonalised by a symplectic matrix $S$ (i.e., a matrix with $S^T \Omega S = \Omega$):
\begin{equation}\label{eq:Williamson}
V = S^T N S, \text{ with } N = {\rm diag}[\nu_1, \nu_1, \nu_2,\nu_2, \dots, \nu_m, \nu_m].
\end{equation}
The values $\nu_1, \dots \nu_m$ are also known as the symplectic spectrum of $V$. From Heisenberg's relation, we then find the additional constraint that $\nu_1, \dots, \nu_m \geqslant 1$, in other words the values in the symplectic spectrum are larger than shot noise. It now becomes straightforward to see the Heisenberg's relation also implies that
\begin{equation}
\det V \geqslant 1.
\end{equation}
It thus becomes apparent that the Heisenberg inequality is saturated when $\det V = 1$. The states for which this is the case must have a covariance matrix $V = S^T S$.

The Gaussian states for which the Heisenberg inequality is saturated turn out to be the pure Gaussian states. Recall that the purity of a quantum state is given by $\mu = \tr [\hat \rho^2]$. This quantity can be directly calculated from the Wigner function via (\ref{eq:wignerProduct}). We then find for an arbitrary Gaussian state
\begin{equation}\label{eq:purityGaussian}
\mu_G = (4\pi)^m \int_{\mathbb{R}^{2m}}{\rm d}\vec x\, W_G(\vec x)^2 = \frac{1}{\sqrt{\det V}}.
\end{equation}
Alternatively, we may use the symplectic spectrum to express $\mu_G = \prod_{k=1}^m \nu_k^{-1}$. This shows us that a Gaussian state is pure if and only if its covariance matrix is a positive symplectic matrix, i.e., it can be written as $V= S^TS$. 

The class of states with a covariance matrix given by $V= S^TS$ is much larger than just the vacuum and coherent states with $V = \mathds{1}$. The additional states turn out to have asymmetric noise in their quadratures, and because the Heisenberg inequality is saturated this implies that some quadratures have less noise than the vacuum state. The states with such covariance matrices are therefore known as squeezed states. To formalise this intuition, we consider the Bloch-Messiah decomposition (which is known in mathematics and classical mechanics as Euler's decomposition) \cite{Arvind1995,PhysRevA.71.055801}. Any symplectic matrix $S$ can be decomposed as follows:
\begin{equation}\label{eq:BlochMessiah}
S = O_1 K O_2, \text{ with } K = {\rm diag}[s_1^{1/2},s_1^{-1/2}, \dots, s_m^{1/2},s_m^{-1/2}],
\end{equation}
where $O_1$ and $O_2$ are orthogonal symplectic matrices, i.e., $O_j^TO_j = \mathds{1}$ and $O_j^T\Omega O_j = \Omega$. We can then see that for any pure Gaussian state, we find
\begin{equation}
V = S^T S = O^TK^2O.
\end{equation}
We have already encountered orthogonal symplectic transformations in (\ref{eq:quadratureTranspform}), where we associated them with transformations of mode bases. Thus, if we find a set of optical modes that are prepared in a pure Gaussian state, we can always find a different mode basis in which the state is given by
\begin{equation}\label{eq:squeezingSpect}
V' = OVO^T = 
\begin{pmatrix}
s_1 &&&&\\
& 1/s_1 & & &\\
& & \ddots & &\\
& & & s_m &\\
& & & & 1/s_m
\end{pmatrix}.
\end{equation}
This means that we can always find a set of symplectic eigenvectors $\{\vec e_1, \Omega \vec e_1, \dots, \vec e_m, \Omega \vec e_m\}$ of a pure Gaussian state's covariance matrix, which have the properties that $\Delta^2 \hat q(\vec e_j) = s_j$ and $\Delta^2 \hat q(\Omega\vec e_j) = 1/s_j$, such that the Heisenberg relation is saturated: $\Delta^2 \hat q(\vec e_j) \Delta^2 \hat q(\Omega\vec e_j) = 1$. At the same time, we find that clearly either $\Delta^2 \hat q(\vec e_j)$ of $\Delta^2 \hat q(\Omega \vec e_j)$ is smaller than one (and thus below shot noise).\\

Gaussian states naturally come with the notion of Gaussian channels \cite{doi:10.114297818609481690002}, they are the completely-positive trace-preserving transformations that map Gaussian states into other Gaussian states. We have already seen that the displacement operators are unitary transformations that fulfil this condition. Because any Gaussian transformation $\Gamma$ preserves the general shape of the Wigner function (\ref{eq:GaussStateWig}), we can simply describe the Gaussian channel $\Gamma$ in terms of its actions on the mean field and the covariance matrix:
\begin{align}
&V \overset{\Gamma}{\mapsto} X V X^T + V_c,\\
&\vec \xi \overset{\Gamma}{\mapsto} X\vec \xi +\vec \alpha.
\end{align}
The vector $\vec \alpha$ simply serves to displace the entire Gaussian to a different location in phase space. On the level of the covariance matrix, $X$ transforms reshapes the initial covariance matrix, whereas $V_c$ describes the addition of Gaussian classical noise. Both can a priori be any real matrices, as long as they satisfy the constraint
\begin{equation}\label{eq:constraintGaussianChannel}
V_c - i \Omega + i X\Omega X^T \geqslant 0.
\end{equation}
This constraint derives from the demand that $X V X^T + V_c$ is a well-defined covariance matrix, and therefore $X V X^T + V_c - i \Omega \geqslant 0$. Because $V$ is a well-defined covariance matrix, $X (V - i \Omega) X^T \geqslant 0$ and thus it can be seen that $X V X^T + V_c$ is also a well-defined covariance matrix whenever \eqref{eq:constraintGaussianChannel} holds. This simple argument proves \eqref{eq:constraintGaussianChannel} is a sufficient condition for $\Gamma$ to transform the covariance matrix of the initial state into a new bona fide covariance matrix.

An important case is obtained when we impose that $\Gamma$ conserves the purity of the state and is thus a unitary transformation. It then immediately follows that $V_c = 0$, since there cannot be any classical noise. The displacement $\vec \alpha$ is simply implemented by a displacement operator, and the constraint (\ref{eq:constraintGaussianChannel}), combined with the demand that purity is conserved implies that $X$ is a symplectic matrix. In other words, a Gaussian unitary transformation $\hat U_G$ satisfies $V \mapsto S^TVS$.  Another relevant example is the case of uniform Gaussian losses, where we set $X = \sqrt{1-\eta} \mathds{1}$, $V_c = \eta \mathds{1}$ and $\vec \alpha = 0$, with the positive value $\eta \leqslant 1$ denoting the amount of loss.

More generally, the action of a Gaussian channel on an arbitrary state can be understood from its action on $\exp[i \hat q(\vec \lambda)]$, which can be proven to take the form
\begin{equation}\label{eq:channelChar}
\exp[i \hat q(\vec \lambda)] \overset{\Gamma}{\mapsto} \exp\left[i \hat q(X^T\vec \lambda) + i \vec \alpha^T \vec \lambda - \frac{1}{2} \vec\lambda^T V_c \vec\lambda \right].
\end{equation}
We can then calculate the quantum characteristic function and use some properties of Fourier transforms to find that the Wigner function transforms as
\begin{equation}\label{eq:GaussianChannel}
W(\vec x) \overset{\Gamma}{\mapsto} \int_{\mathbb{R}^{2m}} {\rm d}\vec y \, W(X^{-1}\vec x - \vec y) \frac{e^{- \frac{1}{2} (\vec y - \vec \alpha)^T V^{-1}_c (\vec y- \vec \alpha)}}{(2\pi)^m \sqrt{\det V_c}}.
\end{equation}
For Gaussian unitary transformations we find the appealing result that $W(\vec x) \mapsto W(S^{-1}(\vec x - \vec \alpha))$. This means that a Gaussian unitary transformation is simply a coordinate transformation on phase space.

Proving that any completely positive Gaussian channel $\Gamma$ is of the form \eqref{eq:GaussianChannel} with condition \eqref{eq:constraintGaussianChannel}, is a challenging task. The result was first obtained in \cite{demoen_completely_1977,demoen_completely_1979}, using the language of $C^*$-algebras. The proof is rather technical and we will not go into details here.

The paradigm of Gaussian channels is also useful to structure general Gaussian states. One may for example wonder which Gaussian channel would transform the vacuum state into the Gaussian state with covariance matrix $V$. In general, there is no unique solution to this question, but there is a straightforward route to find an answer. First, take any symplectic matrix $S$ that satisfies $V - S^TS \geqslant 0$ (the Williamson decomposition guarantees that this is always possible). This implies that there is a positive-definite matrix $V_c$ such that $V = S^TS + V_c$. As such, an arbitrary Gaussian state can always be decomposed as
\begin{equation}\label{eq:GaussianDecomp}
W_G(\vec x) =  \int_{\mathbb{R}^{2m}} {\rm d}\vec y \, W_0(S^{-1}\vec x - \vec y) e^{- \frac{1}{2} (\vec y - \vec \alpha)^T V^{-1}_c (\vec y- \vec \alpha)}.
\end{equation}
The symplectic operation that is applied to the vacuum is known as multimode squeezing in optics. These transformations are fully equivalent to Bogoliubov transformations that are regularly used in condensed matter physics \cite{RevModPhys.84.621,verbeure_many-body_2011}. Combined with a displacement, this operation provides the most general operation that maps quadrature operators into well-defined quadrature operators.\\

Now that we have introduced the basic concepts of Gaussian states, we are equipped to start exploring their non-Gaussian counterparts. Several other important properties of Gaussian states will be introduced along the way to stress just how peculiar these Gaussian states are compared to the rest of state space.
 
\section{Non-Gaussian Quantum States}\label{sec:NonGaussianStates}
Contrary to Gaussian states with their elegant Wigner function and properties that can be nicely deduced from the covariance matrix, the set of non-Gaussian states is vast and wild. Literally all states with Wigner functions that are not Gaussian are contained within this class. To give an idea of the enormous variety, one can consider that highly exotic states such as Gottesman-Kitaev-Preskill states \cite{PhysRevA.64.012310} and Schr\"odinger cat states inhabit the set of non-Gaussian states together with the states that describe single photons and even certain convex mixtures of Gaussian states. Throughout the years, there have been considerable efforts to structure the set of non-Gaussian states. We will introduce the notion of quantum non-Gaussian states \cite{PhysRevLett.106.200401} and then extend it to a hierarchy based on stellar rank \cite{PhysRevLett.124.063605}. A different approach is provided by considering that the negativity of the Wigner function can be used as a genuine signature of non-classicality \cite{Kenfack:2004aa}. However, before we attack these different measures to structure non-Gaussian quantum states, we contrast some properties of Gaussian and non-Gaussian states.

\begin{figure*}
\centering
\includegraphics[width=0.8\textwidth]{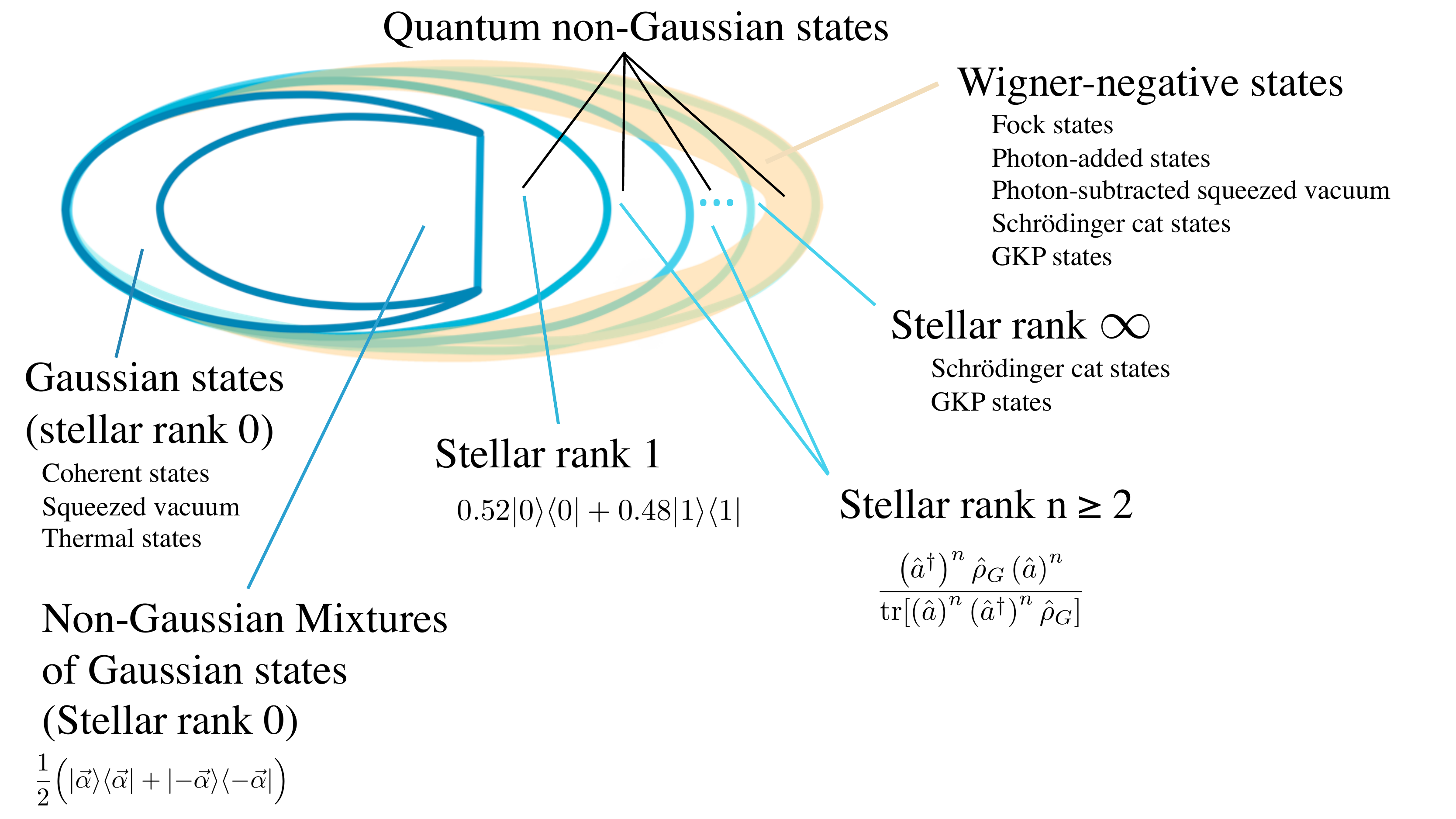}\\
\caption{Overview of the different types of non-Gaussian states that can be found in state space. The different aspects will all be considered throughout Section \ref{sec:NonGaussianStates}. Here we attempt to show the stellar hierarchy and how it differentiates itself from the convex hull (mixtures) of Gaussian states. Furthermore, we emphasise that the stellar rank and Wigner negativity are different quantifiers of non-Gaussianity. It should be noted that all non-Gaussian pure states are Wigner-negative states, but we can find states that are not mixtures of Gaussian states without Wigner negativity. For all the classes, we provide examples of states that belong to this group, the Wigner functions for several of these examples are shown in Fig.~\ref{fig:Examples}.}
 \label{fig:nonGaussianOnion}
\end{figure*}

Fig.~\ref{fig:nonGaussianOnion} provides an overview that can be used as a brief guide to understand the structure of non-Gaussian states. We attempt to highlight how the different quantities used to structure the non-Gaussian part of state space are interconnected.

\subsection{Gaussian vs. Non-Gaussian}
Gaussian states have many extraordinary properties that set them apart from non-Gaussian states.  First of all, pure Gaussian states turn out to be the only quantum states that saturate the uncertainty relation. The easiest way to see this is by describing arbitrary pure states in terms of their wave functions. The wave functions associated with amplitude quadratures $\hat q (\vec f)$ and those associated with phase quadratures $\hat q (\Omega \vec f)$ are related by a Fourier transform. This fact can then be used to show that only Gaussian wave functions saturate the Heisenberg inequality. The extension to arbitrary mixed states can be achieved via Jensen's inequality, which emphasises that no mixed states can saturate the uncertainty relation. Let us consider a mixed state $\hat \rho = \sum_k p_k \ket{\Psi_k}\bra{\Psi_k}$ with variances $\Delta^2 \hat q (\vec f)$. We also introduce the variances $\Delta^2_k \hat q (\vec f)$ for the pure states $ \ket{\Psi_k}$. From Jensen's inequality \cite{10.1007/BF02418571}, it follows that
\begin{equation}
\Delta^2 \hat q (\vec f) \geqslant \sum_k p_k \Delta^2_k \hat q (\vec f).
\end{equation}
For Heisenberg's inequality, we calculate
\begin{equation}\begin{split}
\Delta^2 \hat q (\vec f)\Delta^2 \hat q (\Omega\vec f) \geqslant &\sum_k p_k^2  \Delta^2_k \hat q (\vec f)\Delta^2_k \hat q (\Omega\vec f)\\ 
&+ \sum_{k\neq l} p_k p_l  \Delta^2_k \hat q (\vec f)\Delta^2_l \hat q (\Omega\vec f)\\
& \geqslant 1.
\end{split}\end{equation}
The presence of cross terms highlights that even when all the pure states in the mixture saturate the inequality, the mixture does not. The only possible exception is the case where the state is pure. 

That only pure Gaussian states saturate the Heisenberg inequality may seem like an innocent observation, but it has an important implication for non-Gaussian states. The Heisenberg inequality can be formulated entirely in terms of the covariance matrix. We showed in (\ref{eq:Williamson}) that the inequality is saturated if and only if the covariance matrix is symplectic, i.e., $V=S^TS$. Furthermore, we showed in (\ref{eq:purityGaussian}) that a Gaussian state is pure if and only if its covariance matrix is symplectic $V=S^TS$. The fact that no non-Gaussian states can saturate the inequality thus implies that non-Gaussian states can never have a symplectic covariance matrix $V=S^TS$. This is a first hint of the special role played by Gaussian states.

A more general result along these lines states that for all states $\hat \rho$ with the same covariance matrix $V$, the Gaussian state always has the highest von Neumann entropy \cite{PhysRevA.59.1820}. First of all, note that entropy $-\tr [\hat \rho \log \hat \rho]$ is conserved under unitary transformations. Due to the Williamson decomposition (\ref{eq:Williamson}), we can write any Gaussian state as
\begin{equation}
\hat \rho_G = \hat U_G  \bigotimes_{j=1}^m\hat \rho_{\overline n_j} \hat U_G^{\dag},
\end{equation}
where $\hat \rho_{\overline n_j}$ is a thermal state of the Hamiltonian $\hat a^{\dag}_j \hat a_j$ with average particle number $\overline n_j = (\nu_j - 1)/2$. From statistical mechanics, we know that thermal states are the quantum states that maximise the von Neumann entropy for a given temperature (here fixed by the occupations $\overline n_j$). 

It turns out that Gaussian states are limiting cases for many quantities \cite{PhysRevLett.96.080502}. This result shows that for a range of functionals $f$ on the state space, we find that $f(\hat \rho) \geqslant f(\hat \rho_G)$, where $\hat \rho_G$ is the Gaussian state with the same covariance matrix as $\hat \rho$. Apart from some more technical aspects such as continuity, $f$ must have two important features: it must be conserved under (a certain class of) unitary operations $f(\hat U \hat \rho \hat U^{\dag}) = f(\hat \rho)$ and it must be strongly super-additive $f(\hat \rho)\geqslant f (\hat \rho_1) + f(\hat \rho_2)$ (note that $\hat \rho_1$ and $\hat \rho_2$ are marginals of $\hat \rho$). The equality must be saturated for product states, i.e., $f(\hat \rho_1 \otimes \hat \rho_2) = f (\hat \rho_1) + f(\hat \rho_2)$. For strongly sub-additive functions with $f(\hat \rho)\leqslant f (\hat \rho_1) + f(\hat \rho_2)$ the same result implies $f(\hat \rho) \leqslant f(\hat \rho_G)$, (after all, in that case $-f$ is a strongly super-additive function). It is clear that the von Neumann entropy fulfils the latter conditions and is maximised for Gaussian states. For super-additive entanglement measures, this result can be used to show that for all states with the same covariance matrix, Gaussian states are the least entangled ones (entanglement will be much more extensively discussed in Section \ref{sec:quantumCorr}). However, several common entanglement measures, e.g., the logarithmic negativity \cite{PhysRevA.65.032314} and the entanglement of formation \cite{EntOfFormAdd}, are not super-additive. 

At the heart of these extremal properties lies the central limit theorem \cite{cushen_hudson_1971,QUAEGEBEUR19841,Goderis1989,verbeure_many-body_2011}. There are many versions of the central limit theorem in quantum physics, but we will stick to what is probably the simplest one. As always, we consider our optical phase space $\mathbb{R}^{2m}$, but this time, we will take $N$ copies of it, which implies that we are dealing with a phase space $\mathbb{R}^{2Nm} =\mathbb{R}^{2m} \oplus \dots \oplus \mathbb{R}^{2m}$ for the full system. We can then embed a vector $\vec \lambda  \in \mathbb{R}^{2m}$ in the $j$th of these $N$ copies via $\vec \lambda_j \coloneqq \vec 0\oplus \dots \oplus \vec 0\oplus \vec \lambda \oplus \vec0 \dots \oplus \vec 0$ and introduce the new averaged operator 
\begin{equation}
\overline {q}_N (\vec \lambda) \coloneqq \frac{1}{\sqrt{N}}\sum_{j=1}^N \hat q (\vec \lambda_j).
\end{equation}
It is rather straightforward to see that these observables follow the canonical commutation relation. We can now restrict ourselves to studying the algebra that is generated entirely by such averaged quadrature operators. When we then assume that the different copies of the system are ``independently and identically distributed'' we must set the overall state to be $\hat \rho^{(N)} = \hat \rho^{\otimes N}$. We then find the characteristic function of the algebra of averaged observables by 
\begin{equation}
\chi_N (\vec \lambda) = \tr [\hat \rho^{\otimes N} e^{i \overline {q}_N (\vec \lambda)}]. 
\end{equation}
The following pointwise convergence can be shown:
\begin{equation}
\chi_N (\vec \lambda) \overset{N \rightarrow \infty}{\rightarrow } \chi_G(\vec \lambda),
\end{equation}
where $\chi_G(\vec \lambda)$ is the characteristic function of the Gaussian state $\hat \rho_G$ that has the same covariance matrix as $\hat \rho$. This means that the non-Gaussian features in any state $\hat \rho$ can be coarse grained away by averaging sufficiently many copies of the state. Note that this result considers $N$ copies of an arbitrary $m$-mode state. The single-mode version of this result was proven in \cite{cushen_hudson_1971}, whereas a much more general versions are derived in \cite{QUAEGEBEUR19841,Goderis1989}. In \cite{PhysRevLett.96.080502} the central limit theorem is combined with invariance under local unitary transformations to proof the final extremality result, we will not review these points in detail.\\

The extremality of Gaussian states and the associated central limit theorem highlight why Gaussian states are important in quantum information theory and quantum statistical mechanics. It also shows that Gaussian states have some particular properties compared to non-Gaussian states. It thus should not come as a surprise that some of these properties can be used to measure the degree of non-Gaussianity of the state \cite{PhysRevA.76.042327,PhysRevA.78.060303,PhysRevA.82.052341}. As we mentioned before, for a fixed covariance matrix $V$ the von Neumann entropy is maximised by the Gaussian state. This suggest that we can use the difference in von Neumann entropy as a measure for non-Gaussianity. To formalise things, let us consider an arbitrary state $\hat \rho$ with covariance matrix $V$ and mean field $\vec \xi$ (this quantities can be derived, respectively, for the second and first moments of the quadrature operators). We then construct a Gaussian state $\hat \sigma_V$, has the same covariance matrix and the mean field. In the spirit of extremality, we then define
\begin{equation}\label{eq:nonGauss}
\delta(\hat \rho) =  S(\hat \sigma_V) - S(\hat \rho),
\end{equation}
where $S(\hat \rho) \coloneqq - \tr[\hat \rho \log \hat \rho]$. Because von Neumann entropy is constant under unitary transformations, for a Gaussian state it only depends on the symplectic spectrum $\nu_1, \dots, \nu_m$. In other word, we can calculate $S(\hat \sigma_V)$ directly by using the Williamson decomposition (\ref{eq:Williamson}) on $V$. We find from \cite{PhysRevA.59.1820} that
\begin{equation}
S(\hat \sigma_V) = \sum_{j=1}^m \left[ \frac{\nu_j + 1}{2}\log \frac{\nu_j + 1}{2} - \frac{\nu_j - 1}{2}\log \frac{\nu_j - 1}{2} \right].
\end{equation}
However, it should be noted that the entropy of the non-Gaussian states $S(\hat \rho)$ is generally harder to calculate unless we can accurately approximate the state by a finite density matrix in the Fock basis. Furthermore, if the state $\hat \rho$ is pure, we simply find that $\delta(\hat \rho) =  S(\hat \sigma_V)$.

Due to extremality of Gaussian states it directly follows that $\delta(\hat \rho) \geqslant 0$, but this does not necessarily mean that $\delta(\hat \rho)$ is a good measure for non-Gaussianity. Refs.~\cite{PhysRevA.59.1820,PhysRevA.78.060303} establish that 
\begin{equation}
\delta(\hat \rho) = S(\hat \rho \mid\mid \hat \sigma_V),
\end{equation}
where $S(\hat \rho \mid\mid \sigma_V) \coloneqq \tr[\hat \rho (\log \hat \rho - \log \hat \sigma_V)]$ is the quantum relative entropy between $\hat \rho$ and reference state $\hat \sigma_V$. The quantum relative entropy allows us to connect $\delta(\hat \rho)$ to a range of interesting properties, as shown in \cite{PhysRevA.78.060303}. For example, it directly follow that $\delta(\hat \rho) = 0$ if and only if $\hat \rho = \hat \sigma_V$. Furthermore, the measure $\delta(\hat \rho)$ inherits convexity and monotonicity properties from the relative entropy. These are exactly the properties that made this measure a useful ingredient in the resource theory for quantum non-Gaussianity presented in \cite{PhysRevA.98.052350}. 

Thus, the connection between (\ref{eq:nonGauss}) and relative entropy shows that $\delta(\hat \rho)$ can indeed be used as a measure for non-Gaussianity in the sense that it measures ``entropic distance'' between $\hat \rho$ and $\hat \sigma_V$. Yet, there is one important question that remains to be answered: is $\sigma_V$ indeed the closest Gaussian state to $\hat \rho$? An affirmative answer to this question was provided in Ref.~\cite{PhysRevA.88.012322}, where it was shown that
\begin{equation}
\delta(\hat \rho) = \min_{\hat \rho_G} S(\hat \rho \mid\mid \hat \rho_G),
\end{equation}
where we minimise over all possible Gaussian states $\hat \rho_G$. The main idea of the proof is to show that $S(\hat \rho \mid\mid \hat \rho_G) - S(\hat \rho \mid\mid \hat \sigma_V) = S(\hat \sigma_V \mid\mid \hat \rho_G) \geqslant 0$ such that the smallest relative entropy in indeed achieved for $\hat \sigma_V$. For the technical details, we refer the interested Reader to \cite{PhysRevA.88.012322}. Furthermore, we note that a similar non-Gaussianity measure was introduced by using the Wehrl entropy (based on the Q-function) rather than the von Neumann entropy \cite{WehrlNonGauss}. \\

We have thus shown that Gaussian states are special in the sense that they minimise entanglement and maximise entropy as compared to other states with the same covariance matrix. Another profound distinction can be found when comparing pure Gaussian states to pure non-Gaussian states. In this case, there is a seminal result by Hudson \cite{HUDSON1974249} that was extended by Soto and Claverie to multimode systems \cite{Soto1983} which states that a pure state can have a non-negative Wigner function if and only if the state is Gaussian. In other words, all non-Gaussian pure states exhibit Wigner negativity.

Here, we follow the approach of \cite{PhysRevA.51.3340} to prove this result. First of all, we introduce the function
\begin{equation}\label{eq:Fstellar}
F^{\star}_{\Psi}(\vec \alpha) \coloneqq \langle \vec \alpha \mid \Psi \rangle e^{\frac{1}{8}\norm{\vec\alpha}^2},
\end{equation}
such that the $Q$-function \eqref{eq:Q1} of the state $\ket{\Psi}$ is given by
\begin{equation}\label{eq:QFunction106}
Q(\vec \alpha) = \frac{1}{(4\pi)^m}\abs{F^{\star}_{\Psi}(\vec \alpha)}^2 e^{-\frac{1}{4}\norm{\vec\alpha}^2}.
\end{equation}
From \eqref{eq:Fstellar}, we directly find that
\begin{equation}\label{eq:stellar3}
\abs{F^{\star}_{\Psi}(\vec \alpha)}^2 \leqslant  e^{\frac{1}{4}\norm{\vec\alpha}^2}.
\end{equation}
Next, we observe that a $Q$ function that reaches zero implies a negative Wigner function, which can be seen from \eqref{eq:Q2}. Thus, demanding that the state has a positive Wigner function implies demanding that $Q(\vec \alpha) > 0$, and thus that $F^{\star}_{\Psi}(\vec \alpha)$ has no zeros. Using the equivalence between $2m$-dimensional phase space and a complex $m$-dimensional Hilbert space, we can use the multidimensional but restricted version of the Hadamard theorem \cite{Soto1983}, which states that any entire function $f: \mathbb{C}^m \mapsto \mathbb{C}$ without any zeros and with order of growth \footnote{A function $f$ is said to have an order of growth $r$ when there are constant $a,b \in \mathbb{C}$ such that $\abs{f(z)}\leqslant a \exp(b \abs{z}^r)$ for all $z\in \mathbb{C}$.} $r$ is an exponential $f(z) = \exp g(z)$, where $g(z)$ is a polynomial of degree $s \leqslant r$.  We then note that $F^{\star}_{\Psi}$ is an entire function of maximal growth $r=2$ as given by (\ref{eq:stellar3}). Hadamard's theorem then tells us that $F^{\star}_{\Psi}(\vec \alpha)$ is Gaussian. In other words, if a pure quantum state $\ket{\Psi}$ has a positive Wigner function $F^{\star}_{\Psi}(\vec \alpha)$ must be Gaussian. The only states for which this is the case are Gaussian states.\\

In summary, we have seen that Gaussian states inherit a particular extremal behaviour from the central limit theorem. This should not come as a surprise given that they are the Gibbs states of a free bosonic field at finite temperature. Thus, a non-Gaussian state can be expected to have more ``exotic'' features than the Gaussian state with the same covariance matrix. This also formalises the intuition that Gaussian states are more classical states. This idea is further established by the fact that all pure Gaussian states are the only possible pure states that have a positive Wigner function.

The fact that non-Gaussian states automatically have non-positive Wigner functions no longer holds when mixed states are considered. A simple example is that state $\hat \rho = [\ket{0}\bra{0} + a^{\dag}(\vec f)\ket{0}\bra{0}a(\vec f)]/2$, which is clearly non-Gaussian but also has a positive Wigner function. There have been considerable efforts to extend Hudson's theorem in some form to mixed states \cite{PhysRevA.79.062302}. However, in what follows, we will see that there are many ways for a state to be non-Gaussian. This makes it particularly hard to connect a measure such as (\ref{eq:nonGauss}) to more operational interpretations. In the next section, we start by showing some examples of different non-Gaussian states to make the Reader appreciate their variety.

\subsection{Examples of non-Gaussian states}
\begin{figure*}
\centering
\includegraphics[width=0.8\textwidth]{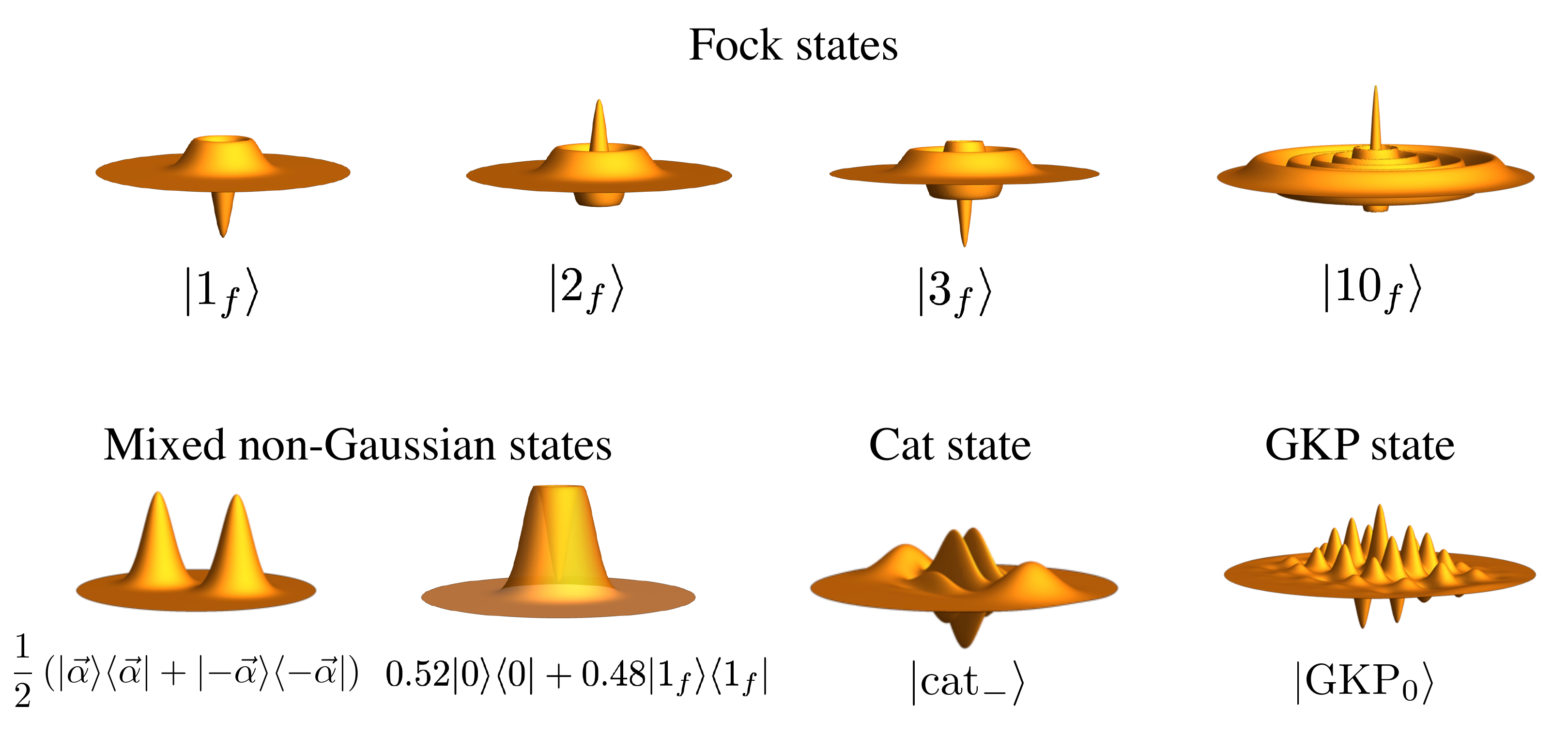}\\
\caption{Several examples of Wigner functions for single-mode non-Gaussian states. The Wigner functions for the Fock states are obtained from (\ref{eq:WigFock}) and the mixture of a Fock state and a vacuum is given by (\ref{eq:WigFockMix}). The mixture of coherent states is expressed in (\ref{eq:coherentMixture}) where we set $\norm{\vec \alpha} = 4$. The Wigner function for the cat state is given by (\ref{eq:WigCat}) where we have chosen $\norm{\vec \alpha} = 6$. Finally, for the GKP state (\ref{eq:GKPstate}), we numerically integrated a wave function expression of the state with $s=2$ and $\delta = 0.3$ to obtain the Wigner function.}
 \label{fig:Examples}
\end{figure*}

 An overview of the different examples discussed in this section if shown in Fig.~\ref{fig:Examples}.\\

A first important class of non-Gaussian states are Fock states, generated by acting with creation operators $a^{\dag}(\vec f)$ on the vacuum state
\begin{equation}\label{def:Fock}
\ket{n_{f}} \coloneqq \frac{1}{\sqrt{n!}} [\hat a^{\dag}(\vec f)]^n \ket{0},
\end{equation}
which is a state of $n$ photons in mode $f$. These states are inherently single-mode, even though they can be embedded in a much larger multimode state space. The Wigner function for such states is commonly found in quantum optics textbooks, but deriving it using (\ref{eq:wigDef}) is a good exercise. Here we simply state the result:
\begin{equation}\label{eq:WigFock}
W_{n_{f}}(\vec{x}) = \sum _{k=0}^n \binom{n}{k}\frac{(-1)^{n+k} \norm{\vec{x}_{\bf f}}^{2k}}{k!} \frac{e^{-\frac{1}{2} \norm{\vec{x}}^2}}{2 \pi}.
\end{equation}
In the most general sense, we write $\vec{x}_{\bf f} = (\vec f^T\vec x) \vec f + (\vec f^T\Omega\vec x) \Omega \vec f$ as the projector of $\vec x$ on the phase space of mode $f$, but it is most practical to use the coordinate representation $\vec{x}_{\bf f} = (x_f,p_f)^T$ where $x_f = \vec f^T\vec x$ and $p_f = \vec f^T\Omega\vec x$ to describe the two-dimensional phase space associated with mode $f$.

Multimode Fock states can be obtained by acting on the vacuum with different creation operators in different modes. Even though these different modes do not necessarily have to be orthogonal \cite{Walschaers_2020}, we here focus on the case where they are. For example, in $m$-mode system, we can chose a basis $\{\vec e_1, \Omega \vec e_1, \dots, \vec e_m, \Omega \vec e_m\} $ of the phase space $\mathbb{R}^{2m}$ and define multimode Fock states as
\begin{equation}\label{def:FockMulti}
\ket{n_{e_1}} \otimes \dots \otimes \ket{n_{e_m}} \coloneqq \frac{1}{\sqrt{n_1! \dots n_m!}} [\hat a^{\dag}(\vec e_1)]^{n_1} \dots [\hat a^{\dag}(\vec e_m)]^{n_m} \ket{0},
\end{equation}
where the $k$th mode in the basis contains $n_k$ photons. Note that for $n_k = 0$ we have a vacuum mode. For the Wigner function, this implies
\begin{equation}\label{eq:WigFockMulti}
W_{n_{e_1}, \dots, n_{e_m} }(\vec{x}) = W_{n_{ e_1}}(\vec{x}_{\bf e_1})\dots W_{n_{e_m}}(\vec{x}_{\bf e_f}).
\end{equation}
Here we have used that the phase space point $\vec x$ can be expressed as $\vec x = \vec x_{\bf e_1} \oplus \dots \oplus \vec x_{\bf e_m}$, where $\vec x_{\bf e_k}$ is the phase space coordinate within the subspace spanned by $\vec e_k$ and $\Omega \vec e_k$. We can note $\vec{x}_{\bf e_k} = (x_k,p_k)^T$, such that we find the coordinate representation $\vec x = (x_1,p_1, \dots, x_m, p_m )^T$ in the chosen basis of phase space.\\

Non-Gaussian states do not necessarily have to be pure, they can also come in the form of statistical mixtures. The most basic example of such as state is a non-Gaussian mixture of Gaussian states. As a simple example, let is consider a mixture of two coherent states
\begin{equation}
\hat \rho = \frac{1}{2} \left( \ket{\vec \alpha}\bra{\vec \alpha} +  \ket{-\vec \alpha}\bra{-\vec \alpha} \right). 
\end{equation}
Even though this is a highly classical state, it is still non-Gaussian as clearly seen from its Wigner function
\begin{equation}\label{eq:coherentMixture}
W(\vec x) = \frac{1}{4\pi} \left(e^{-\frac{1}{2} \norm{\vec x - \vec{\alpha}}^2} + e^{-\frac{1}{2} \norm{\vec x + \vec{\alpha}}^2}\right).
\end{equation}
Another important example of a non-Gaussian mixed state is
\begin{equation}
\hat \rho_{\lambda} = \lambda \ket{0}\bra{0} + (1-\lambda) \ket{1_{f}}\bra{1_{f}}.
\end{equation}
In the mode $\vec f$, the Wigner function of the state behaves as 
\begin{equation}\label{eq:WigFockMix}
W_{\lambda}(\vec{x}) =  [(1-\lambda)\norm{\vec{x}_{\bf f}}^{2} + 2\lambda - 1] \frac{e^{-\frac{1}{2} \norm{\vec{x}}^2}}{2 \pi}.
\end{equation}
This Wigner function reaches negative values as long as $\lambda < 1/2$ and subsequently becomes positive. Nevertheless, it will remain non-Gaussian until $\lambda = 1$. As we will see in Subsection \ref{sec:quantumnongauss}, even when the Wigner function is positive, it is not always possible to describe this state as a mixture of Gaussian states.

Generally speaking, non-Gaussian states can come in a wide variety of shapes which can be much more exotic than the examples discussed above. A popular class of Gaussian states is obtained by taking coherent superpositions of Gaussian states. As we will see in Section \ref{sec:Exp}, there has been a strong experimental focus on two specific types of such states: Schr\"odinger's cat states \cite{PhysRevLett.57.13} and Gottesman-Kitaev-Preskill (GKP) states \cite{PhysRevA.64.012310}. The former are obtained by coherently superposing two coherent states, and often are split in even $\ket{{\rm cat}_+}$ and odd $\ket{{\rm cat}_-}$ cat states:
\begin{equation}
\ket{{\rm cat}_{\pm}} \coloneqq \frac{1}{{\cal N}} \left( \ket{\vec \alpha} \pm  \ket{-\vec \alpha}\right),
\end{equation}
where ${\cal N} = \sqrt{2 (1 \pm \exp[ - \norm{\vec \alpha}^2])}$ is the normalisation coefficient which depends on the displacement $\vec \alpha$. The latter is often referred to as ``the size of the cat''. The Wigner function of these states resembles that of (\ref{eq:coherentMixture}) but has an additional interference term
\begin{equation}\label{eq:WigCat}
W_{{\rm cat}_{\pm}}(\vec x) = \frac{e^{-\frac{1}{2} \norm{\vec x - \vec{\alpha}}^2} + e^{-\frac{1}{2} \norm{\vec x + \vec{\alpha}}^2} \pm \cos( \sqrt{2}\, \vec \alpha^T \vec x )e^{-\frac{1}{2} \norm{\vec x}^2}}{4\pi (1 \pm e^{ - \norm{\vec \alpha}^2})}.
\end{equation}
The appearance of these interference terms creates several regions in phase space where the Wigner function attains negative values. The term ``Schr\"odinger's cat state'' has historically grown from the idea that coherent states describe classical electromagnetic fields and can thus be considered ``macroscopic'', in particular for large values of $\norm{\vec \alpha}$. However, one should honestly admit that they fail to capture an important point of Schr\"odinger's thought experiment \cite{Schrodinger_cat_1935}: the entanglement with a microscopic quantum system (i.e., the decay event that triggers the smashing of the vial of poison). Nevertheless, the term ``cat state'' has established itself firmly in the CV jargon, and now also lies at the basis of derived concepts such ``cat codes'' for error correction \cite{PhysRevA.59.2631,PhysRevLett.111.120501}.

Finally, there are the GKP states. In their idealised form, they rely on eigenvectors of the quadrature operators (sometimes also known as infinitely squeezed states). To keep notation simple, we will restrict to the single mode with quadrature operators $\hat x$ and $\hat p$. The eigenvectors of these operators are then formally written as
\begin{equation}
\hat x \ket{x} = x \ket{x}, \quad \text{and} \quad \hat p \ket{p} = p \ket{p}.
\end{equation}
Furthermore, we have the relations $\langle x' \mid x \rangle = \delta (x'- x),$ $\langle p' \mid p \rangle = \delta (p'- p)$, and $\langle p \mid x \rangle = e^{-ipx}/\sqrt{2\pi}$. GKP states are constructed by considering a grid of such states to create a qubit, by identifying the two following GKP vectors:
\begin{align}
&\ket{GKP_{\tilde 0}} \coloneqq \sum_{k \in \mathbb{Z}} \ket{x = 2k \sqrt{\pi} },\\
&\ket{GKP_{\tilde 1}} \coloneqq \sum_{k \in \mathbb{Z}} \ket{x = (2k+1)\sqrt{\pi}}.
\end{align}
Clearly, these states are not normalisable and not physical as they would require infinite energy to be created. Thus, it is common to construct approximate GKP states, by replacing the states $\ket{x}$ with displaced squeezed states, and by truncating the summation by adding a Gaussian envelope:
\begin{align}\label{eq:GKPstate}
&\ket{GKP_0} \coloneqq {\cal N}_0 \sum_{k \in \mathbb{Z}} e^{-2\pi[k \delta]^2} \hat D[2 k \sqrt{\pi}]\ket{s},\\
&\ket{GKP_1} \coloneqq  {\cal N}_1 \sum_{k \in \mathbb{Z}} e^{-2\pi[(k+1/2) \delta]^2}  \hat D[(2k+1) \sqrt{\pi}] \ket{s}.
\end{align}
here $ {\cal N}_{0,1}$ are normalisation constants and $\ket{s}$ is a single-mode squeezed vacuum, which implies that its Wigner function is given by (\ref{eq:GaussStateWig}) with $\vec \xi = (0,0)^T$ and $V = {\rm diag} [1/s, s]$ for $s > 1$. To get a good GKP state for quantum error correction, we generally need that $s \gg \sqrt{\pi}$. The Wigner function of an ideal GKP state is a grid of delta functions, which again highlights that it is a non-physical state. The more realistic states $\ket{GKP_0}$ and $\ket{GKP_1}$ have well-defined Wigner functions, even though they are not very insightful to write down explicitly. In Fig.~\ref{fig:Examples}, we plot an example that was calculated numerically by taking into account only the first few terms around $k=0$ in the sum.

GKP states may seem a little artificial at first glance, but they have we developed with a very clear purpose: to encode a qubit in a harmonic oscillator \cite{PhysRevA.64.012310}. This encoding implies a notion of fault-tolerance as these states are designed to be very efficient at correcting displacement errors. The more realistic incarnations of these states \eqref{eq:GKPstate} are therefore often proposed as candidates for encoding the information in CV quantum computation protocols  \cite{PhysRevLett.112.120504}. Furthermore, it was shown that these states can also be used as the sole non-Gaussian resource to implement a CV quantum computer \cite{PhysRevLett.123.200502}\\

Once we progress into the realm of multimode states, the class of non-Gaussian states becomes even more vast. In Subsection \ref{sec:PhotonSubtraction}, we will present a dedicated introduction to multimode photon-subtracted states, which is a useful state general concepts in this Tutorial. Furthermore, these states have a particular importance in CV quantum optics experiments. As a final example, we introduce another class of multimode non-Gaussian states, which have been highly relevant for quantum metrology: N00N states \cite{PhysRevA.40.2417,PhysRevLett.85.2733,doi:10.1080/0950034021000011536}. Even though these states very useful for quantum sensing with optical setups, the general idea that underlies these states was first introduced for fermions \cite{PhysRevLett.56.1515} in an attempt to mimic the advantage that is provided by squeezing in optics.

N00N states are two-mode entangled states defined in a pair of orthogonal modes $g_1$ and $g_2$. The state contains exactly $N$ photons, and is a superposition of a state with all photons being mode $g_1$ and a state with all photons in mode $g_2$:
\begin{equation}
\ket{N00N} \coloneqq \frac{1}{\sqrt{2}}\left(\ket{N_{g_1}} + \ket{N_{g_2}}\right).
\end{equation}
Here we recall that the state $\ket{N_{g_1}}$ can be trivially embedded to the full multimode space by adding vacuum in all other modes. We will study these states in more detail for $N=2$ in our discussion of the Hong-Ou-Mandel effect surrounding Eq. \eqref{eq:HOM}. However, here we highlight already that the Wigner function of $\ket{N00N}$ is not simply the sum of Wigner functions of the form \eqref{eq:WigFock}. The entanglement will create additional interference terms, just like we saw in \eqref{eq:WigCat}. In the present case, these interferences are genuinely multimode, and thus related to quantum correlations. 

Experimentally, these states have been created and analysed using a DV approach \cite{PhysRevA.85.022115}. As we highlighted in Subsection \ref{sec:Modes}, the distinction between DV and CV is somewhat subtle and mainly depends on what is measured. Because N00N states are built from Fock states and have a well-defined total photon number, they are most natural to analyse using photon-number resolving detectors.

\subsection{Quantum Non-Gaussianity}\label{sec:quantumnongauss}

Non-Gaussian states come in a wide variety, which means also that some of them are more exotic than others. Non-Gaussian states that are of limited interest, are those which are convex combinations of Gaussian states. Gaussian states do not form a convex set, after all, we can immediately see that, e.g., $[W_0(\vec x -\vec \alpha_1) + W_0(\vec x -\vec \alpha_2)]/2$ is not a Gaussian function even though it is a convex combination of Gaussian states. 

The fact that the set of non-Gaussian states contains mixtures of Gaussian states may lead one to suspect that any mixed state with a positive Wigner function can be written as a well-chosen mixture of Gaussian states. After all, Gaussian states are the only pure states with positive Wigner functions. This intuition turns out to be false \cite{PhysRevLett.106.200401}, which means that the set of states with a non-positive Wigner function is not the same as the set of states that lie outside of the convex hull of Gaussian states ${\cal G}$. More formally, let us define
\begin{equation}
{\cal G} \coloneqq \left\{\hat \rho \mid \hat \rho = \int {\rm d} \gamma \, p(\gamma) \hat \rho_G (\gamma)\right\},
\end{equation} 
where $\gamma$ is some arbitrary way of labelling Gaussian states $\hat \rho_G (\gamma)$ and $p(\gamma)$ is a probability distribution on these labels. Note that (\ref{eq:GaussianDecomp}) tells us that we can generate all Gaussian states by taking convex combinations of displaced squeezed states, and thus we can limit ourselves to $ \hat \rho_G (\gamma) = \ket{\Psi_G(\gamma)}\bra{\Psi_G(\gamma)}$ in the definition of ${\cal G}$.

Any quantum state $\hat \rho$ that is not contained in the convex hull of Gaussian states, i.e., $\hat \rho \notin{\cal G}$, is referred to as a ``quantum non-Gaussian'' state. The intuition behind this terminology is that Gaussian pure states are less quantum than non-Gaussian pure states that boast a non-positive Wigner function. A mixed state that is quantum non-Gaussian may have a positive Wigner function, but it cannot be created without adding states with non-positive Wigner functions into the pure-state decomposition. Hence, these states are more quantum than the states that are in the convex hull of Gaussian states ${\cal G}$.

Next, one may wonder how to differentiate between states that are quantum non-Gaussian and states which are in the convex hull ${\cal G}$. Throughout the last decade, many methods have been developed to answer this question. We start by introducing the main idea of Ref.~\cite{PhysRevA.87.062104} because it is based on the Wigner function. The key idea is that Gaussian distributions have tails, which means that we can take an arbitrary pure Gaussian state $W_0(S^{-1}(\vec x-\vec\alpha))$, and evaluate the Wigner function at to origin of phase space:
\begin{equation}
W_0(S^{-1}\vec\alpha) = \frac{e^{-\frac{1}{2}\norm{S^{-1} \vec\alpha}^2}}{(2\pi)^2}. 
\end{equation}
Clearly, when $\norm{S^{-1} \vec\alpha}^2 \rightarrow \infty$, we do find that $W_0(S^1\vec\alpha) \rightarrow 0$. This limit essentially corresponds to a system with infinite energy. It is thus natural to try to bound the value of the Wigner function in the origin by a function that depends on the energy $\overline N$ of the state. For an arbitrary pure Gaussian state, we find that
\begin{equation}
\overline N = \sum_{j=1}^m \tr[\hat \rho \hat a^{\dag}(\vec e_j) \hat a(\vec e_j)] = \frac{1}{4}\left(\tr [S^T S-\mathds{1}] + \norm{\vec\alpha}^2 \right).
\end{equation}
Using the properties of the operator norm, we write $\norm{S^{-1} \vec\alpha}^2 = \norm{S^{-1} O \vec\alpha}^2$, where $O$ is a symplectic orthogonal transformation. Furthermore, we note that $\norm{\vec\alpha}^2 = \norm{O\vec \alpha}^2$. It is then useful to explicitly write the coordinate representation of the vector $O\vec \alpha = (\alpha^{({x})}_1,\alpha^{({p})}_1, \dots , \alpha^{({x})}_m,\alpha^{({p})}_m)^T$, such that
\begin{align}
\overline N &= \sum_{j=1}^m \frac{1}{4}\left(s_j + \frac{1}{s_j} + (\alpha^{({x})}_j)^2 + (\alpha^{({p})}_j)^2 - 2\right)\\
&=\sum_{j=1}^m \overline n_j.
\end{align}
At the same time, we expand 
\begin{equation}
\norm{S^{-1} \vec\alpha}^2 =\sum_{j=1}^m s_j (\alpha^{({x})}_j)^2 + \frac{ (\alpha^{({p})}_j)^2}{s_j} ,
\end{equation}
and with a little algebra we can show that
\begin{equation}
\frac{1}{2}\norm{S^{-1} \vec\alpha}^2 \leqslant \sum_{j=1}^m 4 \overline n_j(2\overline n_j + 1) \leqslant 4\overline N (2\overline N + 1),
\end{equation}
such that 
\begin{equation}
W_0(S^{-1}\vec\alpha) \geqslant \frac{1}{(2\pi)^m}e^{-4\overline N (2\overline N + 1)}.
\end{equation}
This means that the value of the Wigner function of a pure Gaussian state in the origin of phase space is bounded from below by a function of the average number of particles $\overline N$. However, it is not obvious that we can extend this bound to arbitrary mixtures of Gaussian states. Let us assume that $\hat \rho \in {\cal G}$, then the Wigner function in the origin is given by
\begin{equation}
W(\vec 0) = \int {\rm d}\gamma p(\gamma) W_0(S_{\gamma}^{-1}\vec\alpha_{\gamma}),
\end{equation}
where we saw that $W_0(S_{\gamma}^{-1}\vec\alpha_{\gamma})$ is the value of a pure Gaussian state's Wigner function in the origin and $\gamma$ is some arbitrary label for the Gaussian states in the mixture. Therefore we can bound the states in the convex combination
\begin{equation}
W(\vec 0) \geqslant  \frac{1}{(2\pi)^m} \int_0^{\infty} {\rm d}\overline N_{\gamma} \tilde p(N_{\gamma})e^{-4\overline N_{\gamma} (2\overline N_{\gamma} + 1)},
\end{equation}
where we introduce a probability distribution $\tilde p$ on the average particle numbers of the pure Gaussian states in the mixture. The overall average number of particles in the state $\hat \rho$ is then given by $\overline N = \int_0^{\infty} {\rm d}\overline N_{\gamma} \,\tilde p(\overline N_{\gamma}) \overline N_{\gamma}$. The final element that we require is the fact that $\exp[-4\overline N_{\gamma} (2\overline N_{\gamma} + 1)]$ is a convex function, such that we can apply Jensen's inequality to find that
\begin{equation}\label{quantumNonGausIneq}
\hat \rho \in {\cal G} \implies W(\vec 0) \geqslant  \frac{1}{(2\pi)^m} e^{-4\overline N (2\overline N + 1)}.
\end{equation}
This means that we can simply use the total energy of the state to construct a witness for quantum non-Gaussianity. This clearly shows that there are quantum non-Gaussian states with positive Wigner functions. An explicit example can be constructed by tuning the $\gamma$ in the state $[(1-\gamma) \ket{0}\bra{0} + \gamma \lvert1_{\vec f}\rangle\langle1_{\vec f}\rvert]$ to $1/2 > \gamma > 1/2 - e^{-4\gamma (2\gamma + 1)}$ (where we use that $\gamma$ is also the average particle number in this particular state). 

Nevertheless, there are many quantum non-Gaussian states that do not violate inequality (\ref{quantumNonGausIneq}). After all, why would the origin of phase space be the most interesting point? A first solution is provided in \cite{PhysRevA.87.062104}, where it is argued that 
\begin{equation}
\hat \rho \in {\cal G} \implies W_{\Gamma}(\vec 0) \geqslant  \frac{1}{(2\pi)^m} e^{-4\overline N_{\Gamma} (2\overline N_{\Gamma} + 1)},
\end{equation}
for all Gaussian channels $\Gamma$ that act on the Wigner function of $\hat \rho$ as $W(\vec x) \overset{\Gamma}{\mapsto}W_{\Gamma}(\vec x)$. Recall that the action of a Gaussian channel was defined in (\ref{eq:GaussianChannel}). The quantity $\overline N_{\Gamma}$ then denotes the average number of particles in the state $\Gamma(\hat \rho)$. Further generalisations of this scheme have been worked out in \cite{PhysRevA.90.013810}. Moreover, Ref.~\cite{PhysRevLett.114.190402} has considered combinations of the value of the Wigner function in several points to reach better witnesses for quantum non-Gaussianity. Further progress has been made by identifying observables that are more easily measurable with typical CV techniques \cite{Happ_2018}. Others have considered other phase-space representations of the state to identify witnesses of quantum non-Gaussianity \cite{PhysRevA.97.053823,Bohmann2020probing}.\\

Quantum non-Gaussianity has been investigated with a wide range of tools. In this Tutorial we have limited ourselves to phase space methods in the spirit of the CV approach. However, there is also a significant body of work on quantum non-Gaussianity using techniques that are more typical in DV quantum optics. The earliest works on the subject used photon-statistics to distinguish quantum non-Gaussian states from convex mixtures of Gaussian states \cite{PhysRevLett.106.200401}. This research line has been continued in recent years to uncover new aspects of quantum non-Gaussian states, such as the ``non-Gaussian depth'' \cite{PhysRevLett.113.223603} and techniques to differentiate different types of multi-photon states \cite{Filip-QNonGauss}. Ultimately, these photon-counting techniques were extended to develop a whole hierarchy of quantum non-Gaussian states \cite{PhysRevLett.123.043601}. These ideas have been further formalised and generalised through the notion of the ``stellar rank'' of a quantum state.

\subsection{Stellar Rank}

An interesting starting point to introduce the stellar representation is the method \cite{PhysRevA.51.3340} to prove Hudson's theorem. We recall the definition 
\begin{equation}
F^{\star}_{\Psi}(\vec \alpha) \coloneqq \langle \vec \alpha \mid \Psi \rangle e^{\frac{1}{8}\norm{\vec\alpha}^2}, \tag{\ref{eq:Fstellar}}
\end{equation}
of what we will henceforth refer to as the stellar function. To avoid technical complications, let us now restrict ourselves to single-mode systems such that the optical phase space is $\mathbb{R}^2$. Note that, in a single-mode system, there is only one creation operator $\hat a^{\dag}$ with associated annihilation operators $\hat a$. We can then follow \cite{PhysRevLett.124.063605} to introduce the stellar representation of single-mode quantum states. Note that some similar ideas are also present in other works \cite{PhysRevA.99.053816}.

First, we develop the stellar representation for pure states, which will then be used to generalise the framework to mixed states in \eqref{eq:convexRoofstellar}. We use the definition of the displacement operator to show that
\begin{equation}
\ket{\Psi} = F^{\star}_{\Psi}(\hat a^{\dag})\ket{0},
\end{equation}
which immediately implies that the stellar representation is unique. In other words, if $F^{\star}_{\Psi} = F^{\star}_{\Phi}$ it follows that $\ket{\Psi} = \ket{\Phi}$ up to a phase. In our proof of Hudson's theorem, we have already highlighted that $F^{\star}_{\Psi}$ satisfies the property (\ref{eq:stellar3}) which means that it is an entire function with growth order $r=2$. Because in this single-mode setting $F^{\star}_{\Psi}$ can be interpreted as a function of a single complex variable, the Hadamard-Weierstrass theorem implies that $F^{\star}_{\Psi}$ can be fully represented by its zeros (one can consider this as a generalisation of the fundamental theorem of calculus). This thus implies that a single-mode state is completely determined by the zeros of the $F^{\star}_{\Psi}$, and thus by the zeros of the Q-function in eq.~(\ref{eq:QFunction106}).

It is thus natural to use these zeros in order to classify pure single-mode quantum states and thus the stellar rank is introduced. Ultimately, the stellar rank is simply given by the number of zeros of $F^{\star}_{\Psi}$ or alternatively the number of zeros of the Q-function. Because in practice zeros may coincide, one should also consider the multiplicity of the zeros. We thus define the stellar rank $r^{\star}(\Psi)$ of $\ket{\Psi}$ as the number of zeros of $F^{\star}_{\Psi}:\mathbb{C} \mapsto \mathbb{C}$ counted with multiplicity. Alternatively one may count the zeros of the Q-function with multiplicity and divide by two. 

The fact that a state is fully characterised by its stellar representation $F^{\star}_{\Psi}$ can be made more explicit by considering the roots $\{\vec\alpha_1, \dots, \vec\alpha_{r^{\star}(\Psi)} \}$ of the Q-function which represents $\ket{\Psi}$ (note that we use (\ref{eq:PhaseHilb}) to interchange between phase-space representation and complex Hilbert space representation). The single-mode state $\ket{\Psi}$ can then be used to express
\begin{equation}\label{eq:StateDescZerosQ}
\ket{\Psi} = \frac{1}{{\cal N}} \prod_{j=1}^{r^{\star}(\Psi)} \hat D^{\dag}(\vec\alpha_j) \hat a^{\dag} \hat D(\vec\alpha_j) \ket{\Psi_G},
\end{equation}
where $\ket{\Psi_G}$ is a pure Gaussian state and ${\cal N}$ a normalisation constant. We can then use the stellar rank to induce some further structure in the set of states by defining
\begin{equation}\label{eq:setrankN}
{\cal R}_N \coloneqq \{\ket{\Psi} \mid r^{\star}(\Psi) = N \}.
\end{equation}
that groups all states of stellar rank $N$. Note that the Hadamard-Weierstrass theorem also considers functions with an infinite amount of zeros and the case $N=\infty$ is thus mathematically well-defined. It turns out that this case is not just a pathological limit. An evaluation of the Q-function shows that Gottesman-Kitaev-Preskill states and Schr\"odinger cat states inhabit the set ${\cal R}_{\infty}$.\\

Clearly, all that was introduced so far only works for pure states. We can naturally extend this result via a convex roof construction, by defining 
\begin{equation}\label{eq:convexRoofstellar}
r^{\star}(\hat \rho) \coloneqq \inf_{\{p(\gamma), \ket{\Psi(\gamma)}\}}\sup_{\gamma} \{r^{\star}[\Psi(\gamma)]\},
\end{equation}
where the infimum is considered of all probability distributions on the set of pure states that lead to $\hat \rho = \int {\rm d}\gamma p(\gamma) \ket{\Psi(\gamma)}\bra{\Psi(\gamma)}$. In words, there are many ways to decompose the state $\hat \rho$ in pure states and we consider all of them. For each decomposition, we define the stellar rank as the highest rank of the states in the decomposition. Then we minimize these values over all possible decompositions to arrive at the stellar rank of $\hat \rho$. 

Convex roof constructions are commonly used to treat mixed states as they are easy and natural to formally define. However, they are often much harder to calculate in practice. This is where the stellar representation unveils its most remarkable property: stellar robustness. To formalise this idea, \cite{PhysRevLett.124.063605} introduced the robustness as the trace distance between the state and the nearest possible state of lower stellar rank.
\begin{equation}
R^{\star}(\Psi) \coloneqq \inf_{r^{\star}(\hat \rho) < r^{\star}(\Psi)} \frac{1}{2} \tr \sqrt{(\ket{\Psi}\bra{\Psi} - \hat \rho)^2}.
\end{equation}
And it can be shown that 
\begin{equation}
R^{\star}(\Psi)  = \sqrt{1- \sup_{r^{\star}(\hat \rho) < r^{\star}(\Psi)} \bra{\Psi}\hat \rho \ket{\Psi} },
\end{equation}
where $\bra{\Psi}\hat \rho \ket{\Psi}$ is the fidelity of $\hat \rho$ with target state $\ket{\Psi}$. Remarkably, it can be shown that $R^{\star}(\Psi) > 0$ when $r^{\star}(\Psi) < \infty$ \cite{PhysRevLett.124.063605}. This means that any state that is sufficiently close to a pure state of $r^{\star}(\Psi)$ is also of rank $r^{\star}(\Psi)$ or higher.

The idea of stellar robustness can be generalised \cite{chabaud2020certification} by introducing $k$-robustness. For any $k < r^{\star}(\Psi)$, we define
\begin{equation}
R^{\star}_k(\Psi) \coloneqq \inf_{r^{\star}(\Phi) \leqslant k} \sqrt{1 - \abs{\langle \Phi \mid \Psi \rangle}^2},
\end{equation}
and show subsequently that
\begin{equation}\label{eq:kRob}
R^{\star}_k(\Psi)  = \sqrt{1- \sup_{r^{\star}(\hat \rho) \leqslant k} \bra{\Psi}\hat \rho \ket{\Psi} }.
\end{equation}
The $k$-robustness can thus be interpreted as the nearest distance from a state $\ket{\Psi}$ at which we can find any state of stellar rank $k$, provided $k <  r^{\star}(\Psi)$. Beyond showing that $R^{\star}_k(\Psi)$ is non-zero when $\ket{\Psi}$ is of finite stellar rank, Ref.~\cite{chabaud2020certification} also provides an explicit method to calculate $R^{\star}_k(\Psi)$. Thus, for whichever state $\hat \rho$ is available in an experiment one can attempt to find a pure target state $\ket{\Psi}$ for which $\bra{\Psi}\hat \rho \ket{\Psi} > 1 - R^{\star}_k(\Psi)^2$ to prove that $\hat \rho$ is at least of stellar rank $k$.\\

We note that for pure states $r^{\star}(\Psi) = 0$ implies that the state is Gaussian (this is essentially what is proven in Hudson's theorem). From the definition (\ref{eq:convexRoofstellar}) we can then deduce that
\begin{equation}
r^{\star}(\hat \rho) = 0 \iff \hat \rho \in {\cal G},
\end{equation}
where ${\cal G}$ denotes, again, the convex hull of Gaussian states. On the other hand, states for which $r^{\star}(\hat \rho) > 0$ cannot be written as a mixture of Gaussian states and are thus quantum non-Gaussian. 

This idea can be extended by using the stellar $k$-robustness \eqref{eq:kRob} as a witness of quantum non-Gaussianity. When we want to check whether $\hat \rho$ is quantum non-Gaussian, it suffices to find a pure target state $\ket{\Psi}$ such that $\hat \rho$ is closer to $\ket{\Psi}$ than the $1$-robustness $R^{\star}_k(\Psi)$. More formally written, whenever a pure state $\ket{\Psi}$ exists with the following property:
\begin{equation}
\bra{\Psi}\hat \rho \ket{\Psi} > 1 - R^{\star}_1(\Psi)^2 \implies \hat \rho \notin {\cal G},
\end{equation}
and $\hat \rho$ is quantum non-Gaussian. This may seem like a complicated challenge, but for a single-photon state $\ket{\Psi} = \ket{1}$ we find that $1-R^{\star}_1(\Psi)^2 \approx 0.478$ \cite{chabaud2020certification}. This means that any state which has a fidelity of more than $0.478$ with respect to a Fock state is quantum non-Gaussian. This idea can be extended to higher stellar ranks: whenever we find a target state $\ket{\Psi}$ such that $\bra{\Psi}\hat \rho \ket{\Psi} > 1 - R^{\star}_k(\Psi)^2,$ the state $\hat \rho$ is at least of stellar rank $k$. Note that the fidelity of an experimentally generated state $\hat \rho$ with any pure target state $\ket{\Psi}$ can be calculated from double homodyne measurements on $\hat \rho$ \cite{chabaud2020certification}. There is no need to experimentally create the pure state $\ket{\Psi}$, the latter is just theoretical input needed to analyse the data.

Obviously, the stellar rank imposes a lot of additional structure on the state space. It rigorously orders all states that can be achieved by combining a finite number of creation operators and Gaussian transformations. The creation operator serves as a tool to increase the stellar rank by one and the stellar rank actually corresponds to the minimal number of times the creation operator must be applied to obtain the state, together with Gaussian operations. The stellar rank remains unchanged under Gaussian unitary transformations, which makes sense for a measure of the non-Gaussian character of the state, and thus it falls within the set of intrinsic properties of a state as discussed in Subsection \ref{sec:Modes}. Furthermore, the class of states with infinite stellar rank can be understood as the set that contains the most exotic states. However, it is lonely at the top as it can be shown that $R^{\star}_{\infty}(\Psi) = 0$ for states of infinite rank. This means that we can find states of finite stellar rank arbitrarily close to a state of infinite stellar rank. As stressed in \cite{PhysRevLett.124.063605}, this implies that finite rank states are dense in the full Fock space and any state of infinite-rank can be arbitrarily well approximated by finite rank states. Whereas a finite stellar rank $k$ of any experimental state can be certified by achieving a sufficiently high fidelity to a target state $\ket{\Psi}$ to fall within the range given by its $k$-robustness $R^{\star}_k(\Psi)$, a similar procedure is impossible for infinite stellar ranks. This means, in practice, that genuinely infinite-rank states are impossible to certify in experiments because one never achieves perfect fidelity. Nevertheless, different states of infinite rank may differ significantly in the values $R^{\star}_{k}(\Psi)$ for $k < \infty$.\\

Many of the results on stellar rank rely on the Hadamard-Weierstrass theorem that allows to uniquely factorise $F^{\star}_{\Psi}(\vec \alpha)$ as a Gaussian and a polynomial, where the roots of the polynomial are the roots of $F^{\star}_{\Psi}$ and thus also the roots of the Q-function. Sadly, this theorem cannot be straightforwardly generalised to a multimode setting, which is known in mathematics as Cousin's second problem  \cite{LelongBook}. Notable progress was made in \cite{chabaud2020classical} where one studies multimode stellar functions which are polynomials and it was shown that there is no straightforward generalisation of (\ref{eq:StateDescZerosQ}).


\subsection{Wigner Negativity}\label{sec:WignerNegativity}

Hudson's theorem shows us that all pure non-Gaussian states have non-positive Wigner functions, which sets them apart from normal probability distributions of phase space. For mixed states, this no longer holds and thus we spent the previous two sections developing methods to characterise the non-Gaussian features of these states. Whether it is through quantum non-Gaussianity or the more refined stellar rank, these methods focus on characterising the non-Gaussian resources that are required to generate a certain state. In this subsection, we change the perspective and focus rather on negative values of the Wigner function (``Wigner negativity'' in short) as a resource of interest.

Wigner negativity has the advantage of being a clear quantum feature, it reflects that different quadratures in the same mode cannot be jointly measured and thus goes hand in hand with the principle of complementarity. More formally, it has even been connected to the principle of quantum contextuality \cite{PhysRevLett.101.020401}. Indeed, in Section \ref{sec:Advantages} we will elaborate on the fact that Wigner negativity is a necessary resource for reaching a quantum advantage, i.e., performing a task that cannot be efficiently simulated by a classical computer. However, the idea of using Wigner negativity as a signature of non-classicality was already around before it was connected to a quantum computational advantage. An important step to formalise this idea was the introduction of a measure for Wigner negativity \cite{Kenfack:2004aa}, which lies at the basis of recent resource theories of Wigner negativity \cite{Takagi:2018aa,PhysRevA.98.052350}.

A priori, there are several natural measures than can be used for Wigner negativity. It is therefore useful to consider some desirable properties that are required for a measure of Wigner negativity. First of all, we want the measure to be zero if and only if the Wigner function is positive. It seems natural to demand that, furthermore, Wigner negativity remains unchanged under Gaussian unitary transformations. It is then tempting to simply consider the absolute value of the lowest possible value of the Wigner function, but this would have some unnatural outcomes. It would mean that a single photon state would have more Wigner negativity than a two photon state. We thus need to look for a different measure.

The starting point of \cite{Kenfack:2004aa} is that the normalisation of the Wigner function implies that
\begin{equation}
 \int_{\mathbb{R}^{2m}} {\rm d}\vec x \abs{W(\vec x)} \geqslant 1,
\end{equation}
and that the inequality is strict whenever there is Wigner negativity. Furthermore, Liouville's theorem implies that integrals over phase space are unchanged by Gaussian transformations. A first possible way of measuring Wigner negativity is through the negativity volume
\begin{equation}\label{eq:NegVolume}
{\cal N}(\hat \rho) \coloneqq \int_{\mathbb{R}^{2m}} {\rm d}\vec x \abs{W(\vec x)} - 1.
\end{equation}
This measure has the major advantage of being convex due to the triangle inequality, which means that for $\hat \rho = \int{\rm d}\gamma p(\gamma) \hat \rho(\gamma)$ we find that
\begin{align}
{\cal N}(\hat \rho) \leqslant \int{\rm d}\gamma p(\gamma) {\cal N}[\hat \rho(\gamma)].
\end{align}
However, this measure is not additive, i.e., ${\cal N}(\hat \rho_1 \otimes \hat \rho_2) \neq {\cal N}(\hat \rho_1) +  {\cal N}(\hat \rho_2)$. To circumvent this shortcoming, another measure for Wigner negativity has been introduced \cite{Veitch_2014,Takagi:2018aa,PhysRevA.98.052350}:
\begin{equation}
{\frak N}(\hat \rho) \coloneqq \log  \int_{\mathbb{R}^{2m}} {\rm d}\vec x \abs{W(\vec x)}.
\end{equation}
Clearly, ${\frak N}(\hat \rho_1 \otimes \hat \rho_2) = {\frak N}(\hat \rho_1) +  {\frak N}(\hat \rho_2)$ making this measure additive. However, the introduction of the logarithm destroys the convexity of the measure. Note that the two measures are closely related by ${\frak N}(\hat \rho) = \log [{\cal N}(\hat \rho) + 1]$. Thus when ${\cal N}(\hat \rho_1) > {\cal N}(\hat \rho_2)$, we also find that ${\frak N}(\hat \rho_1) > {\frak N}(\hat \rho_2)$.\\

The single-mode examples that are considered in Ref.~\cite{Kenfack:2004aa} lead to some interesting observations. First of all, they show that for Fock states Wigner negativity increases with the photon number. Furthermore, they show that for Schr\"odinger cat states the integral is bounded from above by a value smaller than the Wigner negativity of a two-photon state. Even though Fock states of increasing stellar rank have increasing Wigner negativity, there is no clear relation between stellar rank and Wigner negativity for more general classes of states. For example, Schr\"odinger cat states are of infinite stellar rank, suggesting that they are in this regard the most exotic states, but they only manifest a limited amount of Wigner negativity.

As a case study, let us briefly concentrate on the Wigner negativity of Fock states (\ref{def:Fock}).  One can now evaluate the Wigner negativity of such states to find that
\begin{align}
&{\cal N}(\ket{1}) \approx 0.42612\quad \text{ and }\quad {\frak N}(\ket{1}) \approx 0.354959,\\
&{\cal N}(\ket{2}) \approx 0.72899 \quad \text{ and }\quad {\frak N}(\ket{2}) \approx 0.547537,\\
&{\cal N}(\ket{3}) \approx 0.97667\quad \text{ and }\quad {\frak N}(\ket{3}) \approx 0.681415,
\end{align}
which shows that the negativity does not simply increase linearly with the number of photons even for the additive measure ${\frak N}$. However, let us now look at a multimode $n$-photon state where each photon occupies a different mode, i.e., a Fock state generated by creation operators in $\vec f_1, \dots, \vec f_n$ with ${\rm span}\{\vec f_j, \Omega\vec f_j\} \neq {\rm span}\{\vec f_k, \Omega \vec f_k\}$ for all $j \neq k$,
\begin{equation}
\hat a^{\dag}(\vec f_1) \dots \hat a^{\dag}(\vec f_n) \ket{0} = \ket{1_{f_1}}\otimes \dots \otimes \ket{1_{ f_n}}.
\end{equation}
The Wigner function for this state can be shown to be (showing this based on (\ref{eq:wigDef}) is again a good exercise)
\begin{equation}
W_{1_{f_1},\dots, 1_{ f_n}}(\vec{x}_{\bf f_1} \oplus \dots \oplus  \vec{x}_{\bf f_n}) = \prod_{k=1}^nW_{1_{ f_k}}(\vec{x}_{\bf f_k}).
\end{equation}
Either by explicitly using the expression of the Wigner function, or by using the additivity property, we find that
\begin{equation}
{\frak N}\left( \ket{1_{f_1}}\otimes \dots \otimes \ket{1_{f_n}} \right) = n {\frak N}(\ket{1}).
\end{equation} 
Numerically, we can show that $n {\frak N}(\ket{1}) > {\frak N}(\ket{n})$ and thus we can generally conclude that $n$ photons in different modes hold more Wigner negativity than $n$ photons in the same mode.

Let us now concentrate on the case where $n = 2$. We showed that two photons in different modes are more Wigner negative than two photons in the same mode, and now we will combine this finding with the idea that Wigner negativity remains unchanged under Gaussian transformations. A particularly simple Gaussian transformation is a balanced beam splitter, which ultimately just implements a change in mode basis that we describe by an orthonormal transformation $O_{BS}$. When we mix two photons, prepared in orthogonal modes $f_1$ and $f_2$ by such a balanced beamsplitter, we will see the Hong-Ou-Mandel effect in action (more details can be found in \cite{Walschaers_2020} where a similar notation is used):
\begin{equation}\label{eq:HOM}
 \ket{1_{ f_1}}\otimes \ket{1_{ f_2}} \overset{O_{BS}}{\mapsto} \frac{1}{\sqrt{2}}\left(\ket{2_{ g_1}} - \ket{2_{ g_2}}\right) \coloneqq \ket{\rm HOM},
\end{equation}
where ${f}_1, {f}_2$ and ${g}_1, {g}_2$ are the input and output modes of the beam splitter, respectively. The Hong-Ou-Mandel output state $\ket{\rm HOM}$ is thus a superposition of two photons in mode $g_1$ and two photons in mode $g_2$. We can now use the simple fact that Wigner-negativity is unchanged under Gaussian unitary transformations to show that
\begin{equation}
{\frak N}\left( \ket{\rm HOM} \right) = {\frak N}\left( \ket{1_{f_1}}\otimes \ket{1_{f_2}} \right) = 2  {\frak N}(\ket{1}) >  {\frak N}(\ket{2}),
\end{equation}
this then also implies that ${\cal N}\left( \ket{\rm HOM} \right) > {\cal N}(\ket{2})$. At first sight, this is somewhat of a peculiar finding: by taking a superposition of two states with the same Wigner negativity one finds a state with a higher Wigner negativity.

An explicit look at the Wigner function of the Hong-Ou-Mandel state $\ket{\rm HOM}$ provides some insight. We find that this Wigner function can be written as (yet again a good exercise to show this explicitly)
\begin{equation}\begin{split}
W_{\rm HOM} (\vec x_{\bf g_1} \oplus \vec x_{\bf g_2}) = &\frac{1}{2}[W_{2_{g_1}}(\vec{x}_{\bf g_1}) + W_{2_{g_2}}(\vec{x}_{\bf g_2})] \\
&+ W_{\rm int} (\vec x_{\bf g_1} \oplus \vec x_{\bf g_2}),
\end{split}
\end{equation}
where $W_{\rm int}$ is the contribution to the Wigner function that contains all the interference terms that are induced by the superposition. We can calculate that
\begin{equation}
{\cal N}\left(\frac{1}{2}[W_{2_{g_1}}(\vec{x}_{\bf g_1}) + W_{2_{g_2}}(\vec{x}_{\bf g_2})]\right) \leqslant {\cal N}(\ket{2}),
\end{equation}
and thus, by additionally applying the triangle inequality, we can understand that the additional negativity in the Hong-Ou-Mandel state is due to the term $W_{\rm int}$. 

In the Hong-Ou-Mandel effect, it is common to talk about interference between particles, but in a more general CV language this interference will be equivalent to some form of entanglement which is exactly described by the Wigner function contribution $W_{\rm int}$. In other words, the superposition between $\ket{2_{g_1}}$ and $\ket{2_{g_2}}$ has more Wigner negativity than each of its two constituents because it creates entanglement between the modes $g_1$ and $g_2$. This is a first indication that there is a connection between quantum correlations and non-Gaussian features of the Wigner function. We will explore this connection in further detail in Section \ref{sec:quantumCorr}.\\

Even though Wigner negativity is an important non-Gaussian feature, it is often hard to witness \cite{chabaud2021witnessing}. The most common experimental technique is homodyne tomography \cite{Ourjoumtsev83} to fully reconstruct the quantum state. These methods come with the inconvenience that it is hard to set good error bars. Techniques to circumvent the need for a full tomography have been developed based on homodyne \cite{PhysRevLett.106.010403} and double-homodyne (or heterodyne) measurements \cite{chabaud2020certification,chabaud2021witnessing}. These methods come with the advantage of permitting to put a degree of confidence on the proclaimed Wigner negativity.

\section{Creating Non-Gaussian States}\label{sec:CreationNonGauss}

In Section \ref{sec:NonGaussianStates} we have discussed the many ways of characterising non-Gaussian quantum states and their properties. In this section, we explore the different theoretical frameworks for creating these states. An overview of some important experimental advances to put these theoretical techniques into practice is left for Section \ref{sec:Exp}. 

Gaussian quantum states can in some sense be understood as naturally occurring states. The foundational work of Planck that lies at the basis of all of quantum mechanics provides a first description of the thermal states of light that describe black body radiation. In a more modern language, we refer to this as the thermal states of an ensemble of quantum harmonic oscillators or a free bosonic field. It has long been understood that these states are Gaussian \cite{doi:10.1063/1.1704002,robinson_ground_1965,verbeure_many-body_2011}. Creating this kind of Gaussian states of light is thus literally as simple as switching on a light bulb. 

When we turn towards more sophisticated light sources such as lasers, we can encounter coherent light that is described by coherent states \cite{PhysRev.131.2766,PhysRevLett.10.277}. Generating squeezed light becomes much harder and typically requires nonlinear optics \cite{doi:10.1080/09500348714550721}. Nevertheless, pumping a nonlinear crystal with a coherent pump generally suffices to deterministically create a squeezed state \cite{PhysRevLett.57.2520}. Recall from the end of subsection \ref{sec:Gaussian} that from a theoretical point of view all these pure Gaussian states can be created by applying Gaussian unitary transformation to the vacuum state.

From an experimental point of view, the creation of non-Gaussian states is much harder than the creation of their Gaussian counterparts. Nevertheless, we start by introducing an ideal theoretical approach that is not too different from Gaussian states. In essence, it suffices to apply a non-Gaussian unitary operation to the state to create a non-Gaussian state. In Subsection \ref{sec:DetnonGaussian} we dig deeper into the desired structure of such non-Gaussian unitary transformations that would in principle allow for the deterministic generation of non-Gaussian quantum states. In experiments (in particular those in optics) such non-Gaussian unitary transformations are hard to come by, which is why one very often uses different preparation schemes. In Subsection \ref{sec:CondnonGaussian}, we provide a general introduction into the conditional preparation of non-Gaussian quantum states, where one measures part of the system and conditions on a certain measurement outcome. This process projects the remainder of the system into a new non-Gaussian state.

\subsection{Deterministic methods}\label{sec:DetnonGaussian}

To introduce some further structure in the sets of Gaussian and non-Gaussian unitary transformations, it is useful to take a quantum computation approach that is inspired by \cite{DiVincenzo255,PhysRevLett.82.1784}. The central idea of this work is that Gaussian unitary transformations are always generated by ``Hamiltonians'' that are at most quadratic in the quadrature operators (or equivalently in the creation and annihilation operators). Let us denote that as
\begin{equation}
\hat U_G = \exp \{ i {\cal P}_2(\hat q)\}
\end{equation}
where the polynomials ${\cal P}_2(\hat q)$ are generated by combining terms of the types $\mathds{1}$, $\hat q(\vec f)$, and $\hat q(\vec f_1)\hat q(\vec f_2)$. A remarkable property of these three types of observables is that they are closed under the action of a commutator. Indeed, using the canonical commutation relation (\ref{eq:CCRHere}) and the general properties of commutators, we can show that
\begin{align}
&[\hat q(\vec f_1), \hat q(\vec f_2)] \sim \mathds{1},\\
&[\hat q(\vec f_1), \hat q(\vec f_2)\hat q(\vec f_3)] \sim \hat q(\vec f'),\\
&[\hat q(\vec f_1)\hat q(\vec f_2),\hat q(\vec f_3)\hat q(\vec f_4)] \sim \sum \hat q(\vec f_1')\hat q(\vec f_2').
\end{align} 
Thus, we can use the Baker-Campbell-Hausdorff formula to show that the combination of two Gaussian unitaries $\hat U_G\hat U'_G$ is again a Gaussian unitary. 

This notion lies at the basis of universal gate sets in the CV approach. Using typical techniques from Lie groups, we can look for a minimal set of Gaussian unitaries than can be combined to generate all possible Gaussian unitary transformations. Generally, such a set is clearly not unique, but there are some natural choices. For example, we previously saw that a Gaussian unitary transformation is a combination of displacement operations and symplectic transformations. Furthermore, the Bloch-Messiah decomposition (\ref{eq:BlochMessiah}) shows us that any symplectic transformation can be decomposed into a combination of multimode interferometers and single-mode squeezing. In turn, interferometers can be decomposed as a combination of beamsplitters and phase shifters \cite{PhysRevLett.73.58}. Indeed, we can choose the set of Gaussian gates to be 
\begin{align}
&\hat U_D(\vec \lambda) \coloneqq \hat D(\vec \lambda),\\
&\hat U_S(\vec \lambda) \coloneqq \exp i [\hat q(\vec \lambda)\hat q(\Omega \vec \lambda) + \hat q(\Omega \vec \lambda)\hat q(\vec \lambda)],\\
&\hat U_P(\vec \lambda) \coloneqq  \exp i [\hat q(\vec \lambda)^2 + \hat q(\Omega \vec \lambda)^2],\\ 
&\hat U_{BS}(\vec \lambda_1,\vec \lambda_2) \coloneqq  \exp i [\hat q(\vec \lambda_1)\hat q(\vec \lambda_2) + \hat q(\Omega \vec \lambda_1)\hat q(\Omega \vec \lambda_2)], 
\end{align}
where we note that $\vec \lambda, \vec \lambda_1, \vec \lambda_2 \in \mathbb{R}^{2m}$ are not normalised and $\vec \lambda_1 \perp \vec \lambda_2$. These unitary operators describe a displacement, a squeezer, a phase shifter, and a beamsplitter, respectively. We note that all these operations act on a single modes, except for the beamsplitter which connects a pair of modes. These transformations are referred to as a Gaussian gate set;  when we can implement all these gates in all the modes of some mode basis, we can generate any multimode Gaussian transformation, and thus any Gaussian state.\\

To generate non-Gaussian unitary transformations and thus non-Gaussian states, we need to add more unitary gates to the Gaussian gate set. The relevant question is thus how many gates one should add and which gates are the best choices. The answer to the first question is surprising: one needs to add just one single gate \cite{PhysRevLett.97.110501}. The argument is simple, when we consider the operators $\hat q(\vec \lambda)^3$, we find that \begin{equation}\begin{split}\label{eq:nonGaussHam1}
[\hat q(\vec \lambda_1)^3,\hat q(\vec \lambda_2)^n]  = &\hat q(\vec \lambda_1)^2[\hat q(\vec \lambda_1),\hat q(\vec \lambda_2)^n] \\
&+ [\hat q(\vec \lambda_1),\hat q(\vec \lambda_2)^n]\hat q(\vec \lambda_1)^2\\
 &+ \hat q(\vec \lambda_1)[\hat q(\vec \lambda_1),\hat q(\vec \lambda_2)^n]\hat q(\vec \lambda_1).
\end{split}\end{equation}
With the canonical commutation relations we can show that $[\hat q(\vec \lambda_1),\hat q(\vec \lambda_2)^n]\sim\hat q(\vec \lambda_2)^{n-1}$, which can be inserted in (\ref{eq:nonGaussHam1}) to obtain
\begin{equation}\begin{split}
[\hat q(\vec \lambda_1)^3,\hat q(\vec \lambda_2)^n]  \sim &\hat q(\vec \lambda_1)^2\hat q(\vec \lambda_2)^{n-1} +\hat q(\vec \lambda_2)^{n-1}\hat q(\vec \lambda_1)^2\\&+ \hat q(\vec \lambda_1)\hat q(\vec \lambda_2)^{n-1}\hat q(\vec \lambda_1) .
\end{split}\end{equation}
This calculation thus shows that commutation with the operators $\hat q(\vec \lambda)^3$ increases the order of the quadrature operators. Thus, with the quadratic Hamiltonians to generate all operations that conserve the order of polynomials of quadrature operators and $\hat q(\vec \lambda)^3$ to increase the order of the polynomial by one, we can ultimately generate the full algebra of observables. On the level of unitary gates, this implies that a full universal gate set is given by
\begin{align}
{\cal U} = &\{\hat U_D(\vec \lambda), \hat U_S(\vec \lambda), \hat U_P(\vec \lambda), \hat U_{BS}(\vec \lambda_1,\vec \lambda_2), \hat U_{C}(\vec \lambda)\},\\
&\text{with } \quad \hat U_{C} \coloneqq \exp i \hat q(\vec\lambda)^3.
\end{align}
In other words, combining sufficiently many of these gates allows us to built any arbitrary unitary transformation generated by a Hamiltonian which is polynomial in the quadrature operators.

The non-Gaussian gate $\hat U_{C}$ is known as the cubic phase gate. The argument above shows that any experiment that can implement Gaussian transformations and a cubic phase gate can in principle generate any arbitrary non-Gaussian state. Even though many protocols have been proposed to experimentally realise a cubic phase gate \cite{gu_quantum_2009,PhysRevA.84.053802,PhysRevA.91.032321,PhysRevA.93.022301,PhysRevLett.124.240503}, any convincing implementations have yet to be demonstrated. One of the key problems is that experimental imperfections and finite squeezing are detrimental for the most commonly proposed methods \cite{arzani_polynomial_2017}.

In principle, there is no particular reason to limit our attention to cubic phase gates. Already in the very first work on the subject it is argued that essentially any Hamiltonian of a higher than quadratic order can be used a a generator \cite{PhysRevLett.82.1784}. Thus, optical processes that performs photon triplet generation can also be used as a non-Gaussian gate, which can even be converted into the cubic phase gate \cite{PRXQuantum.2.010327}. This requires well-controlled high $\chi^{(3)}$ nonlinearities which are generally only achieved by using exotic nonlinear crystals or well-controlled individual atoms. Handling such setups with a sufficient degree of control to actually implement a quantum gate is extremely challenging.

Other CV systems are more appropriate for the implementation of non-Gaussian unitary transformations. In particular the systems used in circuit-QED have such non-Gaussian contributions in their Hamiltonians \cite{Vion886,Devoret1169} which suggests that they may be more capable of deterministically generating non-Gaussian states than their optical counterparts. Still, the characterisation, detection, and control of such states is expected to be challenging. Recently, some important progress was made by demonstrating triplet-generation in these systems \cite{PhysRevX.10.011011}.

\subsection{Conditional methods}\label{sec:CondnonGaussian}

The experimental difficulties that are encountered when trying to implement non-Gaussian unitary transformations can be circumvented by abandoning the demand of unitarity. This implies that we no longer consider operations that can be implemented deterministically, but rather resort to what can be broadly referred to as conditional operations. This idea was formalised by Kraus when characterising the most general ways of manipulating quantum states \cite{KRAUS1971311}.

In the most general sense, we can implement a conditional operation by taking a set of linear operators on Fock space $\hat X_1, \hat X_2, \dots $ and acting on the state in the following way
\begin{equation}
\hat \rho \mapsto \frac{\sum_j \hat X_j \hat \rho \hat X_j^{\dag}}{\tr[ \hat \rho \sum_j \hat X_j^{\dag}\hat X_j]}.
\end{equation}
This formalism is typically implemented by performing some form of generalised measurement on the state $\hat \rho$ \cite{Holevo2001}. When $\hat\rho$ is a deterministically generated Gaussian state, the action of a well-chosen set of operators $\hat X_1, \hat X_2, \dots$ can turn it into a non-Gaussian state. In optics, two of the most well-known examples of this technique are single-photon addition and subtraction. In both cases, there is only a single operator $\hat X_1$. For photon addition, we implement $\hat X_1 = \hat a^{\dag}(\vec f)$, whereas photon subtraction requires the realisation of a case where $\hat X_1 = \hat a(\vec f)$.

In many physical setups, and in particular in optics, the problem is that measurements are destructive and a measurement effectively removes the measured mode from the system. Therefore, it is common to prepare large multimode Gaussian states of which a subset of modes is measured in order to conditionally prepare a non-Gaussian state in the remaining modes. We now introduce a general framework to describe the non-Gaussian Wigner functions that are created accordingly \cite{PRXQuantum.1.020305}.\\

\subsubsection{General framework}

\begin{figure}
\centering
\includegraphics[width=0.49\textwidth]{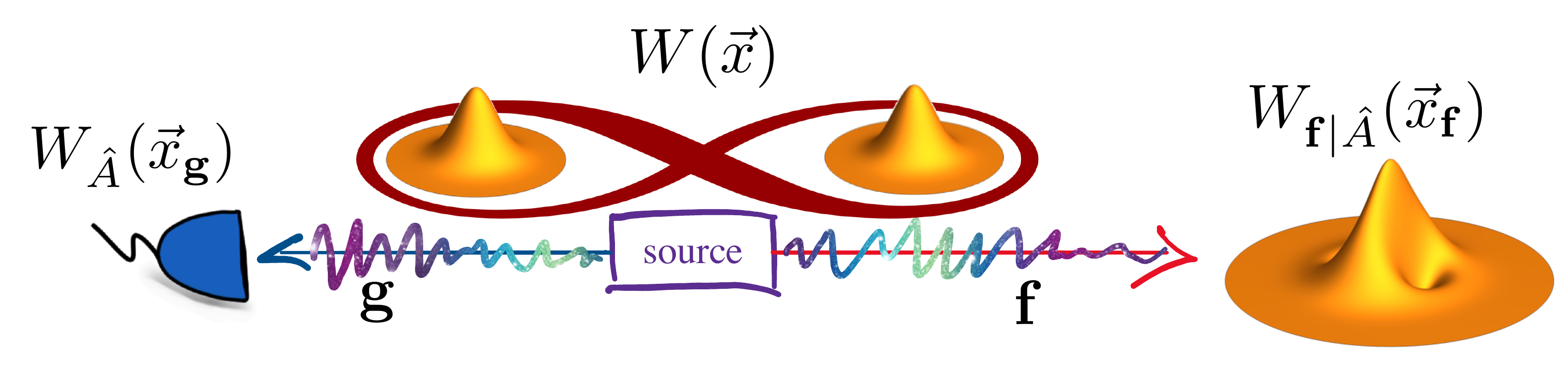}\\
\caption{Sketch representation of the conditional preparation scheme for creating the non-Gaussian states described by (\ref{eq:CondTwo}). Note that both ${\bf f}$ and ${\bf g}$ can be highly multimode. The Wigner function shown on the right was obtained by a conditional protocol shown in \cite{PRXQuantum.1.020305}.
} \label{fig:Conditional}
\end{figure}

First of all, let us consider a general multimode phase space and separate it into two subsystems, i.e., $\mathbb{R}^{2m} = \mathbb{R}^{2l} \oplus \mathbb{R}^{2l'}$, where we will perform some generalised measurement on the $l'$ modes and leave the remaining $l$ modes untouched. This introduces a general structure in the points of phase space $\vec x \in \mathbb{R}^{2m}$, which can now be written as $\vec x = \vec x_{\bf f} \oplus \vec x_{\bf g}$ with $\vec x_{\bf f} \in \mathbb{R}^{2l}$ and $\vec x_{\bf g} \in \mathbb{R}^{2l'}$. The general procedure is schematically outlined in Fig.~\ref{fig:Conditional} and we will present the details step by step.

Any state $\hat \rho$ on this system then comes with a Wigner function $W(\vec x) = W(\vec x_{\bf f} \oplus \vec x_{\bf g})$ that is defined on the global phase space. This state can be reduced to one of the two subsystems by tracing out the other subsystem, which can be described on the level of the Wigner function by the following integrals:
\begin{align}
W_{\bf f}(\vec x_{\bf f}) \coloneqq \int_{\mathbb{R}^{2l'}}{\rm d}\vec x_{\bf g} W(\vec x_{\bf f} \oplus \vec x_{\bf g}),\label{eq:Wf}\\
W_{\bf g}(\vec x_{\bf g}) \coloneqq \int_{\mathbb{R}^{2l}}{\rm d}\vec x_{\bf f} W(\vec x_{\bf f} \oplus \vec x_{\bf g}).\label{eq:Wg}
\end{align}
When the state is Gaussian and the Wigner function is given by (\ref{eq:GaussStateWig}), the structure of the phase space is reflected in the mean field vector $\vec \xi$ and in the covariance matrix $V$:
\begin{align}
&\vec \xi = \vec \xi_{\bf f} \oplus \vec \xi_{\bf g},\\
&V = \begin{pmatrix}V_{\bf f} & V_{\bf fg}\\ V_{\bf gf} & V_{\bf g}\end{pmatrix},\label{eq:CovGaussBipart}
\end{align}
with $V_{\bf fg} = V_{\bf gf}^T$. The matrices $V_{\bf f}$ and $V_{\bf g}$ describe all the variances and correlations of the modes within $\mathbb{R}^{2l}$ and $\mathbb{R}^{2l'}$, respectively. In addition, the submatrix $V_{\bf gf}$ contains all the correlations between the modes in the different subspaces, which will be important for conditional state preparation. One can show that for such Gaussian states, the reduced states are also Gaussian, for the modes in $\mathbb{R}^{2l}$ given by
\begin{equation}
W_{\bf f}(\vec x_{\bf f}) = \frac{e^{-\frac{1}{2}(\vec x_{\bf f} - \vec \xi_{\bf f})^TV_{\bf f}^{-1}(\vec x_{\bf f} - \vec \xi_{\bf f})}}{(2\pi)^m \sqrt{\det V_{\bf f}}},
\end{equation}
and analogously for the modes in $\mathbb{R}^{2l'}$. As Gaussian states are the states that are least challenging to produce, they form the starting point of the conditional state preparation scheme.

As a next step, we must implement some form of operation on the modes that correspond to the phase space $\mathbb{R}^{2l'}$. To do so, we consider the action of a general POVM element $\hat A \geqslant 0$ that corresponds to a specific measurement outcome. We can then obtain a conditional state via
\begin{equation}\label{eq:StateCond}
\hat \rho_{\bf f \mid A} \coloneqq \frac{\tr_{\bf g}[\hat A \hat \rho]}{\tr[\hat A \hat \rho]},
\end{equation}
The partial trace $\tr_{\bf g}[\hat A \hat \rho]$ only runs over the modes in $\mathbb{R}^{2l'}$ because the other modes are left untouched. The denominator $\tr[\hat A \hat \rho]$ renormalises the state and gives the probability of actually obtaining the measurement result that corresponds to $\hat A$. In an actual experiment, this operation is implemented by many repeated measurements of the modes in $\mathbb{R}^{2l'}$ and $\hat A$ corresponds to a specific detector output of these measurements. The non-measured part of the state is only used when the detector indicates this specific output, otherwise it is simply discarded. This conditional selection of the state significantly changes the properties of the state in a way that is strongly influenced by $\hat A$. 

As we described in (\ref{eq:WigA}), the operator $\hat A$ comes with an associated phase-space representation $W_{A}(\vec x_{\bf g})$ which can be used to formally describe the phase-space representation of $\hat \rho_{\bf f \mid A}$:
\begin{equation}\label{eq:CondOne}
W_{\bf f \mid A}(\vec x_{\bf f}) = \frac{\int_{\mathbb{R}^{2l'}}{\rm d}\vec x_{\bf g} W_A(\vec x_{\bf g})W(\vec x_{\bf f} \oplus \vec x_{\bf g}) }{\int_{\mathbb{R}^{2l'}}{\rm d}\vec x_{\bf g} W_A(\vec x_{\bf g})W_{\bf g}(\vec x_{\bf g})}.
\end{equation}
There is a more practical way of expressing this Wigner function by exploiting the fact that the initial multimode Wigner function $W(\vec x_{\bf f} \oplus \vec x_{\bf g})$ is positive and therefore describes a well-defined probability distribution on phase space. This implies that the conditional probability distribution
\begin{equation}\label{eq:CondWigDef}
W( \vec x_{\bf g} \mid \vec x_{\bf f}) \coloneqq \frac{W(\vec x_{\bf f} \oplus \vec x_{\bf g})}{W_{\bf f}(\vec x_{\bf f})},
\end{equation}
is also a well-defined probability distribution, which is obtained when we fix one point in phase space $\vec x_{\bf f} \in \mathbb{R}^{2l}$ and look at the probability distribution for the remaining modes in $\mathbb{R}^{2l'}$. We can then use this conditional probability distribution to write $W(\vec x_{\bf f} \oplus \vec x_{\bf g}) = W( \vec x_{\bf g} \mid \vec x_{\bf f})W_{\bf f}(\vec x_{\bf f})$, which can be inserted in (\ref{eq:CondOne}) to find
\begin{equation}\label{eq:CondTwo}
W_{\bf f \mid A}(\vec x_{\bf f}) = \frac{\<\hat A\>_{{\bf g} \mid \vec x_{\bf f}}}{\<\hat A\>}W_{\bf f}(\vec x_{\bf f}),
\end{equation}
where we define
\begin{align}
&\<\hat A\> \coloneqq (4\pi)^{l'}\int_{\mathbb{R}^{2l'}}{\rm d}\vec x_{\bf g} W_A(\vec x_{\bf g})W_{\bf g}(\vec x_{\bf g}),\\
&\<\hat A\>_{{\bf g} \mid \vec x_{\bf f}} \coloneqq (4\pi)^{l'}\int_{\mathbb{R}^{2l'}}{\rm d}\vec x_{\bf g} W_A(\vec x_{\bf g})W( \vec x_{\bf g}\mid \vec x_{\bf f})\label{eq:Agf}.
\end{align}
The quantity $\<\hat A\>$ is simply the expectation value of the observable $\hat A$ in the state $\hat \rho$. $\<\hat A\>_{{\bf g} \mid \vec x_{\bf f}},$ on the other hand, is the expectation value of the function $W_A(\vec x_{\bf g})$ where $\vec x_{\bf g}$ is distributed according to the distribution $W( \vec x_{\bf g}\mid \vec x_{\bf f})$, which makes $\<\hat A\>_{{\bf g} \mid \vec x_{\bf f}}$ a function of the selected phase space point $\vec x_{\bf f}$. However, even though $W( \vec x_{\bf g}\mid \vec x_{\bf f})$ is a well-defined probability distribution on phase space, it does not necessarily correspond to a quantum state. Indeed, $W( \vec x_{\bf g}\mid \vec x_{\bf f})$ may violate the Heisenberg inequality which will be of vital importance in Section \ref{sec:quantumCorr} as it is narrowly connected to quantum steering. 

In the specific case where $W(\vec x_{\bf f} \oplus \vec x_{\bf g})$ is Gaussian, we find that $W( \vec x_{\bf g} \mid \vec x_{\bf f})$ is also a Gaussian probability distribution, given by
\begin{equation}\label{GaussCond}
    W(\vec{x}_{\bf g} \mid \vec{x}_{\bf f}) = \frac{\exp\left[-\frac{1}{2} (\vec x_{\bf g}-\vec{\xi}_{{\bf g} \mid \vec x_{\bf f}})^T V_{{\bf g}\mid \vec x_{\bf f}}^{-1}(\vec x_{\bf g}- \vec{\xi}_{{\bf g} \mid \vec x_{\bf f}})\right]}{(2\pi)^{l'} \sqrt{\det V_{{\bf g}\mid \vec x_{\bf f}}}}.
\end{equation}
Using the notation of (\ref{eq:CovGaussBipart}), we express its covariance matrix
\begin{equation}\label{eq:schur}
V_{{\bf g}\mid \vec x_{\bf f}} = V_{\bf g} - V_{{\bf gf}}V_{\bf f}^{-1}V_{\bf gf}^T,
\end{equation}
and mean field vector
\begin{equation}\label{eq:schurvec}
\vec{\xi}_{{\bf g} \mid \vec x_{\bf f}} = \vec{\xi}_{\bf g} + V_{\bf gf} V_{\bf f}^{-1} (\vec{x}_{\bf f} - \vec{\xi}_{\bf f}).
\end{equation}
The covariance matrix $V_{{\bf g}\mid \vec x_{\bf f}}$ is known in the mathematics literature \cite{horn_matrix_2017} as the Schur complement of $V$. The Schur complement is interesting properties, for example, $V$ is a positive matrix if and only if the same holds for the Schur complement  $V_{{\bf g}\mid \vec x_{\bf f}}$. This immediately implies that the Gaussian probability distribution in (\ref{GaussCond}) is well-defined. Furthermore, the Schur complement also plays an important role in the theory of Gaussian quantum correlations \cite{Lami_2018}. It should be noted that $V_{{\bf g}\mid \vec x_{\bf f}}$ does not actually depend on the chosen value for $\vec x_{\bf f}$. Thus, the conditional expectation value $\<\hat A\>_{{\bf g}\mid \vec x_{\bf f}}$ only depends on the phase space point $\vec x_{\bf f}$ through the displacement $\vec{\xi}_{{\bf g} \mid \vec x_{\bf f}}$. This is a particular feature of Gaussian states. 

Finally, remark that the derivation of (\ref{eq:CondTwo}) holds true for all initial states with a positive Wigner function. Whenever the initial multimode Wigner function $W(\vec x_{\bf f} \oplus \vec x_{\bf g})$ is positive, it follows that $W_{\bf f}(\vec x_{\bf f})$ is also positive. Furthermore, given that $\<\hat A\>$ is the quantum expectation value of a positive semi-definite operator it clearly also is a positive quantity. Hence, Wigner negativity is entirely contained with $\<\hat A\>_{{\bf g} \mid \vec x_{\bf f}}$. The fact that $\<\hat A\>_{{\bf g} \mid \vec x_{\bf f}}$ can take negative values is exactly due to $W( \vec x_{\bf g}\mid \vec x_{\bf f})$ not being the Wigner function of a quantum state. Furthermore, (\ref{eq:Agf}) teaches us that the conditionally generated Wigner function $W_{\bf f \mid A}(\vec x_{\bf f})$ can only achieve negative value when $W_A(\vec x_{\bf g})$ is non-positive.

Thus, in order to conditionally prepare a state with Wigner negativity, one faces strict requirements, on both the POVM element $\hat A$ that is conditioned upon and on the conditional probability distribution $W( \vec x_{\bf g}\mid \vec x_{\bf f})$ that is obtained from the initial multimode state. We will discuss this point in greater detail in Section \ref{sec:NonGaussByCorr}. For a more experimentally-inclined perspective on the production of non-Gaussian states, we refer to \cite{lvovsky2020production}.\\

Before we move on to consider photon subtraction as an example of conditional creation of non-Gaussian states, let us take a moment the opposite process: Gaussification. The authors of \cite{PhysRevA.67.062320} consider several copies of an initial non-Gaussian state which are mixed through linear optics and subsequently some output modes are measured with on-off detectors. The conditioning is done of the events where no photons are detected, and such that we can interpret $\hat A$ as a projector in vacuum. By repeating several iterations of this scheme (assuming many successful conditioning events), the initial non-Gaussian state is converted into a Gaussian state. The Gaussification process thus relies on starting from a non-Gaussian state and conditioning by projecting on a Gaussian state: the vacuum. This point of view nicely complements our approach to create non-Gaussian states.

\subsubsection{An example: photon subtraction}\label{sec:PhotonSubtraction}

Single-photon subtracted states are theoretically obtained by acting with an annihilation operator on the state. Their density matrices are given by
\begin{equation}
\hat \rho^- = \frac{\hat a (\vec b) \hat \rho \hat a^{\dag} (\vec b)}{\tr [\hat a^{\dag} (\vec b)\hat a (\vec b) \hat \rho]},
\end{equation}
if the photon is subtracted in one specific mode $\bf b$. In practice \cite{PhysRevLett.92.153601,Ourjoumtsev83,zavatta_subtracting_2008}, we can implement this operation on the state $\hat \rho$ through a mode-selective beamsplitter $\hat U = \exp\{\theta [\hat a^{\dag}(\vec g)\hat a(\vec b) - \hat a^{\dag}(\vec b)\hat a(\vec g)]\}$, that couples the mode $\bf b$ to an auxiliary mode $\bf g$ which is prepared in a vacuum state. We thus describe the action of the beamsplitter on the system of interest and the auxiliary mode as $\hat U(\hat \rho \otimes \ket{0}\bra{0})\hat U^{\dag}$. As a next step, we mount a photon-detector on one of the output modes of the beamsplitter and condition on events where the detectors count a single photon. Thus, we generate the state 
\begin{equation}\label{eq:photonSubBS}
\hat \rho^{-}_{\theta} = \frac{ \tr_{\bf g} [ \hat U(\hat \rho \otimes \ket{0}\bra{0})\hat U^{\dag} (\mathds{1} \otimes \ket{1}\bra{1})]}{ \tr [ \hat U(\hat \rho \otimes \ket{0}\bra{0})\hat U^{\dag} (\mathds{1} \otimes \ket{1}\bra{1})]}.
\end{equation}
The reader can now recognise (\ref{eq:StateCond}). As a next step, we assume that the beamsplitter is transmitting nearly all the incoming light, such that $\theta \rightarrow 0$. We can then approximate $\hat U \approx \mathds{1} + \theta[ \hat a^{\dag}(\vec g)\hat a(\vec b) - \hat a^{\dag}(\vec b)\hat a(\vec g)]$. Then, when we insert this approximation in the expression for $\hat \rho^{-}_{\theta}$, we find that only the terms $\sim \theta^2$ survive such that
\begin{equation}
\hat \rho^{-}= \lim_{\theta \rightarrow 0}\hat \rho^{-}_{\theta} = \frac{\hat a (\vec b) \hat \rho \hat a^{\dag} (\vec b)}{\tr [\hat a^{\dag} (\vec b)\hat a (\vec b) \hat \rho]}.
\end{equation}
A much more detailed analysis of multimode photon subtraction with imperfect mode-selectivity can be found in \cite{averchenko_multimode_2016}. We note that through this approach, photon subtraction can be understood as a weak measurement of the number of photons \cite{wiseman_milburn_2009}.\\

\begin{figure}
\centering
\includegraphics[width=0.49\textwidth]{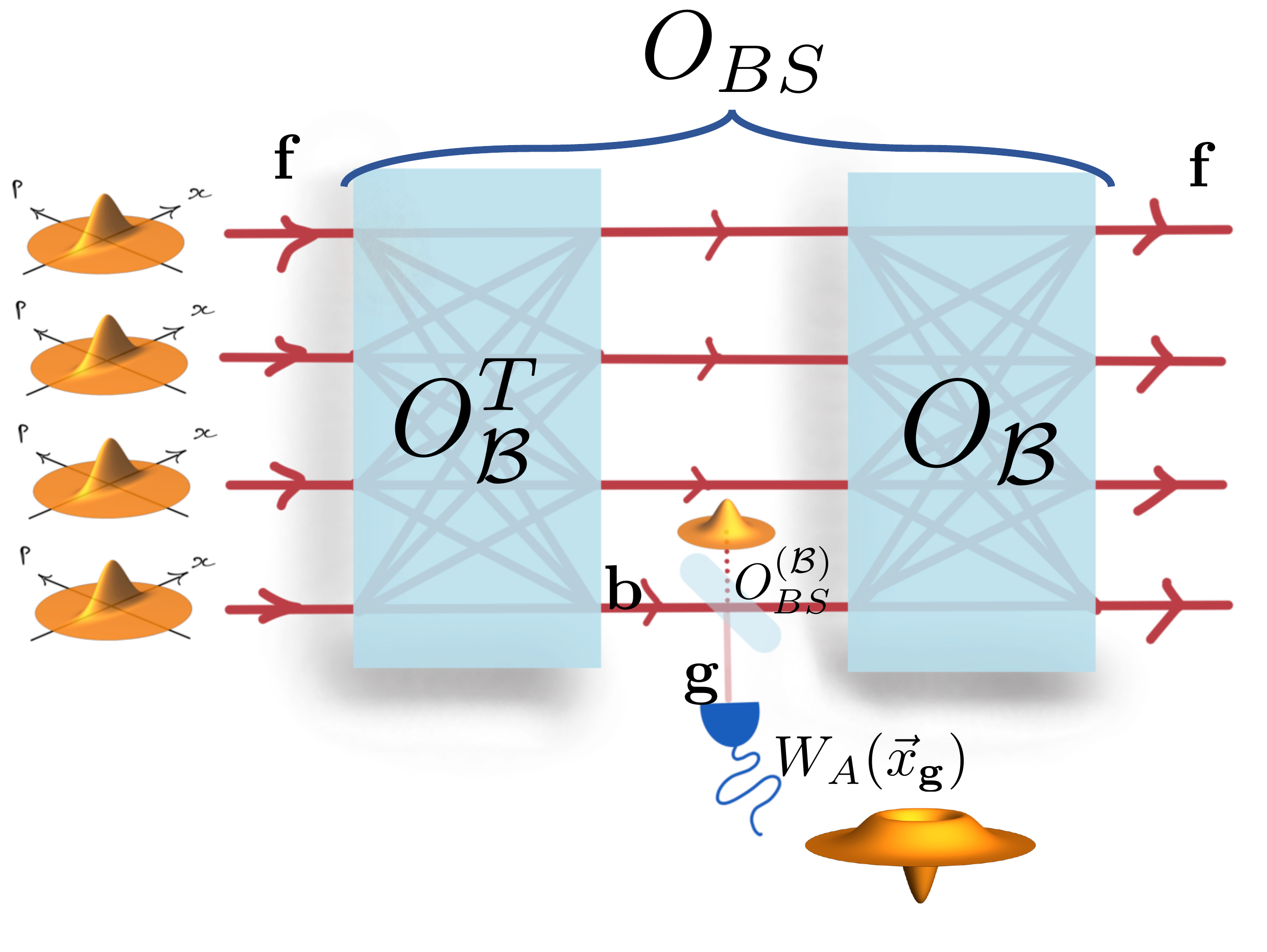}\\
\caption{Schematic representation of an implementation of mode-selective photon subtraction. See main text for details. For illustrtation, the initial state on the left is a product of single-mode squeezed vacuum states, but the protocol can in principle be applied to any Gaussian state.
} \label{fig:PhotonSubtraction}
\end{figure}

We can now derive the Wigner function of a single-photon-subtracted state through (\ref{GaussCond}) by following the idea of (\ref{eq:photonSubBS}). We initially start from a Gaussian state with covariance matrix $V_{\bf f}$ and one auxiliary mode that is prepared in the vacuum
\begin{equation}
V_{\rm ini}=\begin{pmatrix}
V_{\bf f} & 0 \\
0 & \mathds{1}
\end{pmatrix}.
\end{equation}
We then implement a mode-selective beamsplitter that mixes one specific mode ${\bf b}$ with the auxiliary vacuum mode, following the scheme outlined in Fig.~\ref{fig:PhotonSubtraction}. An effective way to describe such a transformation is by designing a new mode basis ${\cal B}$, which has ${\bf b}$ as one of the modes in the mode basis. We complete the basis with complementary modes $b^c_1, \dots b^c_{m-1}$, such that the modes basis of phase space is given by ${\cal B} = \{ \vec b^c_1,  \Omega \vec b^c_1, \dots, \vec b^c_{m-1}, \Omega \vec b^{c}_{m-1},  \vec b, \Omega \vec b\}$. Thus, we can perform such a basis change as
\begin{equation}
\begin{pmatrix}
V_{\bf f} & 0 \\
0 & \mathds{1}
\end{pmatrix} \mapsto
\begin{pmatrix}O^T_{\cal B} & 0\\0& \mathds{1}\end{pmatrix}\begin{pmatrix}
V_{\bf f} & 0 \\
0 & \mathds{1}\end{pmatrix}\begin{pmatrix}O_{\cal B} & 0\\0& \mathds{1}\end{pmatrix},
\end{equation}
where the matrix of basis change is given by
\begin{equation}
    O_{\cal B} = \begin{pmatrix}
     \mid & \mid& &\mid & \mid& \mid & \mid \\
   \vec b^c_1&  \Omega \vec b^c_1& \dots& \vec b^c_{m-1}& \Omega \vec b^{c}_{m-1}& \vec b & \Omega \vec b\\
  \mid & \mid& &\mid & \mid&   \mid & \mid \\
    \end{pmatrix}.
\end{equation}
It is now instructive to explicitly write the rows and columns corresponding to mode $b$:
\begin{equation}
O^T_{\cal B}V_{\bf f} O_{\cal B} = \begin{pmatrix} V^c_{\bf f} & V^{cb}_{\bf f}\\
V^{bc}_{\bf f} & V^{b}_{\bf f}  \end{pmatrix}.
\end{equation}
Note that $V^{b}_{\bf f} $ is the $2 \times 2$ matrix that describes the initial state covariances of mode $b$, while $V^{c}_{\bf f} $ is the $(m-1) \times (m-1)$ that describes all the covariances in the complementary modes. The rectangular matrices $ V^{cb}_{\bf f}$ and $ V^{bc}_{\bf f}$ contain all the correlations between the mode $b$ and the complementary modes in the basis. Now, we mix the mode $b$ and the auxiliary vacuum mode on a beam splitter. This beamsplitter is implemented by the transformation
\begin{equation}\begin{split}
V^{(\cal B)}_{BS}=O_{BS}^{(\cal B)}\begin{pmatrix} V^c_{\bf f} & V^{cb}_{\bf f} & 0\\
V^{bc}_{\bf f} & V^{b}_{\bf f}  & 0 \\
0& 0 & \mathds{1}
\end{pmatrix}
{O_{BS}^{(\cal B)}}^T,
\end{split}
\end{equation}
where $O_{BS}^{(\cal B)}$ is given by
\begin{equation}
O_{BS}^{(\cal B)}=\begin{pmatrix} \mathds{1} & 0 & 0\\
0 & \cos \theta \mathds{1}  & - \sin \theta \mathds{1}\\
0& \sin\theta \mathds{1} &\cos \theta \mathds{1}
\end{pmatrix}.
\end{equation}
As a final step, we change the basis back to the original basis, such that the final state's covariance matrix becomes
\begin{equation}
V = \begin{pmatrix}O_{\cal B} & 0\\0& \mathds{1}\end{pmatrix} V^{(\cal B)}_{BS} \begin{pmatrix}O^T_{\cal B} & 0\\0& \mathds{1}\end{pmatrix}.
\end{equation}
We can now rewrite this entire transformation such that the matrix $V$ in (\ref{eq:CovGaussBipart}) is given by
\begin{equation}
V =  O_{BS} V_{\rm ini} O_{BS}^T,
\end{equation}
with
\begin{equation}\begin{split}
 O_{BS} &= \begin{pmatrix}O_{\cal B} & 0\\0& \mathds{1}\end{pmatrix} O_{BS}^{(\cal B)} \begin{pmatrix}O_{\cal B}^T & 0\\0& \mathds{1}\end{pmatrix}\\
 &=\begin{pmatrix}
 (\cos \theta - 1) \, BB^T +\mathds{1} & \sin \theta \, B\\
 -\sin \theta \, B^T & \cos \theta \mathds{1}
 \end{pmatrix}.
 \end{split}
\end{equation}
We introduce the $2m \times 2$ matrix $B$ which implements the mode-selectivity of the beamsplitter in mode $b$ and is defined as
\begin{equation}
    B = \begin{pmatrix}
    \mid & \mid \\
    \vec b & \Omega \vec b\\
    \mid & \mid \\
    \end{pmatrix}.
\end{equation}
Hence, we can simply use $O_{BS}$ as a mode-selective beamsplitter that mixes one specific mode of a multimode state with the auxiliary mode. We should highlight that $O_{BS}$ ultimately turns out to be independent of the complementary modes $b^c_1, \dots b^c_{m-1}$. This means that the finer details of the interferometer $O_{\cal B}$ are not important for the final $O_{BS}$, the key point is that $O_{\cal B}$ changes towards a mode basis in which $\vec b$ and $\Omega \vec b$ are basis vectors of the phase space.

For the particular case of photon subtraction, we consider a very weak beamsplitter, such that we consider the limit $\theta \rightarrow 0$. In this case, we can express the conditional mean field (\ref{eq:schurvec}) and covariance matrix (\ref{eq:schur}) of the auxiliary mode ${\bf g}$ by
\begin{align}\label{eq:schur2}
&\vec{\xi}_{{\bf g} \mid \vec x_{\bf f}} \approx \theta (V_{\bf bf}-B^T) V_{\bf f}^{-1} (\vec{x}_{\bf f} - \vec{\xi}_{\bf f}) + \theta \vec \xi_{\bf b}\\
&\qquad\qquad= \theta  B^T(\mathds{1} - V_{\bf f}^{-1}) (\vec{x}_{\bf f} - \vec{\xi}_{\bf f}) + \theta \vec \xi_{\bf b}\nonumber\\
&V_{{\bf g}\mid \vec x_{\bf f}} \approx \mathds{1} +\theta^2\left(V_{\bf b} - \mathds{1} - (V_{\bf bf}-B^T) V_{\bf f}^{-1}(V_{\bf fb}-B)\right),\\
&\qquad\qquad= \mathds{1} +\theta^2\left( \mathds{1} - B^T V_{\bf f}^{-1}B\right),\nonumber
\end{align}
where we introduce the matrices $V_{\bf b} = B^TV_{\bf f}B$, $V_{\bf b f} = B^TV_{\bf f}$, and $V_{\bf fb} = V_{\bf f}B$ as well as the vector $\vec{\xi}_{\bf b} = B^T\vec \xi_{\bf f}$. We can then use these quantities to evaluate that
\begin{align}
    &W( \vec x_{\bf g}\mid \vec x_{\bf f})\label{eq:approxExp}\approx\\
    & \frac{e^{-\frac{1}{2}\norm{\vec x_{\bf g}}}}{(2\pi)^m}\Big(1+\theta \, \vec x_{\bf g}^T \cdot B^T(\mathds{1} - V_{\bf f}^{-1}) (\vec{x}_{\bf f} - \vec{\xi}_{\bf f})+\theta \vec x_{\bf g}^T\cdot \vec \xi_{\bf b}\nonumber\\
    &\qquad+\frac{\theta^2}{2}\big[(\vec x_{\bf g}^T \cdot B^T(\mathds{1} - V_{\bf f}^{-1}) (\vec{x}_{\bf f} - \vec{\xi}_{\bf f})+\vec x_{\bf g}^T\cdot \vec \xi_{\bf b})^2 \nonumber\\
    &\qquad - \norm{B^T(\mathds{1} - V_{\bf f}^{-1}) (\vec{x}_{\bf f} - \vec{\xi}_{\bf f})+ \vec \xi_{\bf b}}^2+ \vec{x}_{\bf g}^T  \left( \mathds{1} - B^T V_{\bf f}^{-1}B\right) \vec{x}_{\bf g}\big] \nonumber\\
    &\qquad+ {\cal O}(\theta^3)\Big).\nonumber
\end{align}
As a next step, we must choose a POVM element $\hat A$ to measure. In the case of photon subtraction, we mount a photon counter on the auxiliary mode and for single-photon subtraction we condition on the event where this detector detects exactly one photon. Because we use a very weakly reflective beamsplitter, the probability of obtaining such an event is small but when it occurs, we have created a photon subtracted state on the remaining modes.

On a theoretical level, mounting a photon counter and conditioning on a single photon is translated to choosing $\hat A = \ket{1_{g}}\bra{1_{g}}$. We already encountered the corresponding Wigner function in (\ref{eq:WigFock}), and thus we can combine this with (\ref{eq:approxExp}) to obtain
\begin{equation}\begin{split}
\<\hat A\>_{{\bf g} \mid \vec x_{\bf f}} &= 4\pi \int_{\mathbb{R}^{2}}{\rm d}\vec x_{\bf g} W_{1_{\vec g}}(\vec x_{\bf g})W( \vec x_{\bf g}\mid \vec x_{\bf f})\\
&\approx \frac{\theta^2}{2}\Big(\norm{B^T(\mathds{1} - V_{\bf f}^{-1}) (\vec{x}_{\bf f} - \vec{\xi}_{\bf f})+ \vec \xi_{\bf b}}^2\\
    &\qquad+\tr\left[\mathds{1} - B^T V_{\bf f}^{-1}B\right]\Big) + {\cal O}(\theta^3),
\end{split}
\end{equation}
and in a similar fashion we find that
\begin{equation}\begin{split}
\<\hat A\> &= 4\pi \int_{\mathbb{R}^{2}}{\rm d}\vec x_{\bf g} W_{1_{\vec g}}(\vec x_{\bf g})W_{\bf g}( \vec x_{\bf g})\\
&\approx \frac{\theta^2}{2} \left(\tr\left[V_{\bf b} - \mathds{1}\right]+ \norm{\vec \xi_{\bf b}}^2\right) + {\cal O}(\theta^3).
\end{split}
\end{equation}
The actual evaluation of these integrals is not completely straightforward. As a key idea, we use that the integral takes the form of a polynomial multiplied by a Gaussian. We can thus evaluate the expectation value of the polynomial with respect to this Gaussian distribution. In essence, this boils down to calculating a set of moments of a Gaussian probability distribution. We see rather quickly that the lowest orders in $\theta$ vanish, such that the leading order is $\theta^2$. Putting everything together, we find that
\begin{equation}\begin{split}\label{eq:PhotonSubtractedState}
&\lim_{\theta \rightarrow 0} W_{\bf f \mid A}(\vec x_{\bf f}) = \lim_{\theta \rightarrow 0} \frac{\<\hat A\>_{{\bf g} \mid \vec x_{\bf f}}}{\<\hat A\>}W_{\bf f}(\vec x_{\bf f})\\
&=\frac{\norm{B^T(\mathds{1} - V_{\bf f}^{-1}) (\vec{x}_{\bf f} - \vec{\xi}_{\bf f})+ \vec \xi_{\bf b}}^2 +\tr\left[\mathds{1} - B^T V_{\bf f}^{-1}B\right]}{\tr\left(V_{\bf b} - \mathds{1}\right)+\norm{\vec \xi_{\bf b}}^2} W_{\bf f}(\vec x_{\bf f}).
\end{split}\end{equation}
As such, we obtain the Wigner function for a multimode photon-subtracted state. This Wigner function can be obtained using several different methods, ranging from algebraic \cite{walschaers_entanglement_2017,walschaers_statistical_2017} to analytical \cite{PhysRevA.90.013821}. The difference between those approaches and our method here is that we do not directly use the properties of the annihilation operator, but rather model the exact experimental setup, while relying entirely on phase-space representations.

The methods presented here for treating photon-subtracted states can straightforwardly be extended to the subtraction of multiple photons in different modes and we can easily replace photon-number resolving detection with an on-off detector by setting $\hat A = \mathds{1} - \ket{0}\bra{0}$. The techniques used in the calculations remain essentially the same and it yields the same result in the $\theta \rightarrow 0$ limit (doing this calculation may prove to be a good exercise for the motivated Reader). However, any real implementation of a photon subtraction experiment will use a beamsplitter with finite reflectivity, such that the will be a difference between on-off detectors and photon-number resolving detectors due to the small contributions of higher order terms in $\theta$. In practice, one chooses the reflectivity of the beamsplitter with respect to the energy content of the initial state to effectively suppress all higher order terms in \eqref{eq:approxExp}. In the single-mode case, an early thorough analysis of the implementation of photon-subtraction can be found in \cite{PhysRevA.73.063804}. There are also proposals in literature to use a photon subtraction setup with larger values of $\theta$ to gain an additional advantage in quantum state preparation \cite{dakna_generating_1997,PhysRevA.103.013710,Davis_2021}.

As a final note, we point out that a similar treatment can be used to describe photon-added states, which are also relevant in experiments \cite{Zavatta660,parigi_probing_2007}. It is perhaps surprising that such state can be obtained by performing a measurement on a part of a Gaussian state, but it suffices to replace the beamsplitter in (\ref{eq:photonSubBS}) with a two-mode squeezer. In other words, we set $\hat U = \exp\{\theta [\hat a^{\dag}(\vec g)\hat a^{\dag}(\vec b) - \hat a(\vec b)\hat a(\vec g)] \}$, and consider again the limit where the parameter $\theta$ is small, i.e. weak squeezing. Even though this is a simple step in theory, it is much harder in an actual experimental setting. Photon subtraction can be implemented with a passive linear optics element, while photon addition always requires squeezing and thus a nonlinear optics implementation.

\section{Non-Gaussian States and Quantum Correlations}\label{sec:quantumCorr}

In this Section, we explore the interplay between non-Gaussian effects and quantum correlations. First, in Subsection \ref{sec:QuantumCorrelations}, we provide a crash course to introduce the unfamiliar reader to the most important types of quantum correlations: entanglement, steering, and Bell non-locality. In Subsection \ref{sec:NonGaussByCorr} we will subsequently highlight how certain types of quantum correlations can be used to create certain types of non-Gaussian states via the methods of Subsection \ref{sec:CondnonGaussian}. In Subsection \ref{sec:QunatumCorrThroughnonGauss}, we then explore how non-Gaussian operations can create or enhance quantum correlations by focusing on photon-subtracted states. Finally, we will explore the role that is played by non-Gaussian states in Bell inequalities in Subsection \ref{sec:Bell}.

\subsection{Quantum correlations: a crash course}\label{sec:QuantumCorrelations}

We start by giving a quick introduction to the different kinds of common quantum correlations. Readers who want to get a more thorough overview on these subjects are referred Refs.~\cite{RevModPhys.81.865, RevModPhys.92.015001, RevModPhys.86.419} as natural starting points.\\

In this Tutorial, we solely consider bipartite quantum correlations. This implies that we structure the system in a similar way as in Subsection \ref{sec:CondnonGaussian} and divide the $m$-mode system in two parts, each with their own phase space, i.e., $\mathbb{R}^{2m} = \mathbb{R}^{2l} \oplus \mathbb{R}^{2l'}$. It is noteworthy that the corresponding Fock space takes the structure $\Gamma({\cal H}_m) = \Gamma({\cal H}_l) \otimes \Gamma({\cal H}_{l'})$, where we again use the mapping (\ref{eq:PhaseHilb}) between the phase space $\mathbb{R}^{2k}$ and the $k$-dimension Hilbert space ${\cal H}_k$. These structures are crucial to understand quantum correlations.

\subsubsection{Correlations}

To better understand quantum correlations, it is useful to start by generally defining what a correlations is. In a statistical sense, two stochastic variables $X$ and $Y$ are correlated when the expectation values have the following property
\begin{equation}
\mathbb{E}(XY) \neq \mathbb{E}(X)\mathbb{E}(Y).
\end{equation}
This can be translated to the level of probability distributions by stating that the joint probability distribution for outcomes $X=x$ and $Y=y$ is not the product of the marginals
\begin{equation}\label{eq:classCorrProb}
P(x,y) \neq P(x)P(y),
\end{equation}
where
\begin{equation}
P(x) = \int_{\cal Y}{\rm d}y \,P(x,y), \text{ and } P(y) = \int_{\cal X}{\rm d}x \,P(x,y).
\end{equation}
Here, ${\cal X}$ and ${\cal Y}$ denote the possible outcomes of the stochastic variables $X$ and $Y$, respectively \footnote{One could introduce more general and rigorous notation based on measure theory, but here we restrict to a simpler, though less general, formulation for pedagogical purposes.}. 

When we talk about quantum systems, there are generally many observables that can be considered. When we consider a global multimode system with phase space $\mathbb{R}^{2m}$ and two subsystems with phase spaces $\mathbb{R}^{2l}$ and $\mathbb{R}^{2l'}$, there is a whole algebra of observables involved. The role of the stochastic observables $X$ and $Y$ will be taken up by local observables $\hat X$ and $\hat Y$ that are contained in the observable algebra generated by, respectively, $\hat q(\vec f)$ and $\hat q(\vec g)$, with $\vec f \in \mathbb{R}^{2l}$ and $\vec g \in \mathbb{R}^{2l'}$. These local observables are correlated when
\begin{equation}\label{eq:CorrQuantum}
\tr (\hat X \otimes \hat Y \hat \rho) \neq \tr (\hat X \hat \rho_{\bf f}) \tr(\hat Y \hat \rho_{\bf g}),
\end{equation}
where $\hat \rho_{\bf f}$ and $\hat \rho_{\bf g}$ are the marginals (or reduced states) of $\hat \rho$ for the subsystems $\mathbb{R}^{2l}$ and $\mathbb{R}^{2l'}$.

When we talk about correlated systems rather than correlated observables, we consider that there exists a pair of local observables such that (\ref{eq:CorrQuantum}) holds. Thus, if two systems are not correlated, it follows that for all possible observables $\tr (\hat X \otimes \hat Y \hat \rho) = \tr (\hat X \hat \rho_{\bf f}) \tr(\hat Y \hat \rho_{\bf g})$. This lack of correlations can be expressed on the level of the quantum state by the idenity $\hat \rho = \hat \rho_{\bf f} \otimes \hat \rho_{\bf g}$. On the level of Wigner functions, we can therefore say that a state contains correlations if the Wigner function satisfies
\begin{equation}\label{eq:corrWig}
W(\vec x_{\bf f} \oplus \vec x_{\bf g}) \neq W_{\bf f}(\vec x_{\bf f}) W_{\bf g}(\vec x_{\bf g}), 
\end{equation}
where the marginal Wigner functions are defined as in (\ref{eq:Wf}, \ref{eq:Wg}).

It is clear that correlations between systems can occur, both, in the context of classical probability theory and in quantum theory. However, we already established that quantum physics imposes additional contraints on the statistics of observables, which ultimately make it impossible to describe CV quantum systems in terms of probability distributions on phase space. Similarly, quantum physics leads to new features for the correlations of subsystems. Thus, in our study of quantum correlations we explore correlated systems, in the sense of (\ref{eq:corrWig}), and we seek to differentiate between correlations that are of classical origin and those that can be attributed to a quantum origin.

\subsubsection{Quantum entanglement}\label{eq:explanEntanglement}

Quantum entanglement is probably the most well-known type of quantum correlation. The notion of entanglement derives directly from the structure of the quantum state space and is related to the contrast between pure states in classical and quantum physics. 

To understand this contrast, we loosely follow the idea of \cite{19991657}. Let us be a bit more precise as to what is meant with pure states in classical physics in the context of CV systems. Classically, in a context of statistical mechanics, any CV system can be described by a probability distribution on phase space. From a mathematical point of view, this means that the space containing all the possible classical states is a convex set because any convex combination of two probability distributions is again a probability distribution. Pure states are formally defined as the extreme points of the convex set, i.e., the states that cannot be decomposed as being a convex combination of two other states. In a classical theory, where states can unambiguously be represented by probability distributions on phase space, the pure states are delta functions centred on the different points of phase space. From a physical point of view, this corresponds to the intuition that pure states are ``the least noisy'' states, which simply corresponds to a single point in phase space.

For our phase space $\mathbb{R}^{2m}= \mathbb{R}^{2l} \oplus \mathbb{R}^{2l'}$ these delta functions factorise with respect to the subsystems, i.e., $\delta(\vec x -\vec x') = \delta(\vec x_{\bf f} -\vec x'_{\bf f})\delta(\vec x_{\bf g} -\vec x'_{\bf g})$, with $\vec x, \vec x'\in \mathbb{R}^{2m}$, $\vec x_{\bf f},\vec x'_{\bf f} \in  \mathbb{R}^{2l} $, and $\vec x_{\bf g},\vec x'_{\bf g} \in  \mathbb{R}^{2l'}$. In the light of (\ref{eq:classCorrProb}) we thus conclude that pure states of classical systems are always uncorrelated \footnote{This is true for general ``states'' in classical probability theory, but in the CV context of this Tutorial it is appealing to phrase the argument in terms of probability distributions of phase space.}. Any correlations that are present in classical states are thus obtained by taking a convex combination of uncorrelated pure states.\\

In quantum systems, pure states are represented by state vectors in a Hilbert space (in our case Fock space). They also can be seen as the extreme points of a convex set of states that contains all density matrices $\hat \rho$. As we saw in the example where we discussed the Hong-Ou-Mandel state $\ket{\rm HOM}$ in (\ref{eq:HOM}), pure quantum states can actually be correlated in the sense of (\ref{eq:corrWig}). This crucial difference between classical and quantum pure states lies at the basis of quantum entanglement.

The notion of entanglement derives directly from the structure of the quantum state and is defined as the opposite of a separable state. For pure states, separable states $\ket{\Psi} \in \Gamma({\cal H}_m)$ are the pure states that are uncorrelated and can thus be written as $\ket{\Psi} = \ket{\Psi_l}\otimes \ket{\Psi_{l'}}$ with $\ket{\Psi_l} \in  \Gamma({\cal H}_l) $ and $\ket{\Psi_{l'}} \in  \Gamma({\cal H}_{l'})$. All other pure states are said to be entangled. They possess correlations that are not due to some type of convex combination of uncorrelated states, something which is impossible for classical pure states.

The situation is more subtle when considering mixed states, i.e. convex combinations of pure states. Convex mixtures of classical pure states can also show correlations, and it is therefore crucial to make a distinction between this type of classical correlations and quantum correlations. Due to the structure of classical pure states, we find that any classical joint probability distribution on phase space can be written as a convex combination of local probability distributions
\begin{equation}\label{eq:classCorrProb2}
P(\vec x_{\bf f} \oplus \vec x_{\bf g}) = \int {\rm d}\gamma  p(\gamma)P^{(\gamma)}(\vec x_{\bf f})P^{(\gamma)}(\vec x_{\bf g}),
\end{equation}
where $\gamma$ is some arbitrary way of labelling states, distributed according to distribution $p(\gamma)$. This notion of classical correlations can directly be generalised to quantum states \cite{PhysRevA.40.4277}, and thus a mixed state is said to be separable when all of its correlations are classical, i.e., when it is a convex mixture of product states
\begin{equation}
\hat \rho = \int {\rm d}\gamma \, p(\gamma) \hat \rho^{(\gamma)}_{\bf f} \otimes \hat \rho^{(\gamma)}_{\bf g}.
\end{equation}
In the language of Wigner functions, the separability condition translates to 
\begin{equation}\label{eq:classCorrWig}
W(\vec x_{\bf f} \oplus \vec x_{\bf g}) = \int {\rm d}\gamma  p(\gamma)W_{\bf f}^{(\gamma)}(\vec x_{\bf f})W_{\bf g}^{(\gamma)}(\vec x_{\bf g}),
\end{equation}
where we again use the definitions of (\ref{eq:Wf}, \ref{eq:Wg}). Quantum states that cannot be described by a Wigner function of the form (\ref{eq:classCorrWig}) are not separable and are said to be entangled.\\

Hence, quantum entanglement describes the origin of the quantum correlations rather than their properties. Nevertheless, the set of separable states is a convex set and thus the Hahn-Banach separation theorem \cite{Conway2007BanachSpaces,SOHAIL2021127411} teaches us that it is in principle possible to use observables to distinguish between separable and entangled states. In this sense the difference between entangled and separable states is measurable. For the sake of uniformity, we highlight that separable states lead to the following measurement statistics of local observables $\hat X$ and $\hat Y$:
\begin{equation}\label{eq:LHVEntanglement}
P(x, y) = \int {\rm d}\gamma  p(\gamma)P^{(\gamma)}_{\hat \rho}(x)P_{\hat \rho}^{(\gamma)}(y).
\end{equation}
It is crucial to emphasise that the distributions of measurement outcomes $P^{(\gamma)}_{\hat \rho}(x)$ and $P^{(\gamma)}_{\hat \rho}(y)$ are governed by the laws of quantum physics. Formally, we can use the spectral theorem to write 
\begin{equation}
\hat X = \int_{\cal X} {\rm d}x \, x \hat E_x, \quad \text{and} \quad \hat Y = \int_{\cal Y} {\rm d}y \, y \hat E_y,
\end{equation}
such that $\hat E_{x}$ and $\hat E_{y}$ are the POVM elements that correspond to the measurement outcomes $x$ and $y$ for the measurement of the (generalised) observables $\hat X$ and $\hat Y$, respectively. The probability distribution $P^{(\gamma)}_{\hat \rho}(x)$ is then given by
\begin{equation}\label{eq:localProbQuant}
P^{(\gamma)}_{\hat \rho}(x) = \tr[\hat E_x \hat \rho^{(\gamma)}_{\bf f}] = (4\pi)^l \int_{\mathbb{R}^{2l}}{\rm d}\vec x_{\bf f} \, W_{E_x}(\vec x_{\bf f})W_{\bf f}(\vec x_{\bf f}),
\end{equation}
and analogously for $P^{(\gamma)}_{\hat \rho}(y)$. 

For separable states, the equation (\ref{eq:LHVEntanglement}) with local probability distribution given by (\ref{eq:localProbQuant}) holds for any arbitrary pair of local observables. The model that is described by these equations is known as a local hidden variable model for quantum entanglement, where $\gamma$ is the hidden variable. We may not necessarily know the origins and behaviour of $\gamma$, but the model generally captures two important elements. First, all correlations are induced by the common variable $\gamma$ that governs the convex mixture. Second, the local probability distributions $P^{(\gamma)}_{\hat \rho}(x)$ and $P^{(\gamma)}_{\hat \rho}(y)$ have a quantum origin. For CV systems the latter point for example implies that these local probability distributions must satisfy the Heisenberg inequality. These quantum constraints on the local probability distributions $P^{(\gamma)}_{\hat \rho}(x)$ and $P^{(\gamma)}_{\hat \rho}(y)$ are typically useful for the falsification of the local hidden variable model (\ref{eq:LHVEntanglement}) and thus prove the presence of quantum entanglement \cite{PhysRevA.68.032103,PhysRevA.78.052319}.

\subsubsection{Quantum steering}

In a formal sense, quantum steering is a rather recent addition to the family of quantum correlations. Nevertheless, it is exactly this phenomenon that lies at the basis of the Einstein-Podolsky-Rosen (EPR) paradox \cite{Einstein:1935aa}. Schr\"odinger's response to the EPR paper \cite{Schrodinger:1935aa,schrodinger_1936} lies at the basis of what we now call quantum steering, but the broader implications of these results were only sporadically discovered and formalised \cite{PhysRevA.40.913,Wiseman:2007aa}.

Just like for quantum entanglement, a system is said to be steerable if the measurement statistics cannot be explained in terms of a local hidden variable model. A peculiarity of quantum steering is that it involves a certain directionality, where one of the subsystems is said to "steer" the other subsystem. This asymmetry is represented in the local hidden variable model, which takes the following form:
\begin{equation}\label{eq:LHVSteering}
P(x, y) = \int {\rm d}\gamma  p(\gamma)P^{(\gamma)}(x)P_{\hat \rho}^{(\gamma)}(y),
\end{equation}
where we emphasise the striking resemblance to (\ref{eq:LHVEntanglement}). Note that, contrary to the case of quantum entanglement, we now allow the probability distribution $P^{(\gamma)}(x)$ for the first subsystem to be arbitrary and thus do not impose any constraints of quantum theory on it. If there exist observables $\hat X$ and $\hat Y$ for which the probability distribution is not consistent with the model (\ref{eq:LHVSteering}), the subsystems with phase space $\mathbb{R}^{2l}$ is able to steer the subsystem with phase space $\mathbb{R}^{2l'}$.

Quantum steering is perhaps most logically explained in terms of conditional states and probability distributions. For non-steerable states, the local hidden variable model (\ref{eq:LHVSteering}) must hold for all observables, which in turn imposes conditions on the level of states. Here these conditions manifest on the level of conditional states of the type (\ref{eq:StateCond}). To see this, we consider the conditional probability distribution associated with (\ref{eq:LHVSteering}):
\begin{equation}\label{eq:LHVSteeringCond}
P(y\mid x) = \frac{\int {\rm d}\gamma  p(\gamma)P^{(\gamma)}(x)P_{\hat \rho}^{(\gamma)}(y)}{P(x)},
\end{equation}
where the probability to obtain a certain outcome $\hat X = x$ is given by
\begin{equation}
P(x) = \int {\rm d}\gamma  p(\gamma)P^{(\gamma)}(x).
\end{equation}
Note that for any $x$ the function
\begin{equation} 
\tilde{P}(\gamma\mid x)\coloneqq \frac{p(\gamma)P^{(\gamma)}(x)}{P(x)}
\end{equation}
is a well-defined probability distribution. Furthermore, if we demand that (\ref{eq:LHVSteeringCond}) holds for all observables $\hat Y$, we find the following condition for the conditional state:
\begin{equation}\label{eq:condstatesteering}
\hat \rho_{{\bf g} \mid \hat X = x} = \int {\rm d}\gamma \, \tilde{P}(\gamma\mid x) \hat \rho_{\bf g}^{(\gamma)}. 
\end{equation}
Because quantum steering is a property of the state, we again require (\ref{eq:condstatesteering}) to hold for all observables $\hat X$ for a state to not be steerable.\\

The local hidden variable model (\ref{eq:LHVSteering}) and the consequence for the conditional state (\ref{eq:condstatesteering}) may seem stringent, but it is often intricate to formally prove that such a model cannot explain observed data. It turns out that computational methods based on semidefinite programming \cite{0034-4885-80-2-024001} are well suited to prove that the set of all possible conditional states is inconsistent with (\ref{eq:condstatesteering}). A more physical point of view is based on developing steering inequalities \cite{PhysRevA.80.032112}. As a notable example, one can derive a type of conditional Heisenberg inequality for states of the form (\ref{eq:condstatesteering}). The local hidden variable model (\ref{eq:LHVSteering}) assumes that the laws of quantum physics constrain the measurement statistics in the second subsystem. We can then define the conditional variance of an arbitrary observable $\hat Y$
\begin{equation}
\Delta^2(\hat Y \mid \hat X = x) \coloneqq \tr[\hat Y^2 \hat \rho_{{\bf g} \mid \hat X = x}] - \tr[\hat Y \hat \rho_{{\bf g} \mid \hat X = x}]^2,
\end{equation}
which leads to the ``average inference variance''
\begin{equation}
\Delta^2_{\rm inf}(\hat Y) \coloneqq \int_{\cal X} {\rm d}x P(x) \Delta^2(\hat Y \mid \hat X = x),
\end{equation}
that characterises the precision with which we can infer the measurement outcome of $\hat Y$, given a measurement outcome of $\hat X$. Under the assumption that (\ref{eq:condstatesteering}) holds, we can than prove the inference Heisenberg inequality \cite{PhysRevA.80.032112}
\begin{equation}\label{eq:condHeisenberg}
\Delta^2_{\rm inf}(\hat Y_1)\Delta^2_{\rm inf}(\hat Y_2) \geqslant \frac{1}{2} \int_{{\cal X}_3}{\rm d}x P(x) \abs{\tr\left([\hat Y_1, \hat Y_2]  \hat \rho_{{\bf g} \mid \hat X_3 = x} \right)}^2,
\end{equation}
where $\Delta^2_{\rm inf}(\hat Y_1)$ and $\Delta^2_{\rm inf}(\hat Y_2)$ can be conditioned on any observables $\hat X_1$ and $\hat X_2$, respectively.

Thus, whenever one performs a series of conditional measurements that violate the inference Heisenberg inequality (\ref{eq:condHeisenberg}), the assumption (\ref{eq:condstatesteering}) cannot hold and thus the measurements in the subsystem with phase space $\mathbb{R}^{2l}$ have steered those in the subsystem with phase space $\mathbb{R}^{2l'}$. In more colloquial terms, the inequality (\ref{eq:condHeisenberg}) sets a limit on the precision with which classical correlations between observables can be used to infer measurement outcomes of one quantum subsystem, based on measurement outcome of the other subsystem (regardless of whether it is quantum or not). Quantum correlations allow us to outperform these bounds and provide better inference than classically possible, and this phenomenon is the essence of quantum steering.

Now let us now express (\ref{eq:condHeisenberg}) for quadrature operators:
\begin{equation}\label{eq:HeisenbergCondQuadratures}
\Delta^2_{\rm inf}[\hat q(\vec g_1)]\Delta^2_{\rm inf}[\hat q(\vec g_2)] \geqslant \abs{\vec g_1^T \Omega \vec g_2}^2,
\end{equation}
where $\vec g_1, \vec g_2 \in \mathbb{R}^{2l'}$. As a next step, we must understand the properties of the average inference variance $\Delta^2_{\rm inf}[\hat q(\vec g_1)]$, which we obtain by conditioning on a quadrature observable in the other subsystem's phase space $\mathbb{R}^{2l}$. More specifically let us assume that we condition on measurements of $\hat q(\vec f_1)$, such that we must evaluate the conditional variance $\Delta^2(\hat q(\vec g_1) \mid \hat q(\vec f_1) = x)$. The conditional variance $\Delta^2(\hat q(\vec g_1) \mid \hat q(\vec f_1) = x)$ is then given by the matrix element of the covariance matrix that describes $W(\vec x_{\bf g} \mid x \vec f_1)$ as defined in (\ref{GaussCond}):
\begin{equation}
\Delta^2(\hat q(\vec g_1) \mid \hat q(\vec f_1) = x) = \vec g_1^T \left[ V_{\bf g} - \frac{V_{\bf g f}\vec f_1 \vec f_1^T V_{\bf fg}}{ \vec f^T_1 V_{\bf f} \vec f_1}  \right]\vec g_1,
\end{equation}
because the quantity does not depend on the actual outcome that is post-selected upon, we find that 
\begin{align}\label{eq:DeltaInf1}
&\Delta^2_{\rm inf}[\hat q(\vec g_1)] = \vec g_1^T \left[ V_{\bf g} - \frac{V_{\bf g f}\vec f_1 \vec f_1^T V_{\bf fg}}{ \vec f^T_1 V_{\bf f} \vec f_1}  \right]\vec g_1,\\
&\Delta^2_{\rm inf}[\hat q(\vec g_2)] = \vec g_2^T \left[ V_{\bf g} - \frac{V_{\bf g f}\vec f_2 \vec f_2^T V_{\bf fg}}{ \vec f^T_2 V_{\bf f} \vec f_2}  \right]\vec g_2\label{eq:DeltaInf2}
\end{align}
From \eqref{eq:DeltaInf1} and \eqref{eq:DeltaInf2} we can deduce that $\Delta^2_{\rm inf}[\hat q(\vec g)] \geqslant \vec g^T V_{{\bf g}\mid \vec x_{\bf f}} \vec g$ for all $\vec g \in \mathbb{R}^{2l'}$ regardless of the $\hat q(\vec f)$ that is conditioned upon. Thus, if $V_{{\bf g}\mid \vec x_{\bf f}}$ satisfies the Heisenberg inequality the inference Heisenberg inequality (\ref{eq:HeisenbergCondQuadratures}) is also satisfied. 

The setting with homodyne measurements, or more general Gaussian measurements, is close to the system that is discussed in \cite{Einstein:1935aa}. For this reason, we will refer to quantum steering with Gaussian measurements as EPR steering in contrast to more general quantum steering. This type of steering has been studied extensively in literature, e.g., \cite{Wiseman:2007aa,Kogias:2015aa,PhysRevLett.114.060402,Deng:2017aa} and will be a key element in Section \ref{SteeringWignerNeg}.\\

Note that both subsystems clearly play a very different role in this setting. The first subsystem simply produces measurement results of different observables. 
The information of these measurements in the first subsystem is then used to infer measurement results in the second subsystem, which is assumed to be a quantum system. In a quantum communication context, this asymmetry corresponds to a level of trust: we position ourselves in the steered system and trust that our system is a well-behaved quantum system, but we do not trust the party that controls the other subsystem (up to a point where we do not even want to assume that the data that are communicated to us come from an actual quantum system). The violation of a steering inequality practically allows to verify in such a setting that there is indeed a quantum correlation between the two subsystems \cite{Branciard:2012aa}.\\

The inference Heisenberg inequality (\ref{eq:condHeisenberg}) shows that quantum steering describes certain properties of the quantum correlations. States that can perform quantum steering thus possess correlations that can be used to infer measurement outcomes better than any classical correlations could. These correlations cannot be described by a hidden variable model of the form (\ref{eq:LHVSteering}), which is more general than the model (\ref{eq:LHVEntanglement}). Thus, all states that produce statistics consistent with (\ref{eq:LHVEntanglement}) are also consistent with (\ref{eq:LHVSteering}) such that states that can perform steering must be entangled. However, there are states that produce statistics that is consistent with (\ref{eq:LHVSteering}), but inconsistent with (\ref{eq:LHVEntanglement}). In other words, not all entangled states can be used to perform quantum steering. In this sense, quantum steering can be said to be ``stronger'' than quantum entanglement.

\subsubsection{Bell nonlocality}\label{sec:BellIntro}

To date, the seminal work of John S. Bell on the Einstein-Podolsky-Rosen paradox \cite{PhysicsPhysiqueFizika.1.195} is probably one of the most remarkable findings on the foundations of quantum physics. What most had long taken for granted, the existence of local hidden variables to explain the probabilistic nature of quantum physics, turned out to be inconsistent with the theoretical quantum formalism. It is here that we find the real historical origin of the concept of quantum correlations as something fundamentally different from classical ones.

As for quantum entanglement and steering, the story of Bell nonlocality starts from a local hidden variable model that bears a strong resemblance to (\ref{eq:LHVEntanglement}) and (\ref{eq:LHVSteering}). In this case, the model attempts to describe the joint measurement statistics of $\hat X$ and $\hat Y$ as 
\begin{equation}\label{eq:LHVBell}
P(x, y) = \int {\rm d}\gamma  p(\gamma)P^{(\gamma)}(x)P^{(\gamma)}(y).
\end{equation}
The key observation is that now all the quantum constraints on the probability distributions have been dropped and both the $P^{(\gamma)}(x)$ and $P^{(\gamma)}(y)$ can be any mathematically well-defined probability distributions. Even though the difference between (\ref{eq:LHVEntanglement}) and (\ref{eq:LHVSteering}) on the one hand, and (\ref{eq:LHVBell}) on the other hand, may appear small, the impact of dropping the constraints on the local distributions is enormous. Think for example of the Hahn-Banach separation theorem that is invoked to define entanglement witnesses, this crucially relies on the Hilbert space structure of the state space. Think for example of (\ref{eq:condHeisenberg}) which crucially depends on the fact that quantum probabilities are constrained by the Heisenberg inequality. Abandoning all connections that tie probabilities to operator algebras on Hilbert spaces deprives quantum mechanics of their toolbox. Nevertheless, it turns out that some quantum states induce statistics that is inconsistent with (\ref{eq:LHVBell}).

Again, we note that states that can be described by the models (\ref{eq:LHVEntanglement}) or (\ref{eq:LHVSteering}) can also be described by the model (\ref{eq:LHVBell}). Bell's local hidden variable model (\ref{eq:LHVBell}) is thus the most general one and the class of states that lead to measurement statistics that cannot be described by it is the smallest. Therefore, we say that the correlations that lead to a violation of the model (\ref{eq:LHVBell}), also known as Bell nonlocality, are the strongest types of quantum correlations.\\

The key insight of Bell's work \cite{PhysicsPhysiqueFizika.1.195, RevModPhys.38.447} is that (\ref{eq:LHVBell}) puts constraints on the correlations of different combinations of observables in the subsystems. These constraints, cast in the form of Bell inequalities can be violated by certain quantum states. The inconsistency of quantum physics with the model (\ref{eq:LHVBell}) can in itself be seen as a special case of contextuality \cite{RevModPhys.65.803}. Over the decades, many different kinds of Bell inequalities have been derived (see for example \cite{Peres1999}). Here we restrict to presenting one of the most commonly used incarnations: the Clauser-Horne-Shimony-Holt (CHSH) inequality \cite{PhysRevLett.23.880}. This inequality relies on the measurement of four observables: $\hat X$ and $\hat X '$ on the first subsystem and $\hat Y$ and $\hat Y '$ on the second subsystem. Furthermore, we consider that the observables can take two possible values: $-1$ or $1$. Assuming the model in (\ref{eq:LHVBell}) it is then possible to derive
\begin{equation}\label{eq:CHSH}
\abs{\<\hat X \hat Y\> - \<\hat X \hat Y'\> + \<\hat X' \hat Y\> + \<\hat X' \hat Y'\>} \leqslant 2,
\end{equation}
where $\<.\>$ denotes the expectation value. In this Tutorial we skip the derivation of this result, but the interested reader is referred to \cite{PhysRevD.10.526} for a detailed discussion. Remarkably, certain highly entangled states can violate this inequality.\\

The experimental violation of Bell's inequalities formally shows that quantum correlations are profoundly different than classical correlations \cite{PhysRevLett.28.938,PhysRevLett.47.460,PhysRevLett.49.1804,PhysRevLett.81.5039}. However, one needs clever combinations of several observables in both subsystems to actually observe the difference. With most experimental loopholes now closed \cite{HansonBell,PhysRevLett.115.250401,PhysRevLett.115.250402,BIGBell,PhysRevLett.121.080403}, Bell inequalities can now in principle be used to impose a device independent level of security on various quantum protocols \cite{PhysRevLett.98.230501}. 

As a concluding remark, it is interesting to highlight the existence of a semi-device independent framework for testing quantum correlations \cite{PhysRevLett.108.200401, PhysRevLett.110.060405}. The key idea is that nothing is assume about the measurements devices nor about the states, much like in the scenario of Bell inequalities. Yet, in the framework of \cite{PhysRevLett.108.200401, PhysRevLett.110.060405} one does add an additional level of trust in the sense that one assumes that the inputs of the measurement device can be controlled and trusted. In a way, this additional intermediate level of trust is somewhat reminiscent of quantum steering. This framework was very recently extended to the CV setting \cite{PhysRevLett.126.190502}.

\subsection{Non-Gaussianity through Quantum Correlations}\label{sec:NonGaussByCorr}

In Subsection \ref{sec:CondnonGaussian}, we explained how conditional operations can be used to create non-Gaussian quantum states. The presence of correlations plays an essential role in this framework. Indeed, in the absence of correlations the combination of (\ref{eq:classCorrWig}) and (\ref{eq:CondWigDef}) implies that $W(\vec x_{\bf g} \mid \vec x_{\bf f}) = W_{\bf g}(\vec x_{\bf g})$. As a consequence, we see from (\ref{eq:Agf}) for the conditional expectation value $\<\hat A\>_{{\bf g}\mid \vec x_{\bf f}} = \<\hat A\>$, and thus from (\ref{eq:CondTwo}) that $W_{{\bf f}\mid \hat A}(\vec x_{\bf f}) = W_{{\bf f}}(\vec x_{\bf f})$. In other words, the conditional operation has no effect whatsoever and gives the same result as tracing out the modes in $\mathbb{R}^{2l'}$.

A closer look at the explicit expressions 
\begin{equation}\tag{\ref{eq:CondTwo}}
W_{\bf f \mid A}(\vec x_{\bf f}) = \frac{\<\hat A\>_{{\bf g} \mid \vec x_{\bf f}}}{\<\hat A\>}W_{\bf f}(\vec x_{\bf f}),
\end{equation}
and 
\begin{equation}
\<\hat A\>_{{\bf g} \mid \vec x_{\bf f}} \coloneqq (4\pi)^{l'}\int_{\mathbb{R}^{2l'}}{\rm d}\vec x_{\bf g} W_A(\vec x_{\bf g})W( \vec x_{\bf g}\mid \vec x_{\bf f})\tag{\ref{eq:Agf}},
\end{equation}
shows that whenever there are correlations, and thus $\<\hat A\>_{{\bf g} \mid \vec x_{\bf f}} \neq \<\hat A\>$, the conditional Wigner function is a priori non-Gaussian. When we use explicitly that the initial state is Gaussian and thus that $W(\vec x_{\bf g} \mid \vec x_{\bf f}) $ is given by (\ref{GaussCond}), this condition can be translated to the existence of non-zero components in $V_{\bf gf}$ in (\ref{eq:CovGaussBipart}). The precise properties of the resulting non-Gaussian quantum state depend on the conditional expectation value $\<\hat A\>_{{\bf g} \mid \vec x_{\bf f}}$.

\begin{figure*}
\centering
\includegraphics[width=0.8\textwidth]{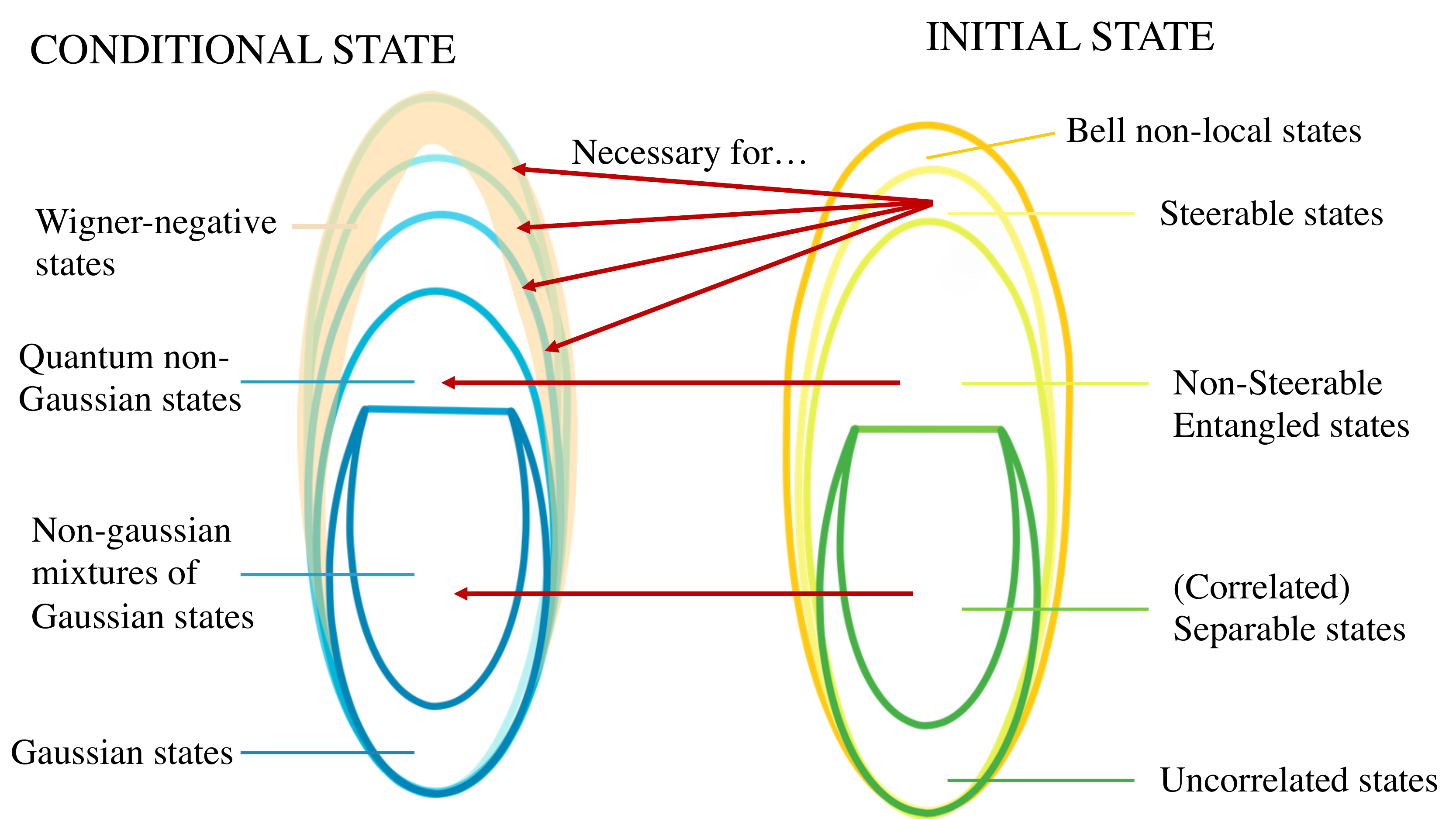}\\
\caption{Different types of quantum correlations are required to be present in the initial Gaussian state $W(\vec x)$ to create conditional states $W_{\bf f \mid A}(\vec x_{\bf f})$, as described in (\ref{eq:CondTwo}), that belong to a certain class. We thus show how the typical hierarchy of quantum correlations (right) can be connected to the structure of the CV state space that was introduced previously in Fig.~\ref{fig:nonGaussianOnion}. Throughout Subsection \ref{sec:NonGaussByCorr}, we prove that these different types of quantum correlations are necessary resources to achieve different types of states.}
 \label{fig:nonGaussianCorr}
\end{figure*}

In literature, some attention has been devoted to proposing different types of measurements for such heralding procedures. One may think of using on-off detectors \cite{PhysRevLett.56.58}, photon-number resolving detectors \cite{dakna_generating_1997}, parity detectors \cite{Thekkadath2020engineering}, and more exotic multimode setups \cite{Eaton_2019,su2019generation}. However, these works usually assume that the initial quantum state is a pure Gaussian state obtained by an idealised source of multimode squeezed vacuum states. As we saw in Subsection \ref{eq:explanEntanglement}, for pure states correlations automatically imply entanglement, and it even turns out that all correlated pure states violate a Bell inequality \cite{PhysRevLett.109.120402}. In other words, for pure states all correlations are quantum correlations and all these quantum correlations are of the strongest type. When we no longer make such assumptions on the initial multimode Gaussian state, we will see that $\<\hat A\>_{{\bf g} \mid \vec x_{\bf f}}$ will not only depend on the chosen POVM $\hat A$, but also on the properties of $W(\vec x_{\bf g} \mid \vec x_{\bf f})$. In the Subsections \ref{sec:EntQuantmNonGauss} and \ref{SteeringWignerNeg}, we explain that certain types of non-Gaussian features can only be achieved through certain types of quantum correlations in the initial Gaussian state. An overview of the results of this section is provided in Fig.~\ref{fig:nonGaussianCorr}.

\subsubsection{Quantum non-Gaussianity and Entanglement}\label{sec:EntQuantmNonGauss}

To understand the role of quantum entanglement in a conditional preparation scheme, we contrast it to a system with only classical correlations. In that regard, let us suppose that the initial quantum state is separable such that its Wigner function can be cast in the form (\ref{eq:classCorrWig}). By inserting this form in (\ref{eq:CondOne}), we find that
\begin{equation}
W_{\bf f \mid A}(\vec x_{\bf f}) = \int {\rm d}\gamma  p(\gamma) \frac{\int_{\mathbb{R}^{2l'}}{\rm d}\vec x_{\bf g} W_A(\vec x_{\bf g})W_{\bf g}^{(\gamma)}(\vec x_{\bf g})}{\int_{\mathbb{R}^{2l'}}{\rm d}\vec x_{\bf g} W_A(\vec x_{\bf g})W_{\bf g}(\vec x_{\bf g})} W_{\bf f}^{(\gamma)}(\vec x_{\bf f}).
\end{equation}
As a next step, we define 
\begin{equation}\label{eq:ptilde}
\tilde{p}_A(\gamma) \coloneqq p(\gamma) \frac{\int_{\mathbb{R}^{2l'}}{\rm d}\vec x_{\bf g} W_A(\vec x_{\bf g})W_{\bf g}^{(\gamma)}(\vec x_{\bf g})}{\int_{\mathbb{R}^{2l'}}{\rm d}\vec x_{\bf g} W_A(\vec x_{\bf g})W_{\bf g}(\vec x_{\bf g})},
\end{equation}
and show that $\tilde{p}_A(\gamma)$ is a well-defined probability distribution. First, we use the definition of the reduced state
\begin{align}
W_{\bf g}(\vec x_{\bf g}) &= \int_{\mathbb{R}^{2l}}{\rm d}\vec x_{\bf f} W(\vec x_{\bf f} \oplus \vec x_{\bf g})\\
&= \int {\rm d}\gamma  p(\gamma)W_{\bf g}^{(\gamma)}(\vec x_{\bf g}),
\end{align}
and thus we immediately find that $\int {\rm d}\gamma  \tilde p_A(\gamma) = 1$. Furthermore, we note that $W_{\bf g}^{(\gamma)}(\vec x_{\bf g})$ is the Wigner function of a well-defined quantum state $\hat \rho_{\bf g}^{(\gamma)}$ and thus 
\begin{equation}
\int_{\mathbb{R}^{2l'}}{\rm d}\vec x_{\bf g} W_A(\vec x_{\bf g})W_{\bf g}^{(\gamma)}(\vec x_{\bf g}) = \tr[ \hat \rho_{\bf g}^{(\gamma)} \hat A] \geqslant 0.
\end{equation}
The final inequality follows from the fact that $\hat A$ is a positive semi-definite operator. As a consequence, we find that $\tilde{p}_A(\gamma) \geqslant 0$ for every possible $\gamma$. Thus, we find that for a separable initial state
\begin{equation}\label{eq:mixtureseparable}
W_{\bf f \mid A}(\vec x_{\bf f}) = \int {\rm d}\gamma  \tilde{p}_A(\gamma) W_{\bf f}^{(\gamma)}(\vec x_{\bf f}).
\end{equation} 
Up to this point, we only assumed that the initial state is separable. As we saw in Subsection \ref{sec:quantumnongauss}, a mixed quantum state with a positive Wigner function cannot necessarily be decomposed in states with positive Wigner functions. Therefore, we can generally not infer much about the properties of the Wigner function $W_{\bf f}^{(\gamma)}(\vec x_{\bf f})$ in (\ref{eq:mixtureseparable}).\\

As a next step, we use the fact that the initial state is also a Gaussian state. Recall from (\ref{eq:GaussianDecomp}) that any mixed Gaussian state can be decomposed as a mixture of pure Gaussian states. A priori, however, it is not trivial that this decomposition is consistent with decomposition in separable states (\ref{eq:classCorrWig}). Thus, it remains to show that for Gaussian separable states the Wigner functions $W_{\bf f}^{(\gamma)}(\vec x_{\bf f})$ and $W_{\bf g}^{(\gamma)}(\vec x_{\bf g})$ in (\ref{eq:classCorrWig}) are also Gaussian.

We start from a crucial observation on covariance matrices that was made in \cite{PhysRevLett.86.3658}: whenever an $m$-mode state with covariance matrix $V$ is separable, there are covariance matrices $V_{\bf f}'$ and $V_{\bf g}'$ such that
\begin{equation}\label{eq:boundEnt}
V \geqslant \begin{pmatrix} V_{\bf f}' & 0 \\ 0 & V_{\bf g}' \end{pmatrix} = V_{\bf f}' \oplus V_{\bf g}'.
\end{equation}
Note that $V_{\bf f}' $ and $V_{\bf g}'$ are covariance matrices on the phase spaces $\mathbb{R}^{2l}$ and $\mathbb{R}^{2l'}$, respectively. Nevertheless, $V_{\bf f}' $ and $V_{\bf g}'$ are generally not the same as the covariance matrices $V_{\bf f}$ and $V_{\bf g}$ of (\ref{eq:CovGaussBipart}) that describe the marginal distributions. We should emphasise that the Williamson (\ref{eq:Williamson}) and Bloch-Messiah (\ref{eq:BlochMessiah}) decompositions offer the necessary tools to explicitly construct $V_{\bf f}' $ and $V_{\bf g}'$ (we will come back to this point in Subsection \ref{sec:QunatumCorrThroughnonGauss}). This allows us to use similar techniques as in (\ref{eq:GaussianDecomp}). Let us first define 
\begin{align}
W'_{\bf f}(\vec x_{\bf f}) \coloneqq \frac{e^{-\frac{1}{2}\vec x_{\bf f}^T{V'_{\bf f}}^{-1}\vec x_{\bf f}}}{(2\pi)^m \sqrt{\det V'_{\bf f}}},\\
W'_{\bf g}(\vec x_{\bf g}) \coloneqq \frac{e^{-\frac{1}{2}\vec x_{\bf g}^T{V'_{\bf g}}^{-1}\vec x_{\bf g}}}{(2\pi)^m \sqrt{\det V'_{\bf g}}}.
\end{align}
We can then use (\ref{eq:boundEnt}) to define a positive definite matrix $V_c \coloneqq V - V_{\bf f}' \oplus V_{\bf g}',$ such that a decomposition of the type (\ref{eq:GaussianDecomp}) gives us
\begin{equation}\label{eq:GaussianDecompSep}\begin{split}
W&(\vec x_{\bf f} \oplus \vec x_{\bf g}) = \\
&\int_{\mathbb{R}^{2m}} {\rm d}\vec y \, W'_{\bf f}(\vec x_{\bf f}- \vec y_{\bf f})W'_{\bf g}(\vec x_{\bf g} - \vec y_{\bf g}) \frac{e^{- \frac{1}{2} (\vec y - \vec \xi)^T V^{-1}_c (\vec y- \vec \xi)}}{(2\pi)^m \sqrt{\det V_c}},
\end{split}
\end{equation}
where we again impose the structure of the bipartition on $\vec y = \vec y_{\bf f} \oplus \vec y_{\bf g}$, with $ \vec y_{\bf f} \in \mathbb{R}^{2l}$ and $\vec y_{\bf g} \in \mathbb{R}^{2l'}$. Furthermore, recall that $\vec \xi$ is the mean field of the initial Gaussian state $W(\vec x_{\bf f} \oplus \vec x_{\bf g})$. The structure we obtain in (\ref{eq:GaussianDecompSep}) exactly corresponds to (\ref{eq:classCorrWig}), where $\vec y$ now labels the states and thus plays the role of the abstract variable $\gamma$.

We can then use the structure (\ref{eq:GaussianDecompSep}) in the derivation (\ref{eq:mixtureseparable}) and then we find that 
\begin{equation}\label{eq:mixtureseparableGauss}
W_{\bf f \mid A}(\vec x_{\bf f}) = \int_{\mathbb{R}^{2l}} {\rm d}\vec y_{\bf f} W_{\bf f}'(\vec x_{\bf f} -  \vec y_{\bf f})  \tilde{p}_A(\vec y_{\bf f}).
\end{equation} 
In any concrete choice of $\hat A$, one can use (\ref{eq:ptilde}) to derive an explicit expression for $\tilde{p}_A(\vec y_{\bf f})$, which will generally be a non-Gaussian probability distribution, such that $W_{\bf f \mid A}(\vec x_{\bf f})$ describes a non-Gaussian state. However, the resulting conditional state (\ref{eq:mixtureseparableGauss}) is clearly a statistical mixture of Gaussian states and thus lies in the convex hull of Gaussian states. In the language of Subsection \ref{sec:quantumnongauss} this means that the conditional state is non-Gaussian but not quantum non-Gaussian and has a stellar rank $0$.\\

In summary, we have assumed that our initial state with Wigner function $W(\vec x_{\bf f} \oplus \vec x_{\bf g})$ is a separable Gaussian state. Without making any assumptions on the POVM element $\hat A$ of the conditional operation, we retrieve that the conditional state always is a convex combination of Gaussian states, given by (\ref{eq:mixtureseparableGauss}). Thus, when the initial state is Gaussian, entanglement is a necessary resource to produce quantum non-Gaussian states via conditional operations. 

\subsubsection{Wigner Negativity and Einstein-Podolsky-Rosen Steering}\label{SteeringWignerNeg}

In Subsection \ref{sec:WignerNegativity}, we explained that Wigner negativity is a ``stronger'' non-Gaussian feature than quantum non-Gaussianity. Here, we show that also stronger types of quantum correlations are required to conditionally create Wigner negativity. To understand how Wigner negativity can be achieved through a conditional preparation scheme, it suffices to understand when the conditional expectation value $\<\hat A\>_{{\bf g} \mid \vec x_{\bf f}}$ in (\ref{eq:Agf}) reaches negative values.

Regardless of the chosen POVM, $W_A(\vec x_{\bf g})$ is the Wigner function of a positive semi-definite operator $\hat A$ as defined by (\ref{eq:WigA}). Thus, whenever there is a quantum state $\hat \rho'$ that has $W(\vec x_{\bf g}\mid \vec x_{\bf f})$ as associated Wigner function, (\ref{eq:wignerProduct}) implies that $\<\hat A\>_{{\bf g} \mid \vec x_{\bf f}} = \tr[\hat \rho'\hat A] \geqslant 0$. Hence, to conditionally create a non-positive Wigner function (\ref{eq:CondTwo}) the conditional probability distribution $W(\vec x_{\bf g}\mid \vec x_{\bf f})$ cannot be a well-defined Wigner function. This observation holds whenever the initial state has a positive Wigner function.\\

When in addition we assume that the initial state is Gaussian, we find that $W(\vec x_{\bf g}\mid \vec x_{\bf f})$ is a Gaussian distribution (\ref{GaussCond}). Whether the conditional probability distribution $W(\vec x_{\bf g}\mid \vec x_{\bf f})$ describes a Gaussian quantum state depends entirely in the properties of its covariance matrix, i.e.,  the Schur complement $V_{{\bf g}\mid \vec x_{\bf f}}$. Indeed, $W(\vec{x}_{\bf g} \mid \vec{x}_{\bf f}) $ describes a quantum state if and only if $V_{{\bf g}\mid \vec x_{\bf f}}$ satisfies the Heisenberg inequality. Because $V_{{\bf g}\mid \vec x_{\bf f}}$ does not depend on the choice $\vec x_{\bf f} \in \mathbb{R}^{2l}$, it follows that $W(\vec{x}_{\bf g} \mid \vec{x}_{\bf f})$ corresponds to a quantum state either for all $\vec x_{\bf f} \in \mathbb{R}^{2l}$ (if the Schur complement (\ref{eq:schur}) satisfies the Heisenberg inequality) or for none of the $\vec x_{\bf f} \in \mathbb{R}^{2l}$ (if the Schur complement (\ref{eq:schur}) violates the Heisenberg inequality).

If $V_{{\bf g}\mid \vec x_{\bf f}}$ satisfies the Heisenberg inequality, the conditional state's Wigner function $W_{\bf f \mid A}(\vec x_{\bf f})$ must thus be positive. To better understand the physical resources required to conditionally create Wigner negativity, one must comprehend what it means for $V_{{\bf g}\mid \vec x_{\bf f}}$ to violate Heisenberg's inequality in terms of quantum correlations. It turns out that this condition is closely related to the original argument of the EPR paper \cite{Einstein:1935aa}. The violation of Heisenberg's inequality by the Schur complement $V_{{\bf g}\mid \vec x_{\bf f}}$ corresponds to Gaussian quantum steering in the state $W(\vec x_{\bf g} \oplus \vec x_{\bf f})$.


To understand the connection between the conditional covariance matrix $V_{{\bf g}\mid \vec x_{\bf f}}$ and quantum steering, we first express the Wigner function obtained by conditioning on a Gaussian measurement, such that the associated POVM element has a Wigner function $W_{G} (\vec x_{\bf f})$:
\begin{equation}\begin{split}
W_{{\bf g} \mid G}(\vec x_{\bf g}) &= \frac{\int_{\mathbb{R}^{2l}} {\rm d}\vec x_{\bf f}W_{G} (\vec x_{\bf f}) W(\vec x_{\bf f} \mid \vec x_{\bf g})}{\int_{\mathbb{R}^{2l}} {\rm d}\vec x_{\bf f}W_{G} (\vec x_{\bf f}) W_{\bf f}(\vec x_{\bf f})}W_{\bf g}(\vec x_{\bf g}).
\end{split}\end{equation}
In a very similar way, we can also show that
\begin{equation}\begin{split}\label{eq:GaussianMeasurementSteering}
&W_{{\bf g} \mid G}(\vec x_{\bf g})= \int_{\mathbb{R}^{2l}} {\rm d}\vec x_{\bf f}  \frac{W_{G} (\vec x_{\bf f})W_{\bf f}(\vec x_{\bf f}) }{\int_{\mathbb{R}^{2l}} {\rm d}\vec x_{\bf f} W_{G} (\vec x_{\bf f})W_{\bf f}(\vec x_{\bf f})}W(\vec x_{\bf g} \mid \vec x_{\bf f}).
\end{split}\end{equation}
Hence, when $W(\vec x_{\bf g} \mid \vec x_{\bf f})$ is a bona fide Wigner function for every $\vec x_{\bf f}$ this expression is an explicit manifestation of the local hidden variable model (\ref{eq:condstatesteering}). In other words, whenever $W(\vec x_{\bf g} \mid \vec x_{\bf f})$ describes a quantum state, the modes in ${\bf g}$ cannot be steered by Gaussian measurements on the modes ${\bf f}$. Note that we can generalise Gaussian measurements to any measurement with a positive Wigner function.

The remarkable feature of EPR steering is that the inverse statement also holds: when $W(\vec x_{\bf g} \mid \vec x_{\bf f})$ is not a bona fide Wigner function Gaussian measurements can steer the state. Let us assume that (\ref{eq:condstatesteering}) holds for Gaussian measurements. It then follows that a well-defined covariance matrix $U$ exists such that the covariance matrix $V_{{\bf g} \mid G}$ of the conditional state $W_{{\bf g} \mid G}(\vec x_{\bf g})$ satisfies $V_{{\bf g} \mid G} \geqslant U$ for all Gaussian measurements. Furthermore, $U$ is physical and satisfies the Heisenberg inequality. Ref.~\cite{Wiseman:2007aa} shows that the existence of such a $U$ implies that the full covariance matrix of the system satisfies $V + 0_{\bf f} \oplus i\Omega_{\bf g} \geqslant 0$, which in turn implies that $V_{{\bf g}\mid \vec x_{\bf f}}$, the Schur complement of $V$, satisfies the Heisenberg inequality.

This shows that we can only generate Wigner negativity through (\ref{eq:CondTwo}) if the initial state can be steered by Gaussian measurements on the subsystem associated with phase space $\mathbb{R}^{2l}$. Note that the creation of Wigner negativity occurs in the opposite direction as the steering: We can produce Wigner negativity in the modes ${\bf f}$ by performing a measurement on the modes ${\bf g}$ if the modes ${\bf g}$ can be steered by performing Gaussian measurements on the modes ${\bf f}$. Somewhat counterintuitively, it turns out that the created Wigner negativity volume \eqref{eq:NegVolume} is not directly proportional to the strength of EPR steering \cite{xiang2021quantification}.

As a final remark, we note that, in a multimode context, EPR steering is constrained by monogamy relations \cite{PhysRevA.88.062108,Ji_2015,PhysRevLett.118.230501}. Notably, this implies that when a single mode $g$ can be steered by a single other mode $f$, it is impossible for any other mode to also steer $g$. This naturally has profound consequences for the conditional generation of Wigner negativity that we discussed in this section. The monogamy relations for quantum steering can be used to derive similar monogamy relations \cite{xiang2021quantification} for the created Wigner negativity volume \eqref{eq:NegVolume}.

\subsection{Quantum Correlations through non-Gaussianity}\label{sec:QunatumCorrThroughnonGauss}
In Subsection \ref{sec:NonGaussByCorr}, we extensively considered the use of quantum correlations as a resource to create non-Gaussian effects. In this subsection, we focus on the opposite idea where non-Gaussian operations increase or even create quantum correlations. The subject of entanglement in non-Gaussian states is generally difficult to study, for some states it may be sufficient to evaluate lower order moments \cite{PhysRevLett.95.230502} and when the density matrix in the Fock representation is available one can apply DV approaches to characterise entanglement \cite{PhysRevA.79.052313}. However, these methods cannot always be applied and there are no universally applicable entanglement criteria that are practical to evaluate for arbitrary CV quantum states. 

\subsubsection{Entanglement measures on phase space}

In Subsection \ref{eq:explanEntanglement}, we argued that any pure state that manifests correlations between subsystems contains entanglement. Measuring entanglement in this case becomes equivalent to measuring the amount of correlation within the pure state. In particular for pure states, one finds a wide range of entanglement measures in literature \cite{RevModPhys.81.865}. In the case of CV systems, some measures are more appropriate than others, and here we will focus on one particularly intuitive measure that is based on purity.

When we consider an arbitrary bi-partite pure quantum state with Wigner function $W( \vec x_{\bf f} \oplus \vec x_{\bf g})$ (with $\vec x_{\bf f} \in \mathbb{R}^{2l}$ and $\vec x_{\bf g} \in \mathbb{R}^{2l'}$), we find that its purity is $\mu = 1$ by definition. However, this is not necessarily true for the subsystems ${\bf f}$ and ${\bf g}$. We can use (\ref{eq:purityGaussian}) to evaluate the purity of any state based on its Wigner function, and we define
\begin{equation}
\mu_{\bf f} = \int_{\mathbb{R}^{2l}} {\rm d}\vec x_{\bf f} [W_{\bf f}(\vec x_{\bf f})]^2, \text{ and} \quad \mu_{\bf g} = \int_{\mathbb{R}^{2l'}} {\rm d}\vec x_{\bf g} [W_{\bf g}(\vec x_{\bf g})]^2,
\end{equation}
where we again use the definition (\ref{eq:Wf}, \ref{eq:Wg}). Because the global state with Wigner function $W( \vec x_{\bf f} \oplus \vec x_{\bf g})$ is pure, we always find that $\mu_{\bf f} = \mu_{\bf g}$ (this is a general consequence of the existence of a Schmidt decomposition for pure states). Furthermore, if the pure state is separable, we find $W( \vec x_{\bf f} \oplus \vec x_{\bf g}) = W_{\bf f}( \vec x_{\bf f})W_{\bf g}(\vec x_{\bf g})$ and as a consequence we obtain that $\mu_{\bf f} = \mu_{\bf g} = 1$. However, when $\mu_{\bf f} = \mu_{\bf g} < 1$ there must be correlations between the subsystems ${\bf f}$ and ${\bf g}$ and the smaller the purity of the subsystems, the stronger these correlations are. Without delving into the details, we stress that the opposite notion also holds: when there is a correlation between the subsystems, the purity of the subsystems is smaller than one.

To convert this quantity into an entanglement measure \cite{Kim_2010}, it is useful to define the R\'enyi-2 entropy for subsystem ${\bf f}$
\begin{equation}
S_R \coloneqq -\log \mu_{\bf f}. 
\end{equation}
We then find that $S_R \geqslant 0$ and $S_R = 0$ if and only if the state is separable. Furthermore, it should be clear that $S_R$ cannot be increased by local unitary operations on the subsystems ${\bf f}$ and ${\bf g}$. We can thus define an entanglement measure for the pure state $\ket{\Psi}$ with Wigner function $W( \vec x_{\bf f} \oplus \vec x_{\bf g})$ by setting
\begin{equation}
{\cal E}_{R}(\ket{\Psi}) \coloneqq S_R.
\end{equation}
This constitutes a well-defined entanglement measure for any chosen bi-partition and any pure state on the phase space.

To extend this measure to mixed states, we follow Ref.~\cite{Kim_2010} and construct a convex roof. Any mixed state $\hat \rho$ can be decomposed in pure states as $\hat \rho = \int {\rm d}\gamma \, p(\gamma) \ket{\Psi^{(\gamma)}} \bra{\Psi^{(\gamma)}}$, we abbreviate this decomposition as the ensemble $\{p(\gamma), \ket{\Psi^{(\gamma)}}\}$. For each pure state in this ensemble, we can evaluate the entanglement ${\cal E}_{R}(\ket{\Psi^{(\gamma)}})$ and subsequently average all of these values according to $p(\gamma)$. However, the decomposition of $\hat \rho$ in pure states is far from unique and different ensembles $\{p(\gamma), \ket{\Psi^{(\gamma)}}\}$ generally lead to a different value of entanglement even though they are all constrained to produce the same state $\hat \rho$. Therefore, it is common to define
\begin{equation}\label{eq:RenyiEnt}
{\cal E}_R(\hat \rho) \coloneqq \inf_{\{p(\gamma), \ket{\Psi^{(\gamma)}}\}} \int {\rm d}\gamma\, p(\gamma) {\cal E}_{R}\left(\ket{\Psi^{(\gamma)}}\right)
\end{equation}
as the general ``R\'enyi-2 entanglement'' of the state $\hat \rho$.

Formally, this is an elegant definition that can in principle be calculated directly from the Wigner function. However, in practice it is nearly impossible to actually identify all possible decompositions $\{p(\gamma), \ket{\Psi^{(\gamma)}}\}$ which makes this measure notoriously hard to evaluate for mixed states. This has sparked some alternative definitions of entanglement measures for Gaussian states, where any Gaussian state can be decomposed in an ensemble of Gaussian states (\ref{eq:GaussianDecomp}). Thus, one can define ``Gaussian R\'enyi-2 entanglement'' by restricting (\ref{eq:RenyiEnt}) to only Gaussian decompositions \cite{PhysRevLett.109.190502}. In this sense, Gaussian R\'enyi-2 entanglement is by construction an upper bound to the general R\'enyi-2 entanglement.\\

As an alternative to entanglement measures, it is common to use entanglement witnesses. These have been particularly successful for Gaussian states \cite{PhysRevLett.84.2726,PhysRevLett.84.2722,PhysRevA.67.052315,Hyllus_2006,PhysRevLett.114.050501,Lami_2018}, where one commonly applies methods based on the covariance matrix of the state. Due to the extremality of Gaussian states \cite{PhysRevLett.96.080502} these results also provide witnesses for entanglement if the state is non-Gaussian. However, there are several examples of non-Gaussian entangled states for which no entanglement can be detected from the covariance matrix. Notable progress was made by developing entanglement witnesses for non-Gaussian states with specific structure in their Wigner function \cite{PhysRevA.90.052321}. 

It is noteworthy to emphasise that the positive-partial transpose (PPT) criterion of \cite{PhysRevLett.84.2726} can in principle be implemented on the level of Wigner functions. To make this apparent, let us first define the transposition operator $T$ that implements $\hat \rho \mapsto \hat \rho^T$. When $W(\vec x)$ with $\vec x \in \mathbb{R}^{2m}$ denotes the Wigner function of the state $\hat \rho$, we can write the Wigner function of $\hat \rho^T$ as  $W(T\vec x)$. The matrix $T$ can be written as
\begin{equation}\label{eq:TranspositionOperator}
T = \bigoplus^m \begin{pmatrix}1 & 0 \\ 0 & -1\end{pmatrix},
\end{equation}
which can be derived from the definition of the Wigner function \footnote{Note that there is also a more profound connection between the transpose and time-reversal through Wigner's theorem. The latter is here effectively implemented by changing the sign of the momentum variables. However, this connection is beyond the scope of this tutorial and interested readers are invited to indulge in the literature on quantum chaos instead.}. The concept of partial transposition in entanglement theory relies on the simple idea that one can apply a transpose only on one of the two subsystems in the bi-partition. In our context, this means that the Wigner function changes as $W( \vec x_{\bf f} \oplus \vec x_{\bf g}) \mapsto W( \vec x_{\bf f} \oplus T\vec x_{\bf g}) $ (where $T$ is now taken only on the $l'$ modes of subsystem $\bf g$). The PPT criterion is based on the idea that, in absence of entanglement, the function $W( \vec x_{\bf f} \oplus T\vec x_{\bf g})$ still gives a well-defined Wigner function of a quantum state. However, there are entangled states for which this is no longer true and $W( \vec x_{\bf f} \oplus T\vec x_{\bf g})$ becomes unphysical. This lack of physicality is expressed by the fact there exist positive semidefinite operators $\hat A$ for which 
\begin{equation}
(4\pi)^m \int_{\mathbb{R}^{2m}} {\rm d} \vec x \, W_{A}(\vec x) W( \vec x_{\bf f} \oplus T\vec x_{\bf g}) < 0.
\end{equation}
Finding such observables $\hat A \geqslant 0$ for a non-Gaussian state $W( \vec x_{\bf f} \oplus \vec x_{\bf g})$ is generally a very hard task. For Gaussian states, on the other hand, the physicality of $W( \vec x_{\bf f} \oplus T\vec x_{\bf g})$ is simply checked through Heisenberg's inequality. For more general non-Gaussian states, this is insufficient and one should check a full hierarchy of inequalities instead \cite{PhysRevLett.95.230502}. Nevertheless, one may yet uncover more direct methods to check the properties of $W( \vec x_{\bf f} \oplus T\vec x_{\bf g})$. 

\subsubsection{Entanglement increase}\label{sec:entDist}

One of the most well-known protocols for increasing entanglement is entanglement distillation. In this protocol, one acts with local operations on a large number of mixed entangled states that are shared by two parties and concentrate the entanglement in a smaller number of maximally entangled pairs \cite{PhysRevLett.76.722}. When the initial states are pure and the local operation only serves to increase the entanglement and not the purity, we speak of entanglement concentration \cite{PhysRevA.53.2046}. Conditional operations play an important role in these protocols, and we can alternatively think of entanglement distillation as the idea that a conditional operation can increase the entanglement of a state. For Gaussian quantum states, there is a notorious no-go theorem that states that Gaussian measurements (or Gaussian operations in general) cannot increase bi-partite entanglement \cite{PhysRevLett.89.137903,PhysRevLett.89.137904,PhysRevA.66.032316}. It was quickly realised that these no-go results can be circumvented by even the most basics non-Gaussian states: those created through a non-Gaussian noise process \cite{Dong-2008,Hage-2008}. On the other hand, if one wants to distill entanglement in a CV system starting from initial Gaussian states one really requires non-Gaussian operations. One such example is given in \cite{PhysRevLett.84.4002,PhysRevA.62.032304}, where the authors propose to use a Kerr-nonlinearity to distill entanglement for mixed Gaussian states. In contrast, conditional schemes have also been proposed \cite{PhysRevA.67.062320, PhysRevA.82.042331,PhysRevA.87.042330}, avoiding the need for optical nonlinearities. In those protocols, one first uses conditional operations to create non-Gaussian states and subsequently uses Gaussification to obtain states with higher entanglement. A narrowly related protocol \cite{PhysRevA.95.022312} relies on the implementation of noiseless linear amplification \cite{NLA}, where the non-Gaussian element is injected in the form of auxiliary Fock states.

The realisation that photon subtraction and addition can be used to increase the entanglement of a Gaussian input state was developed reasonably early \cite{PhysRevA.61.032302,PhysRevA.65.062306,PhysRevA.67.032314} and was further formalised in works such as \cite{PhysRevA.73.042310,PhysRevA.80.022315,PhysRevA.86.012328}. Remarkably, all of these works explicitly assume that the initial state under consideration is a two-mode squeezed state and the approach strongly relies on the structure of this type of state in the Fock basis. Beyond the two mode setting, the class of CV graph states has also been studied in the context of entanglement increase \cite{PhysRevA.93.052313,Walschaers:2018aa}. Here we will provide an alternative approach, based on phase-space representations to understand entanglement increase due to the subtraction of a single photon.\\

Our approach relies on the fact that we can easily apply the entanglement measure (\ref{eq:RenyiEnt}) when the global state is pure. This means that we are focusing on a context of entanglement concentration. Furthermore, when we perform photon subtraction on a pure Gaussian state, the resulting photon-subtracted state is also pure, as we saw in Subsection \ref{sec:PhotonSubtraction}. The starting point is the Wigner function of the photon-subtracted state (\ref{eq:PhotonSubtractedState}) which we rewrite as
\begin{equation}\begin{split}\label{eq:PhotSubDist}
W^-(\vec x) =&\frac{W(\vec x)}{\tr\left(V_{\bf b} - \mathds{1}\right)+\norm{\vec \xi_{\bf b}}^2} \Big(\norm{B^T(\mathds{1} - V^{-1}) (\vec{x} - \vec{\xi})+ \vec \xi_{\bf b}}^2\\
    &\quad+ \tr\left[\mathds{1} - B^TV^{-1}B\right]\Big).
\end{split}\end{equation}
The state $W^-(\vec x)$ is thus obtained by subtracting a photon from the Gaussian state $W(\vec x)$. As we consider a pure two-mode state we assume that the state has a $4 \times 4$ covariance matrix of the form $V= S^TS$, where $S$ is a symplectic matrix. We assume that the photon is locally subtracted in one of the modes of the mode basis, such that
\begin{equation}
B = \begin{pmatrix}
0&0\\
0&0\\
1&0\\
0&1 \end{pmatrix}.
\end{equation}
However, to assess the entanglement in the system, we must obtain the Wigner function for the reduced state associated to either of the two modes. When we focus on the mode $\bf b$ where the photon is subtracted, we can simply obtain the reduced photon subtracted state $W_{\bf b}^-(\vec x_{\bf b})$ by subtracting a photon from the reduced Gaussian state $W_{\bf b}(\vec x_{\bf b})$. As such, we obtain 
\begin{equation}\begin{split}
W_{\bf b}^-(\vec x_{\bf b}) =&\frac{\norm{(\mathds{1} - V_{\bf b}^{-1})\vec{x}_{\bf b} + V_{\bf b}^{-1} \vec \xi_{\bf b}}^2+ \tr\left[\mathds{1} - V_{\bf b}^{-1}\right]}{\tr\left(V_{\bf b} - \mathds{1}\right)+\norm{\vec \xi_{\bf b}}^2}W_{\bf b}(\vec x_{\bf b}).
\end{split}\end{equation}
This is now a single-mode photon subtracted state, but it is no longer pure. This lack of purity is notably reflected by $V_{\bf b}$ which is no longer symplectic. Nevertheless, we can use the Williamson decomposition (\ref{eq:Williamson}) and write
\begin{equation}\label{eq:generalSingleMode}
V_{\bf b} = \nu\begin{pmatrix}r & 0\\
0 & r^{-1}
\end{pmatrix},
\end{equation}
where we set the phase such that the squeezing coincides with one of the axes of phase space. What remains is for us to calculate the purity
\begin{equation}
\mu^-_{\bf b} = 4\pi \int_{\mathbb{R}^2}{\rm d}\vec x_{\bf b} [W_{\bf b}^-(\vec x_{\bf b})]^2.
\end{equation} 
The final expression for the purity is not very insightful. When on top we use that the purity $\mu_{\bf b}$ of the Gaussian state $W_{\bf b}(\vec x_{\bf b})$ is given by $\mu_{\bf b} = 1/\nu$, an explicit calculation of $\mu_{\bf b}^-$ makes it possible to prove (the motivated reader can use a combination of patience and software for symbolic algebra to do so) that
\begin{equation}\label{eq:purityStuff}
\frac{\mu^-_{\bf b}}{\mu_{\bf b}} \leqslant \frac{1}{2}.
\end{equation}
In other words, photon subtraction reduces the purity at most by a factor of two.

When we use (\ref{eq:RenyiEnt}) to define the entanglement of the two-mode photon-subtracted state (\ref{eq:PhotSubDist}), we find that it is given by
\begin{equation}
{\cal E}_R(\ket{\Psi^-}) = - \log \mu^-_{\bf b},
\end{equation}
because the two-mode state is pure. The entanglement of the initial Gaussian state is given by ${\cal E}_R(\ket{\Psi_G}) =  - \log \mu_{\bf b} $, such that we can use (\ref{eq:purityStuff}) to find that
\begin{equation}\label{eq:EntIncrease}
\Delta{\cal E}_R \coloneqq {\cal E}_R(\ket{\Psi^-}) - {\cal E}_R(\ket{\Psi_G}) \leqslant \log 2.
\end{equation}
In other words, photon subtraction can increase the R\'enyi-2 entanglement of an arbitrary Gaussian state, but at most by an amount $\log 2$. It turns out that this result can be generalised to all bipartitions of Gaussian pure states of an arbitrary number of modes \cite{zhang2021maximal}. Furthermore, the same work shows that when the entanglement measure ${\cal E}_R(\ket{\Psi_G})$ is replaced with the Gaussian R\'enyi-2 entropy of \cite{PhysRevLett.109.190502}, the result holds for all bipartitions of all Gaussian states (including mixed ones).\\

\begin{figure}
\centering
\includegraphics[width=0.49\textwidth]{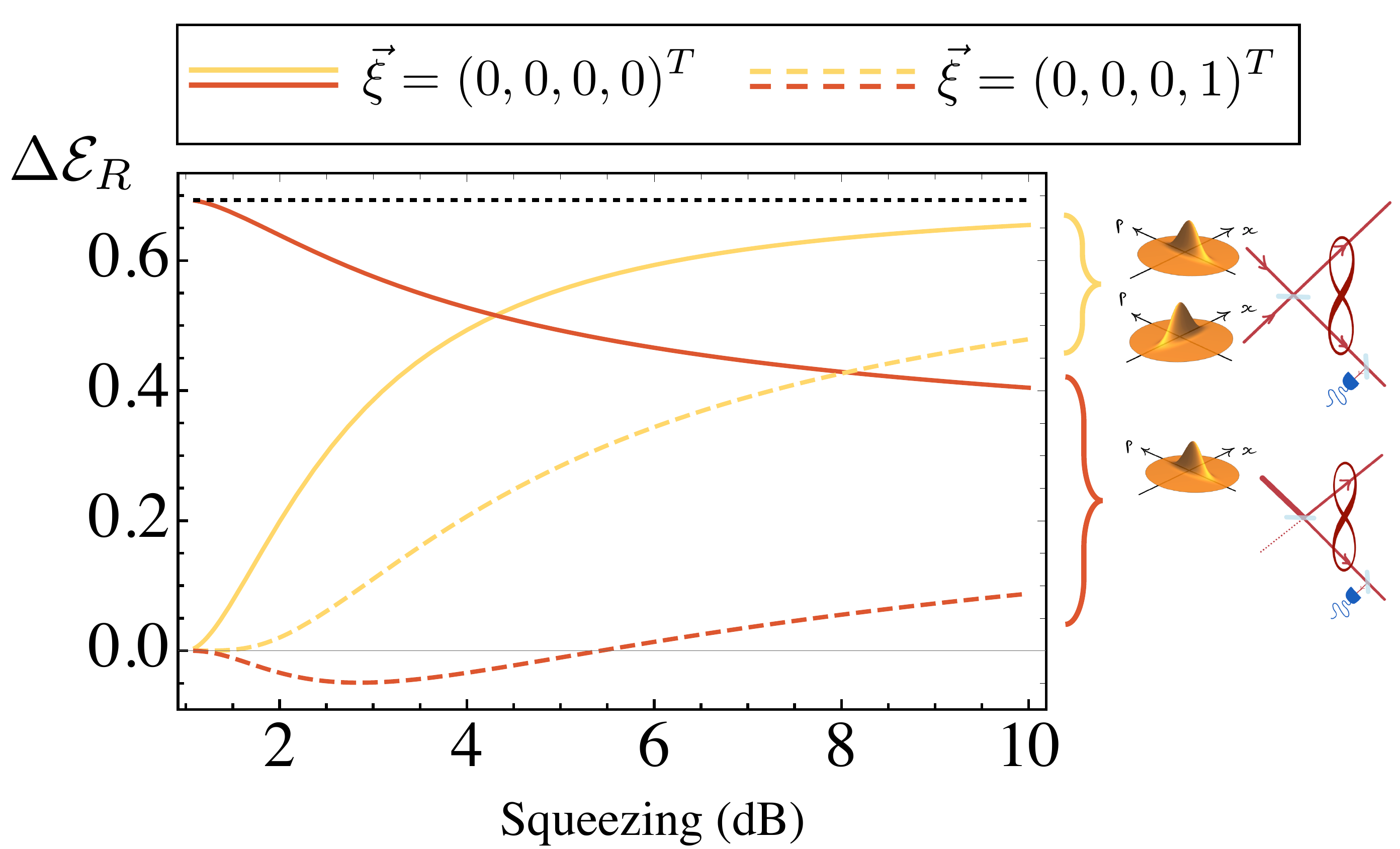}\\
\caption{Entanglement increase (\ref{eq:EntIncrease}) through photon subtraction in one mode of a pair of entangled modes. The initial Gaussian states are obtained by mixing either two equally squeezed modes (yellow curves) or one squeezed mode and one vacuum mode (red curves) on a beam splitter (see also sketches on the right). We show how a variation of squeezing (in dB compared to shot noise level) in these initial squeezed vacuum states influences the entanglement increase due to photon subtraction. We consider cases without mean field (solid curves) and with a mean field $\vec \xi = (0,0,0,1)^T$ (dashed curves). }
 \label{fig:EntIncrease}
\end{figure}

For the particular case of a two-mode pure Gaussian state, we can directly evaluate $\Delta{\cal E}_R$ for some important examples. Say, for example, that we consider the EPR state that is obtained by mixing two oppositely squeezed vacuum states on a balanced beamsplitter. In this case $\vec \xi = 0$ and $V$ is given by
\begin{equation}\begin{split}
V &= \frac{1}{2}\begin{pmatrix} 1& 0 & 1 & 0 \\ 0 & 1 & 0 & 1 \\ -1 & 0 & 1 & 0 \\ 0 & -1 & 0 & 1\end{pmatrix}^T \begin{pmatrix} s &&& \\ &s^{-1} && \\ &&s^{-1}&\\&&&s\end{pmatrix}\begin{pmatrix} 1& 0 & 1 & 0 \\ 0 & 1 & 0 & 1 \\ -1 & 0 & 1 & 0 \\ 0 & -1 & 0 & 1\end{pmatrix}\\
&= 
\frac{1}{2 s}\begin{pmatrix}
s^2+1 & 0 & s^2-1& 0 \\
 0 &s^2+1 & 0 & 1-s^2 \\
s^2-1& 0 &s^2+1 & 0 \\
 0 & 1-s^2 & 0 & s^2+1 \\
\end{pmatrix}.
\end{split}
\end{equation}
We then extract directly that
\begin{equation}
V_{\bf b} =  \frac{s^2+1}{2 s}\, \mathds{1},
\end{equation}
such that we find that the parameters in (\ref{eq:generalSingleMode}) are set to $r = 1$ and $ \nu= (s^2+1)/(2s)$. And thus we directly obtain
\begin{equation}
\Delta{\cal E}_R = \log (2)-\log \left(\frac{s^4+6 s^2+1}{\left(s^2+1\right)^2}\right).
\end{equation}
We clearly see that the entanglement increase vanishes in absence of squeezing, whereas we achieve the $\log(2)$ limit for $s \rightarrow \infty$. Adding a mean field with $\vec \xi_{\bf b} \neq 0$ immediately complicates the problem. As can be in Fig.~\ref{fig:EntIncrease}, where we plot the case $\vec \xi_{\bf b} = (0,1)^T$, the presence of a mean field in the mode of photon subtraction lowers the entanglement increase $\Delta{\cal E}_R$. Nevertheless, in the limit $s\rightarrow \infty$ we reach the limit $\log(2)$ regardless of the displacement.

This example clearly shows that photon subtraction can be used as a tool to increase entanglement. The setting corresponds to the case that is typically studied in most works on CV entanglement distillation such as \cite{PhysRevA.73.042310,PhysRevA.80.022315,PhysRevA.86.012328}. It turns out that one can further increase entanglement in such systems by  subtracting more photons. Furthermore, photon addition and the combination of addition and subtraction on both modes have also been considered. The methods we use in this Tutorial are not easily generalised to the subtraction and addition of many photons, but in return they can be applied to a much wider class of initial Gaussian states.\\

As a second example, we consider a single-mode squeezed state that is split in two on a balanced beamsplitter. This means that the Gaussian state is given by
\begin{equation}\begin{split}
V &= \frac{1}{2}\begin{pmatrix} 1& 0 & 1 & 0 \\ 0 & 1 & 0 & 1 \\ -1 & 0 & 1 & 0 \\ 0 & -1 & 0 & 1\end{pmatrix}^T \begin{pmatrix} s &&& \\ &s^{-1} && \\ &&1&\\&&&1\end{pmatrix}\begin{pmatrix} 1& 0 & 1 & 0 \\ 0 & 1 & 0 & 1 \\ -1 & 0 & 1 & 0 \\ 0 & -1 & 0 & 1\end{pmatrix}\\
&= 
\frac{1}{2}\begin{pmatrix}
 s+1 & 0 & s-1 & 0 \\
 0 & \frac{s+1}{s} & 0 & \frac{1}{s}-1 \\
 s-1 & 0 & s+1 & 0 \\
 0 & \frac{1}{s}-1 & 0 & \frac{s+1}{s} \\
\end{pmatrix},
\end{split}
\end{equation}
such that we get 
\begin{equation}
V_{\bf b} =  \frac{1}{2}\begin{pmatrix} s+ 1 & 0 \\ 0 & \frac{s+1}{s}\end{pmatrix},
\end{equation}
such that we find that we identify the parameters of (\ref{eq:generalSingleMode}) as $\nu = \sqrt{2 + s + s^{-1}} / 2$ and $r = (1+s) / \sqrt{2 + s + s^{-1}}$. In absence of any mean field, i.e. with $\vec \xi = 0$, we then find an entanglement gain given by
\begin{equation}
\Delta {\cal E}_R = \log (2)-\log \left(\frac{3 + 2s+3 s^2}{2 (s+1)^2}\right).
\end{equation}
Interestingly, in this case we reach the maximal entanglement gain for vanishing squeezing $s \rightarrow 1$, where we reach $\Delta {\cal E}_R \rightarrow \log (2)$. This case may seem somewhat counter-intuitive, but it should be emphasised that the success probability of photon subtraction also vanishes in this case. Yet, our conditional approach assumes that we are in the scenario where a photon was subtracted and the negligible fraction of the state that is not in vacuum is enhance. In the limit of vanishing squeezing, the photon subtracted state converges to the Bell state $(\ket{1,0} + \ket{0,1})/\sqrt{2}$. On the other hand, in the limit where squeezing is high we still find a finite entanglement increase as $\Delta {\cal E}_R \rightarrow  \log (4/3)$.

When we add a mean field given by $\vec \xi_{\bf b} \neq 0$, there is an importance of the phase because our state locally has some remaining asymmetry (which can be seen from $r \neq 1$). In Fig.~\ref{fig:EntIncrease} we particularly show the case where $\vec \xi_{\bf b} = (0,1)^T$ such that the direction of the displacement coincides with the quadrature where the noise is minimal. In this case we observe that for some values of initial squeezing, the entanglement decreases due to photon subtraction. Note that this quite remarkably implies that in some cases photon subtraction can actually be used to increase the purity of a state.\\

We thus showed that photon subtraction is a useful non-Gaussian operation to increase entanglement. However, in the presence of a mean field in the subtraction mode, it is also possible to decrease entanglement. Even though this subject has been studied for nearly two decades, for arbitrary Gaussian input states, there are still many open questions. Notably, there has not been much work on the effect of photon subtraction on multipartite entanglement, nor on stronger types of quantum correlations. Our discussion in Subsection \ref{SteeringWignerNeg} suggests an important interplay between EPR steering and Wigner negativity, and thus it is intriguing to wonder whether well-chosen non-Gaussian operations can increase quantum steering. Since all steerable states are also entangled, it is reasonable some of the protocols that can increase quantum entanglement should also increase quantum steering. 

We have followed the terminology found in literature and referred to this process as entanglement distillation, because our conditional operation has only a finite success probability. This implies that we can use a large number of Gaussian entangled states and use photon subtraction to obtain a much smaller number of more entangled states. Yet, it must be stressed that there is a more subtle process happening: the entanglement is increased by adding non-Gaussian entanglement on top of the existing Gaussian entanglement. To get a better grasp of this non-Gaussian entanglement, it is useful to go to a setting where no other type of entanglement is present as we do in Subsection \ref{sec:nonGaussianEnt}.

\subsubsection{Purely non-Gaussian quantum entanglement}\label{sec:nonGaussianEnt}

In this subsection, we explore an idea that is in many ways complementary to the previous subsection: rather than using a local non-Gaussian operation to increase already existing entanglement, we now use a non-local non-Gaussian operation to create entanglement between unentangled modes.

Let us again assume that our state is initially Gaussian as described by (\ref{eq:GaussStateWig}), and we will induce the non-Gaussian features through the conditional methods of Subsection \ref{sec:CondnonGaussian}. The mean field of the initial state is given by $\vec \xi = \vec \xi_{\bf f} \oplus \vec \xi_{\bf g}$, and 
\begin{equation}
V = \begin{pmatrix} V_{\bf f} & V_{\bf fg} \\ V_{\bf gf} & V_{\bf g} \end{pmatrix}, \quad \text{ with } \quad V_{\bf f} = V_{\bf f_1} \oplus V_{\bf f_2}.
\end{equation}
Here, we have introduced the modes of interest, labeled by ${\bf f}$ and a set of auxiliary modes ${\bf g}$ upon which a measurement will be performed to induce non-Gaussian features in the modes ${\bf f}$. In the initial state, we consider a bi-partition in the modes ${\bf f}$ without any direct correlations, hence $V_{\bf f} = V_{\bf f_1} \oplus V_{\bf f_2}$. In other words, the modes in ${\bf f_1}$ are completely uncorrelated to the modes in ${\bf f_2}$.

To induce non-Gaussian effects, we resort to the conditional framework by acting with a POVM element $\hat A$ upon the auxiliary modes ${\bf g}$, and we rewrite \eqref{eq:CondTwo} as
\begin{equation}
W_{\bf f \mid A}(\vec x_{\bf f_1}\oplus \vec x_{\bf f_2}) = \frac{\<\hat A\>_{{\bf g} \mid \vec x_{\bf f_1}\oplus \vec x_{\bf f_2}}}{\<\hat A\>}W_{\bf f_1}(\vec x_{\bf f_1})W_{\bf f_2}(\vec x_{\bf f_2}),
\end{equation}
and the conditional expectation value $\<\hat A\>_{{\bf g} \mid \vec x_{\bf f_1}\oplus \vec x_{\bf f_2}}$ is again given by (\ref{eq:Agf}). The entanglement in the resulting state thus crucially depends on the exact properties of $\<\hat A\>_{{\bf g} \mid \vec x_{\bf f_1}\oplus \vec x_{\bf f_2}}$.

First of all, note that $W_{\bf f_1}(\vec x_{\bf f_1})$ and $W_{\bf f_2}(\vec x_{\bf f_2})$ are generally not pure states and as a consequence $W_{\bf f \mid A}(\vec x_{\bf f_1}\oplus \vec x_{\bf f_2})$ is not a pure state either. Even though the specific structure of the Wigner function makes it a suitable case to apply the methods of \cite{PhysRevA.90.052321}, we follow a different route in this Tutorial by focusing on a particular example for which we can assume that $W_{\bf f_1}(\vec x_{\bf f_1})$ and $W_{\bf f_2}(\vec x_{\bf f_2})$ are pure.

\begin{figure}
\centering
\includegraphics[width=0.49\textwidth]{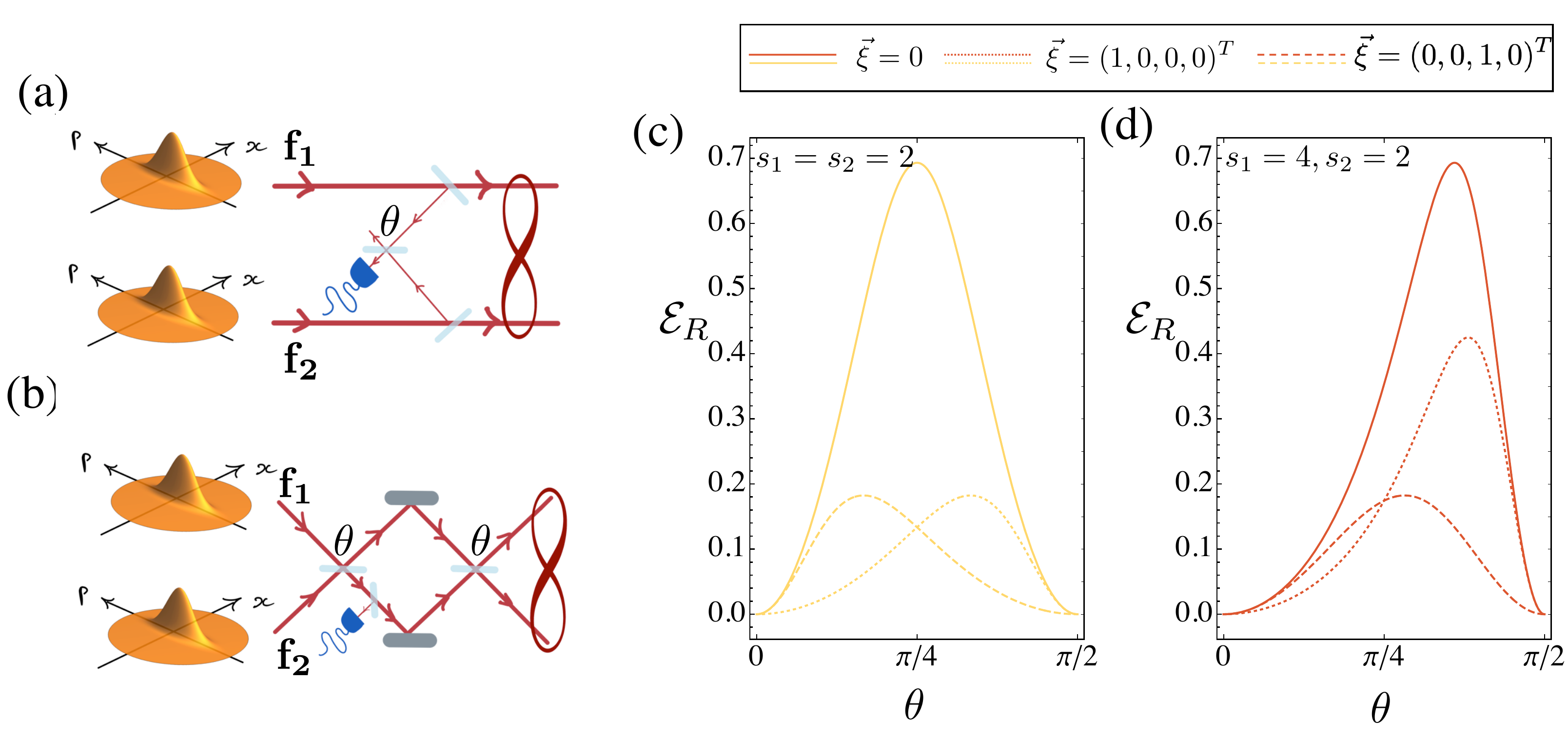}\\
\caption{\footnotesize Entanglement creation through photon subtraction in a superposition of uncorrelated modes ${\bf f_1}$ and ${\bf f_2}$. Panels (a) and (b) sketch two equivalent setups to implement a photon subtraction in the mode ${\bf b}$, with $\vec b = \cos \theta \vec f_1 + \sin \theta \vec f_2$. In panels (c) and (d), we show the created entanglement, as measure through the R\'enyi entropy (\ref{eq:RenyiEnt}) for varying values of $\theta$. The initial Gaussian states are pure, with covariance matrices $V_{\bf f_1} = {\rm diag}[s_1,1/s_1]$ and $V_{\bf f_2} = {\rm diag}[s_2,1/s_2]$ for modes  ${\bf f_1}$ and ${\bf f_2}$, respectively. The global mean field, i.e., displacement, is varied $\vec \xi = 0$ (solid curves), $\vec \xi = (1,0,0,0)^T$ (dotted curves), and $\vec \xi = (0,0,1,0)^T$ (dotted curves). Panel (c) shows the particular case where the squeezing is balanced, i.e., $s_1 = s_2 = 2$. Panel (d) shows an unbalanced example where $s_1 = 4$ and $s_2 = 2$. All squeezing values $s_1$ and $s_2$ are measured in units of vacuum noise.
} \label{fig:EntCreation}
\end{figure}

Just as in Subsection \ref{sec:entDist}, we concentrate on photon subtraction. To get a conceptual idea of such a setup in this specific scenario, we present two equivalent schemes in Panels (a) and (b) of Fig.~\ref{fig:EntCreation}. Note that the equivalence stems from the fact that the beamsplitters that subtract the light from the signal beams to send it to the photodetector are of extremely low reflectivity. In this limit, we can be sure that there is at most one photon in the path and when it is detected, we herald a single-photon-subtracted state. In Fig. \ref{fig:EntCreation}(a), the combination of this heralding process and the presence of at most one photon avoids that the unmeasured output causes any losses or impurities. Nevertheless, the unmeasured output will practically change the success probability of the heralding process, such that for practical implementations Fig. \ref{fig:EntCreation}(b) may be the preferential setup. Recall that the Wigner function for a state with a photon subtracted in a particular mode ${\bf b}$ was given by \eqref{eq:PhotSubDist}, which here becomes
\begin{equation}\begin{split}
W^-(\vec x_{\bf f_1} \oplus \vec x_{\bf f_2}) =&\frac{W_{\bf f_1}(\vec x_{\bf f_1})W_{\bf f_2}(\vec x_{\bf f_2})}{\tr\left(V_{\bf b} - \mathds{1}\right)+\norm{\vec \xi_{\bf b}}^2} \\&\Big(\norm{B^T(\mathds{1} - V_{\bf f_1}^{-1}\oplus V_{\bf f_2}^{-1}) (\vec x_{\bf f_1} \oplus \vec x_{\bf f_2}- \vec{\xi}_1 \oplus \vec{\xi}_2)+ \vec \xi_{\bf b}}^2\\
    &\quad+ \tr\left[\mathds{1} - B^T(V_{\bf f_1}^{-1}\oplus V_{\bf f_2}^{-1})B\right]\Big).
\end{split}\end{equation}
Because we consider a limit where the state is completely transmitted by the beamsplitter and only a negligible amount is sent to the photon counter to subtract the photon, we can indeed assume that the state is pure. For simplicity, we also assume that ${\bf f_1}$ and ${\bf f_2}$ are single modes. As we did before, we now calculate the reduced state
\begin{equation}
W^-_1(\vec x_{\bf f_1}) = \int_{\mathbb{R}^{2}} {\rm d}\vec x_{\bf f_2} W^-(\vec x_{\bf f_1} \oplus \vec x_{\bf f_2}).
\end{equation}
The integral is rather tedious to evaluate, therefore we immediately jump to the result (see \cite{walschaers_statistical_2017} for an alternative method that circumvents the explicit calculation of integrals):
\begin{equation}\begin{split}
&W^-_1(\vec x_{\bf f_1}) = \\
&\frac{W_{\bf f_1}(\vec x_{\bf f_1})}{\tr\left( V_{\bf b}- \mathds{1}\right)+\norm{\vec \xi_{\bf b}}^2}\Big(\norm{B^TF_1(\mathds{1} - V_{\bf f_1}^{-1}) (\vec x_{\bf f_1} - \vec{\xi}_1)+ \vec \xi_{\bf b} }^2\\
    &\quad+ \tr\left[B^T F_1 (\mathds{1} - V_{\bf f_1}^{-1} )F_1^T B\right] + \tr\left[B^T F_2 (V_{\bf f_2} - \mathds{1})F_2^TB \right]\Big),
\end{split}
\end{equation}
where we introduce the matrices $F_k$, given by
\begin{equation}
F_k = \begin{pmatrix}\mid & \mid \\
\vec{f}_k& \Omega \vec{f}_k \\
\mid & \mid 
\end{pmatrix},
\end{equation}
such that we can use the properties of the symplectic form $\Omega$ to obtain
\begin{equation}
B^TF_k = \begin{pmatrix}\vec{b}^T\vec{f_k} &\vec{b}^T\Omega \vec{f_k}\\ -\vec{b}^T\Omega \vec{f_k} & \vec{b}^T\vec{f_k} \end{pmatrix}.
\end{equation}
If the mode ${\bf b}$ is orthogonal to the mode ${\bf f}_1$, we find that $B^TF_1 = 0$ such that $W^-_1(\vec x_{\bf f_1}) = W_{\bf f_1}(\vec x_{\bf f_1})$. On the other hand, when the mode ${\bf b}$ is exactly the same as ${\bf f}_1$ we find that $B^TF_1 = \mathds{1}$ such that the photon is only subtracted there. In this case $W^-_1(\vec x_{\bf f_1})$ is a pure state and no entanglement is created. In this case, one can check that $W^-_2(\vec x_{\bf f_2}) = W_2(\vec x_{\bf f_2})$. 

To create entanglement, we are thus interested in the case where ${\bf b}$ is a superposition of the two modes ${\bf f}_1$ and ${\bf f}_2$. To keep things simple, let us assume that $\vec b = \cos \theta \, \vec f_1 + \sin \theta \, \vec f_2$. Because the modes ${\bf f}_1$ and ${\bf f}_2$ are orthogonal, we can use that $\vec f_1^T \vec f_2 = 0$ and thus we find that $B^TF_1 = \cos \theta \, \mathds{1}$ and $B^TF_2 = \sin \theta \, \mathds{1}$. Nevertheless, the general expression for $W^-_1(\vec x_{\bf f_1})$ does not simplify much.

To acquire additional insight, let us now assume that both modes ${\bf f_1}$ and ${\bf f_2}$ have exactly the same squeezing in the same quadrature:
\begin{equation}
V_{\bf f_1} = V_{\bf f_2} = \begin{pmatrix}s & 0 \\0 & \frac{1}{s}\end{pmatrix}.
\end{equation}
Furthermore, let us assume that there is no mean field, such that $\vec \xi = 0$. In this particular case, we find the expression
\begin{equation}\begin{split}
W^-_1(x_{\bf f_1},p_{\bf f_1}) &= W_1(x_{\bf f_1},p_{\bf f_1}) \times\\
& \left[ p_{\bf f_1}^2 s+\frac{x_{\bf f_1}^2}{s} + \cos (2 \theta) \left( p_{\bf f_1}^2 s+ \frac{x_{\bf f_1}^2}{s} -2\right)\right].
\end{split}\end{equation}
In particular, it turns out that the purity takes a simple form, such that we can quantify the entanglement for this state as
\begin{equation}\label{eq:BalancedEntCre}
{\cal E}_R = \log (2)-\log \left(\frac{\cos (4 \theta)+3}{2}\right).
\end{equation}
This shows that the maximal entanglement is reached for $\theta = \pi / 4$ and --as expected-- the entanglement vanishes when $\theta = 0$ and $\theta = \pi / 2$, i.e., when we subtract entirely in either mode ${\bf f_1}$ and ${\bf f_2}$.

More general settings are shown in Fig.~\ref{fig:EntCreation}, where we show the entanglement creation for unbalanced squeezing, by setting
\begin{equation}
V_{\bf f_1} = \begin{pmatrix}s_1 & 0 \\0 & \frac{1}{s_1}\end{pmatrix}, \quad \text{and} \quad V_{\bf f_2} = \begin{pmatrix}s_2 & 0 \\0 & \frac{1}{s_2}\end{pmatrix}.
\end{equation}
We compare the case with $s_1 = s_2$ to the case with $s_1 \neq s_2$ and find that in absence of a mean field one can reach the same maximal amount of entanglement. However, the maximum is attained at a different value of $\theta$ when the squeezing is unbalanced. From (\ref{eq:BalancedEntCre}) we know that in absence of a mean field, the curve for  $s_1 = s_2$ does not depend on the actual value of squeezing.

Fig.~\ref{fig:EntCreation} also shows the effect of an existing mean field, by probing a mean field in mode ${\bf f_1}$ with $\vec{\xi} = (1,0,0,0)^T$ and in mode ${\bf f_2}$ with $\vec{\xi} = (0,0,1,0)^T$. Generally speaking, we observe that the mean field reduces the created entanglement. Nevertheless, the unbalance of squeezing ($s_1 \neq s_2$) also unbalances the effect of the mean field. The higher squeezing in mode ${\bf f_1}$ makes the entanglement creation more resilient to displacements, but a mean field in mode ${\bf f_2}$ will reduce the maximal attainable amount of entanglement to the same level as in the balanced case (because in both panels (c) and (d) the mode ${\bf f_2}$ is squeezed with $s_2 = 2$). In the presence of a mean field, we also find that unbalanced squeezing shifts the value $\theta$ for which most entanglement is created. In other words, to achieve maximal entanglement upon photon subtraction in two modes with unequal squeezing, one must subtract in an unbalanced superposition of these modes.\\

Through this example, we showed that entanglement between previously uncorrelated Gaussian states can be created by a non-Gaussian operation. This entanglement has some additional peculiarities. For example, a quick glance at how this procedure affects (\ref{eq:CondTwo}) shows that we can split the state in a Gaussian, i.e., $W_{\bf f}(\vec x_{\bf f})$, and a non-Gaussian part, i.e., $\<\hat A\>_{{\bf g}\mid \vec x_{\bf f}}/\<\hat A\>$. In this case of (\ref{eq:PhotSubDist}) the Gaussian part of the state clearly remains fully separable. This means that, in this representation, all entanglement is originating from the non-Gaussian part of the state. Nevertheless, the decomposition (\ref{eq:CondTwo}) of the state into a Gaussian and a non-Gaussian part most probably not unique for mixed states, making it challenging to study such non-Gaussian entanglement in its most general sense.

Yet, common tools that rely on the covariance matrix, such as \cite{PhysRevLett.84.2726,PhysRevLett.84.2722}, to characterise entanglement in the photon subtracted states (\ref{eq:PhotSubDist}) are doomed to fail. In \cite{walschaers_entanglement_2017} it is explicitly shown that the covariance matrix of a photon-subtracted state is given by the covariance matrix of the initial Gaussian state with a positive matrix added to it. This means that photon subtraction just adds correlated noise to the covariance matrix and if we consider a Gaussian state that has exactly this covariance matrix we can decompose it using (\ref{eq:GaussianDecomp}). In other words, when there is no entanglement visible in the covariance matrix of the initial Gaussian state, we will not witness any entanglement based on the covariance matrix of the photon subtracted state. In this case, the non-Gaussian entanglement is thus genuinely non-Gaussian in the sense that it cannot be detected through Gaussian witnesses. Hence, rather than decomposing the states in a Gaussian and non-Gaussian part, as was done in (\ref{eq:CondTwo}), it may be more fruitful to define non-Gaussian entanglement as any entanglement that cannot be witness based solely on the covariance matrix of the state. This approach also offers a natural connection to the framework of Gaussian passivity on quantum thermodynamics \cite{Brown_2016}.

Another peculiarity that was presented in \cite{walschaers_entanglement_2017,walschaers_statistical_2017} is the intrinsic nature of this non-Gaussian entanglement. When we transform the system into a different mode basis, there will still be entanglement in the system. The entanglement is said to be intrinsic because the state is entangled in every possible mode basis. As we saw in (\ref{eq:GaussianDecompSep}) Gaussian entanglement is never intrinsic as there always exists a basis in which a Gaussian state is separable.

Panel (b) of Fig.~\ref{fig:EntCreation} gives a rather interesting approach to understanding the intrinsic nature of non-Gaussian entanglement. In this sketch, the second beam splitter is intended to undo the superposition $\theta$ and return to the initial mode basis with modes ${\bf f_1}$ and ${\bf f_2}$. Changing this beam splitter thus implies a basis change. If we remove this beam splitter entirely, we find ourselves in the entanglement distillation scenario of Subsection \ref{sec:entDist}. In this case, the photon subtraction is fully local, but it happens on a state with Gaussian entanglement. The photon subtraction can then increase the R\'enyi entanglement by a maximal  amount of $\log 2$. When we change to a mode basis where there is no Gaussian entanglement and the entanglement is created through a non-local photon subtraction, we create a maximal amount of R\'enyi entanglement given by $\log 2$. Changing the mode basis in a different way will combine the physics of these two extreme case such that there will always be entanglement, regardless of the basis. 

Extending these ideas to more general non-Gaussian operations on more general Gaussian mixed states is a hard and currently open problem. This reflects the general status of entanglement theory in CV systems: we lack a structured theoretical understanding of this phenomenon and as a consequence we also lack good tools to detect it.


\subsection{Non-Gaussianity and Bell inequalities}\label{sec:Bell}

In this final Subsection of our study of quantum correlations in non-Gaussian states, we study Bell inequalities. First of all, we argue that it is impossible to violate Bell inequalities when both all the states and all the measurements involved can be described by positive Wigner functions. Then, we show that the Wigner function of the state can itself be used to formulate a Bell inequality when we allow for non-positive Wigner functions.\\

The general setup for studying non-locality in CV revolves around a multimode state with Wigner function $W(\vec x_{\bf f} \oplus \vec x_{\bf g} )$ defined on a phase space $\mathbb{R}^{2m} = \mathbb{R}^{2l} \oplus \mathbb{R}^{2l'}$. Bell non-locality entails that some measurements on this state cannot be described by a local hidden variable model of the type (\ref{eq:LHVBell}). In a quantum framework, the local measurements with POVM elements $\{\hat A_j\}$ (on the modes in {\bf f}) and $\{\hat B_j\}$ (on the modes in {\bf g}) can also be described by Wigner functions $W_{A_j}(\vec x_{\bf f})$ and $W_{B_j}(\vec x_{\bf g})$. Because we are dealing with a POVM, we find that
\begin{equation}\label{eq:GoingToProbs}
(4\pi)^{l} \sum_j W_{A_j}(\vec x_{\bf f}) = (4\pi)^{l'} \sum_j W_{B_j}(\vec x_{\bf g})  = 1.
\end{equation}
Note that this equality holds for all possible coordinates $\vec x_{\bf f}$ and $\vec x_{\bf g}$. Here we assume that the measurement outcomes $A_j$ and $B_j$ are discrete, but by correctly defining resolutions of the identity we can also deal with more general probability distributions, e.g., homodyne measurements.

The probability to get the joint measurement result $(A_j,B_k)$ is given by
\begin{equation}\begin{split}\label{eq:PJointPos}
&P(A_j,B_k) = \\ &(4\pi)^m\int_{\mathbb{R}^{2l}}\int_{\mathbb{R}^{2l'}} {\rm d}\vec x_{\bf f} {\rm d}\vec x_{\bf g} W(\vec x_{\bf f} \oplus \vec x_{\bf g} )W_{A_j}(\vec x_{\bf f})W_{B_k}(\vec x_{\bf g})
\end{split}
\end{equation}
Now let us assume that all these Wigner functions are positive. Because they are normalised, this implies that $W(\vec x_{\bf f} \oplus \vec x_{\bf g})$ is a probability distribution on the entire phase space $\mathbb{R}^{2m}$, and $W_{A_j}(\vec x_{\bf f})$ and $W_{B_j}(\vec x_{\bf g})$ are probability distributions on the reduced phase spaces $\mathbb{R}^{2l}$ and $\mathbb{R}^{2l'}$, respectively. However, the model (\ref{eq:LHVBell}) does not require probability distributions on phase space, but rather on the possible measurement outcomes. 

This is where (\ref{eq:GoingToProbs}) comes into play. Because $W_{A_j}(\vec x_{\bf f})$ and $W_{B_j}(\vec x_{\bf g})$ are positive, more than just treat them as probability distributions in phase space we can also consider $P_{\vec x_{\bf f}}(A_j) = (4\pi)^{l}W_{A_j}(\vec x_{\bf f})$ and $P_{\vec x_{\bf g}}(B_j) = (4\pi)^{l'}W_{B_j}(\vec x_{\bf g})$ as the probability of getting the measurement outcomes $A_j$ and $B_k$, respectively. Because of (\ref{eq:GoingToProbs}) we find that these probabilities are correctly normalised
\begin{equation}
\sum_j P_{\vec x_{\bf f}}(A_j) = \sum_j P_{\vec x_{\bf g}}(B_j) = 1,
\end{equation}
and because the Wigner functions are positive, we also find that $P_{\vec x_{\bf f}}(A_j), P_{\vec x_{\bf g}}(B_j) \geqslant 0$. Note that the phase space coordinates $\vec x_{\bf f}$ and $\vec x_{\bf g}$ are no longer treated as the variable, but rather as a label. The set $\{P_{\vec x_{\bf f}}(A_j) \mid \vec x_{\bf f} \in \mathbb{R}^{2l} \}$ denotes a family of different probability distributions on the space of measurement outcomes $\{A_1, A_2, \dots \}$. The set $\{P_{\vec x_{\bf g}}(B_j) \mid \vec x_{\bf g} \in \mathbb{R}^{2l'} \}$ can be interpreted analogously. 

We can thus recast (\ref{eq:PJointPos}) in the following form 
\begin{equation}\begin{split}\label{eq:PJointPos2}
&P(A_j,B_k) =\int {\rm d}\vec x_{\bf f} {\rm d}\vec x_{\bf g} W(\vec x_{\bf f} \oplus \vec x_{\bf g} )P_{\vec x_{\bf f}}(A_j)P_{\vec x_{\bf g}}(B_k).
\end{split}
\end{equation}
Because $W(\vec x_{\bf f} \oplus \vec x_{\bf g} )$ is a positive and normalised Wigner function, it is a joint probability distribution on the coordinates $\vec x_{\bf f}$ and $\vec x_{\bf g}$. These coordinates label families of probability distributions $\{P_{\vec x_{\bf f}}(A_j) \mid \vec x_{\bf f} \in \mathbb{R}^{2l} \}$ and $\{P_{\vec x_{\bf g}}(B_j) \mid \vec x_{\bf g} \in \mathbb{R}^{2l'} \}$ for the measurement outcomes. The expression (\ref{eq:PJointPos2}) is thus fully consistent with Bell's local hidden variable model (\ref{eq:LHVBell}). As a consequence, we cannot violate any Bell inequalities when the system is prepared in a state with a positive Wigner function and when we only have access to POVM that have Wigner representations with positive Wigner functions.

Let us emphasise that there is generally no reason to assume that the probabilities $P_{\vec x_{\bf f}}(A_j)$ and $P_{\vec x_{\bf g}}(B_j)$ are also consistent with quantum mechanics. In other words, there is not necessarily any state $\hat \rho$ such that $P_{\vec x_{\bf f}}(A_j) = \tr[\hat \rho \hat A_j]$. However, because we are dealing with Bell non-locality, we do not need this to be the case, since (\ref{eq:LHVBell}) allows for arbitrary local probability distributions.

To make a long story short, we have shown that Wigner negativity is necessary for witnessing Bell non-locality. The interested reader can consult works such as \cite{PhysRevLett.101.020401} that relate Wigner negativity to the more general concept of quantum contextuality. However, the topic of contextuality in CV systems is still a matter of scientific debate \cite{barbosa2019continuousvariable}.\\

There has been a significant body of work about the violation of Bell inequalities in CV setups \cite{PhysRevA.79.012112,PhysRevLett.93.130409,KLYSHKO1993399}. It is evident that this is an arduous task once one approaches a realistic experimental setting \cite{PhysRevA.98.062101}. Here, we focus on one particular suggestion to test Bell non-locality based on a state's Wigner function \cite{PhysRevA.58.4345,PhysRevLett.82.2009}.

The starting point of this approach is the CHSH inequality 
\begin{equation}\tag{\ref{eq:CHSH}}
\abs{\<\hat X \hat Y\> - \<\hat X \hat Y'\> + \<\hat X' \hat Y\> + \<\hat X' \hat Y'\>} \leqslant 2.
\end{equation}
As we discussed in Subsection \ref{sec:BellIntro}, this inequality relies on some assumptions for the observables $X, X', Y,$ and $Y'$. In particular, we must assume that the measurement outcomes are either $-1$ or $1$. In a CV setting, where we generally deal with a continuum of possible measurement outcomes, this seems like a serious constraint. Nevertheless, we have already encountered some natural examples during this Tutorial. For example, photon counters yield a discrete number of possible measurement outcomes. Here, we choose a related observable that takes us all the way back Subsection \ref{sec:Phase}, where we encountered the observable
\begin{equation}\tag{\ref{eq:Parity}}
\hat \Delta (\vec x) = \hat D(-\vec x) (-\mathds{1})^{\hat N} \hat D(\vec x).
\end{equation}
This displaced parity operator has a rich structure, but when it comes to actual measurement outcomes is will return either $-1$ or $1$. This means that we can  choose $X, X', Y,$ and $Y'$ to be parity operators. First of all, let us note that
\begin{equation}
\hat \Delta (\vec x_{\bf f} \oplus \vec x_{\bf g}) = \hat \Delta(\vec x_{\bf f}) \otimes \hat \Delta( \vec x_{\bf g}) . 
\end{equation}
To see this, one can first show that $(-\mathds{1})^{\hat N_m} = (-\mathds{1})^{\hat N_l + \hat N_{l'}} =  (-\mathds{1})^{\hat N_l} \otimes (-\mathds{1})^{\hat N_{l'}}$ and subsequently use $\hat D(\vec x_{\bf f} \oplus \vec x_{\bf g}) = \hat D(\vec x_{\bf f}) \otimes \hat D( \vec x_{\bf g})$ (displacements in different modes are independent from each other). 

Now we can identify the observables as follows:
\begin{equation}
\begin{split}
X = \hat \Delta (\vec x_{\bf f}), \quad X' = \hat \Delta (\vec x'_{\bf f})\\
Y = \hat \Delta (\vec x_{\bf g}), \quad Y' = \hat \Delta (\vec x'_{\bf g})
\end{split}
\end{equation}
and therefore the CHSH inequality is transformed into 
\begin{equation}\begin{split}\label{eq:CHSHPar1}
&\abs{\<\hat \Delta(\vec x_{\bf f} \oplus \vec x_{\bf g})\> - \< \hat \Delta(\vec x_{\bf f} \oplus \vec x'_{\bf g})\> + \< \hat \Delta(\vec x'_{\bf f} \oplus \vec x_{\bf g}) \> + \<\hat \Delta(\vec x'_{\bf f} \oplus \vec x'_{\bf g})\>}\\
& \leqslant 2.
\end{split}
\end{equation}
As a next step, we use (\ref{eq:wigexpvalueparity}) to write
\begin{equation}
\<\hat \Delta(\vec x_{\bf f} \oplus \vec x_{\bf g})\> = (2\pi)^m W(\vec x_{\bf f} \oplus \vec x_{\bf g}),
\end{equation}
such that the inequality (\ref{eq:CHSHPar1}) can be recast as
\begin{equation}\begin{split}\label{eq:CHSHWig}
&\abs{W(\vec x_{\bf f} \oplus \vec x_{\bf g}) - W(\vec x_{\bf f} \oplus \vec x'_{\bf g}) + W(\vec x'_{\bf f} \oplus \vec x_{\bf g}) + W(\vec x'_{\bf f} \oplus \vec x'_{\bf g})}\\
& \leqslant \frac{2}{(2\pi)^m }.
\end{split}
\end{equation}
Any state with a Wigner function that violates this inequality for some choice of coordinates $\vec x_{\bf f},\vec x'_{\bf f}, \vec x_{\bf g},$ and $\vec x'_{\bf g}$ possesses some form of Bell non-locality. In the works \cite{PhysRevA.58.4345,PhysRevLett.82.2009} it is argued that the inequality (\ref{eq:CHSHWig}) can be violated by sending a single photon through a beam splitter, but also by an EPR state. The fact that a Gaussian Wigner function suffices to violate (\ref{eq:CHSHWig}) sometimes comes as a surprise, because we previously argued that one needs Wigner negativity to violated Bell inequalities. The reason why one can detect Bell non-locality with this inequality even when the Wigner function is positive stems from our choice of observable $\hat \Delta(\vec x_{\bf f} \oplus \vec x_{\bf g})$. The POVM elements that correspond to the measurement outcomes $1$ and $-1$ have Wigner functions that are strongly Wigner negative. As a consequence the necessary Wigner negativity is baked into (\ref{eq:CHSHWig}) by construction. \\

In practice, the inequality (\ref{eq:CHSHWig}) is highly sensitive to impurities and can often be hard to violate with experimentally reconstructed Wigner functions. The hunt for good new techniques to show Bell non-locality in CV systems is therefore still open. However, this Subsection clearly showed us that Wigner negativity is necessary to observe one of the most exotic features in quantum physics. This negativity might be baked into the state, but it could just as well be induced by measurements. The conditional methods of Subsection \ref{sec:CondnonGaussian} also highlight this duality, where Wigner negativity in the measurement is used to induce Wigner negativity in the state. It should come a no surprise that Wigner negativity is also a necessary ingredient for the most exotic quantum protocols. However, it should also be highlighted that even with a little extra trust, it is possible to design protocols that do not require Wigner-negativity to witness quantum correlations \cite{PhysRevLett.126.190502}. In the next section, we discuss its importance for reaching a quantum advantage with CV systems.

\section{Non-Gaussian Quantum Advantages}\label{sec:Advantages}

It has been long known that systems that are entirely built with Gaussian building blocks are easy to simulate \cite{PhysRevLett.88.097904}. It should perhaps not come as a surprise that efficient numerical tools exist to sample numbers from a multivariate Gaussian distribution. The discrete variable analog of this result comes across as less intuitive and goes by the name ``Gottesman-Knill theorem'' \cite{PhysRevA.70.052328}. Yet, it turns out that something stronger than mere non-Gaussian elements is required to render a system hard to simulate.

In Subsection \ref{sec:Bell}, we encountered the power of Wigner negativity by realising that it is a necessary requirement for Bell non-locality. This connection between Wigner negativity and the most exotic types of quantum correlations shows us that Wigner negativity is key to giving CV systems their most prominent quantum features. It is then perhaps not a surprise that such Wigner negativity is also a necessary requirement for implementing any type of protocol that cannot be efficiently simulated by a classical computer \cite{mari_positive_2012,Veitch_2013,rahimi-keshari_sufficient_2016}. We thus start our discussion of quantum advantages by explaining the result of \cite{mari_positive_2012}. To show the necessity of Wigner negativity, we show an explicitly simulation algorithm for general quantum protocols without Wigner negativity.\\

Any quantum protocol ultimately relies on the measurement of a certain set of measurement operators $\{ \hat E_j \}$ (typically a POVM) of a system prepared on a state $\hat \rho$. In a Wigner function formalism, we then find
\begin{equation}\label{eq:pj}
p_j = (4\pi)^m \int_{\mathbb{R}^{2m}} {\rm d}\vec x\, W_{E_j}(\vec x)W(\vec x).
\end{equation}
Furthermore, the fact that the set $\{ \hat E_j \}$ forms a POVM implies that 
\begin{equation}\label{eq:samplingprobs}
(4\pi)^m \sum_j W_{E_j}(\vec x) = 1.
\end{equation}
As we already discussed in Subsection \ref{sec:Bell}, surrounding eq.~(\ref{eq:GoingToProbs}), it is crucial that the normalisation condition (\ref{eq:samplingprobs}) holds for any phase space coordinate $\vec x$. In the present context, we want to show that there is an efficient method for a classical device to sample values from the probability distribution $\{p_j\}$ when all involved Wigner functions are positive.

Let us start by assuming that the Wigner functions that describe the POVM elements are all positive. When combined with the POVM condition (\ref{eq:samplingprobs}), this implies that we can identify a set of probabilities $P_{\vec x}(e_j) = (4\pi)^m W_{E_j}(\vec x)$ as the probability to obtain the measurement outcome $e_j$, associated with the POVM element $\hat E_j$. These $P_{\vec x}(e_j)$ depend on a parameter $\vec x$, we can thus form a family of probability distributions $\{P_{\vec x}(e_j) \mid \vec x \in \mathbb{R}^{2m}\}$ that describe the probability of obtaining the different results $e_j$, depending on a chosen phase space point. The normalisation condition (\ref{eq:samplingprobs}) now states that $\sum _j P_{\vec x}(e_j) = 1$ for all $\vec x$. Let us emphasise that this family of probabilities would not be well-defined if $W_{E_j}(\vec x)$ were not positive Wigner functions, as some of the probabilities would be negative. 

Going back to the initial equation (\ref{eq:pj}), we now find that 
\begin{equation}\label{eq:pj2}
p_j = \int_{\mathbb{R}^{2m}} {\rm d}\vec x\, P_{\vec x}(e_j) W(\vec x).
\end{equation}
To find the actual probability of getting the $j$th outcome is thus given by ``averaging'' the probabilities $P_{\vec x}(e_j)$ over the different phase space coordinates. Generally speaking, this is not a real average, unless the Wigner function $W(\vec x)$ of the state is an actual probability distribution on phase space. The latter is exactly the case when $W(\vec x)$ is positive. Then, we can simply think of the probability $p_j$ for obtaining event $e_j$ as $p_j =\mathbb{E}_W [P_{\vec x}(e_j)] $, where $\mathbb{E}_W$ is the expectation value over the probability distribution $W(\vec x)$.

Hence, when all Wigner functions are positive, the algorithm to simulate our relevant quantum process can simply be expressed by the following steps:
\begin{enumerate}
\item Sample a phase space coordinate $\vec x$ from the probability distribution $W(\vec x)$.
\item Construct the probability distribution $P_{\vec x}(e_j)$ for the sampled value $\vec x$.
\item Sample an outcome $e_j$ from the probability distribution $P_{\vec x}(e_j)$.
\end{enumerate}
Even though this is the general idea behind our sampling protocol, there are some major hidden assumptions. First, we assume here that the Wigner function for the state and the measurement are known. Furthermore, we also assume that we can simply sample points from any distribution on phase space and from any distribution of measurement outcomes $P_{\vec x}(e_j)$. In particular for the sampling aspects it is not at all clear that these are reasonable assumptions to make. Standard sampling protocols for multivariate probability distributions tend to get highly inefficient once the probability distributions become too exotic such that it is dangerous to assume that we can ``just sample''. \\

To address this point Ref.~\cite{mari_positive_2012} makes more assumptions on the exact setup we are trying to simulate. First, we assume that the detection is done by a series of single-mode detectors, such that our label $j$ now become a tuple ${\bf j} = (j_1, j_2, \dots, j_m)$ where $j_k$ denotes the outcome $e^{(k)}_{j_k}$ for the detector on the $k$th mode. We can thus write the POVM element as $\hat E_{\bf j} = \hat E^{(1)}_{j_1} \otimes \dots \otimes \hat E^{(m)}_{j_m}$, such that 
\begin{equation}
W_{E_{\bf j}}(\vec x) = W_{E^{(1)}_{j_1}} (x_1,p_1)W_{E^{(2)}_{j_2}} (x_2,p_2) \dots W_{E^{(m)}_{j_m}} (x_m,p_m).
\end{equation}
We assume that each detector has been accurately calibrated, such that all the individual Wigner functions are known. This implies that for a given point in phase space $\vec x = (x_1, p_1, \dots, x_m,p_m)^T$, we can simply evaluate the probabilities for each detector to produce a certain outcome. Thus, we calculate the probability distributions $P_{(x_k,p_k)}(e^{(k)}_{j}) = 4 \pi W_{E^{(k)}_j}(x_k,p_k)$ for all the possible measurement outcomes for that specific mode. We will assume that sampling outcomes $e^{(k)}_{j}$ from these probability distributions $P_{(x_k,p_k)}(e^{(k)}_{j})$ is a feasible task. For typical detectors in quantum optics experiments this is a very reasonable assumption.

The Wigner function $W(\vec x)$ that describes the state is more subtle as it also includes all correlations between modes. If the state is Gaussian, a measurement of the covariance matrix would be sufficient to know the full Wigner function. Because of its Gaussian features, there are efficient tools to directly sample phase space points. Yet, for more general non-Gaussian positive Wigner functions this sampling may be much harder. Therefore, Ref.~\cite{mari_positive_2012} makes an essential assumption: it assumes that we know a protocol that combines local operations to design the state $W(\vec x)$ from a known initial state with no correlations between the modes. The notion of ``locality'' should here be understood in the sense of acting on a small set of modes while leaving the others fully untouched. These local operations are also supposed to be represented by positive Wigner functions which only depend on the phase space coordinates of the subset of modes on which they act. 

Generally speaking, such Wigner positive operations $\Xi: {\cal H}^{\rm in} \rightarrow {\cal H}^{\rm out}$ map a state $\hat \rho$ to a new state $\Xi[\hat \rho]$. In Ref.~\cite{mari_positive_2012}, the Choi representation \cite{CHOI1975285,doi:10.1063/1.3581879} is used to represent $\Xi$ as a state on a larger Hilbert space ${\cal H}^{\rm in}\otimes{\cal H}^{\rm out}$. This becomes particularly appealing when we go to a Wigner representation, where the Choi representation of $\Xi$ is given by a Wigner function $W_{\Xi} ( \vec x^{\, \rm in} \oplus\vec x^{\, \rm out})$. The action of $\Xi$ on a state with Wigner function $W(\vec x^{\, \rm in})$ is then given by
\begin{equation}\label{eq:in-out}
W_{\rm out}(\vec x^{\,\rm out}) = (4 \pi)^m \int_{\mathbb{R}^{2m}} {\rm d}\vec x^{\,\rm in}\, W_{\Xi} (\vec x^{\,\rm in} \oplus T\vec x^{\,\rm out})W(\vec x^{\,\rm in}),
\end{equation}
where $m$ is the number of modes of the input state. For technical reasons, we must include the transposition operator $T$ (\ref{eq:TranspositionOperator}) in the action of the channel. Because this operation must be trace-preserving, we on top get the property that 
\begin{equation}\label{eq:normalisationOperations}
 (4 \pi)^m \int_{\mathbb{R}^{2m}} {\rm d}\vec x^{\, \rm out}\, W_{\Xi} (\vec x^{\, \rm in} \oplus T\vec x^{\, \rm out}) = 1.
\end{equation}
When we now assume that the operation $\Xi$ has a Wigner-Choi representation $W_{\Xi} ( \vec x^{\, \rm in} \oplus\vec x^{\, \rm out})$ which is a positive function, it immediately follows that the operation $\Xi$ turns a Wigner positive initial state $W(\vec x^{\, \rm in})$ into a Wigner positive output state $W_{\rm out}(\vec x^{\, \rm out})$. 

It is useful to note that such operations (\ref{eq:in-out}) can be trivially embedded in a larger space. Let us assume that we consider a state $W(\vec x_{\bf f} \oplus \vec x_{\bf g})$, we can simply let the operation act on the modes $\bf g$ by taking
\begin{equation}\label{eq:local_operations}
W_{\rm out}(\vec{x}_{\bf f} \oplus \vec x^{\, \rm out}_{\bf g}) = (4 \pi)^{l'} \int_{\mathbb{R}^{2l'}} {\rm d}\vec x_{\bf g}^{\rm in}\, W_{\Xi} (\vec x^{\, \rm in}_{\bf g} \oplus T\vec x^{\, \rm out}_{\bf g})W(\vec x_{\bf f} \oplus \vec x^{\, \rm in}_{\bf g}),
\end{equation}
Notationally, this may seem a little complicated, but, in essence, we just carry out the integration over a subset of the full phase space. We will call these operations local Wigner positive operations.

In our simulation protocol, we thus assume that $W(\vec x)$ is created by a series of such local Wigner positive operations of $\Xi_1, \dots, \Xi_t$ on a non-correlated input state $W_{\rm in}(\vec x_{\rm in}) = W^{(1)}_{\rm in}(x_1,p_1)W^{(2)}_{\rm in}(x_2,p_2) \dots W^{(m)}_{\rm in}(x_m,p_m)$. 
\begin{equation}\begin{split}
W(\vec x) = (4\pi)^{m t}\int_{\rm \mathbb{R}^{2m}} &{\rm d} \vec x_t \dots \int_{\rm \mathbb{R}^{2l}} {\rm d} \vec x_t W_{\Xi_t}(\vec x_t \oplus T \vec x)\dots 
\\ &\times W_{\Xi_2}(\vec x_1 \oplus T\vec x_2 ) W_{\Xi_1}( \vec x_{\rm in} \oplus T\vec x_1)\\
&\times W_{\rm in}(\vec x_{\rm in}).
\end{split}
\end{equation}
We assume on top that each operation is local over a small number of modes $l \ll m$. To model this with (\ref{eq:local_operations}), it suffices to split $\vec x_{t_{k-1}} = \vec x_{t_{k-1}}^l \oplus \vec x_{t_{k-1}}^{l'}$ and $\vec x_{t_{k}} = \vec x_{t_{k}}^l \oplus \vec x_{t_{k}}^{l'}$, such that 
\begin{equation}\label{embeddingLocalOperation}
(4\pi)^m W_{\Xi_{t_k}}(\vec x_{t_{k-1}} \oplus T \vec x_{t_k}) = (4\pi)^l W_{\Xi_{t_k}}(\vec x^{\, l}_{t_{k-1}} \oplus T \vec x^{\, l}_{t_k}) \delta (\vec x^{l'}_{t_{k-1}} - \vec x^{l'}_{t_k}).
\end{equation}
Even though the notation is complicated, it simply describes that we act on an $l$-mode subspace with the operation $\Xi_{t_k}$ and leave the other $l'$ modes untouched. 

The normalisation condition (\ref{eq:normalisationOperations}) now has an important consequence, since it allows us to identify a probability distribution on phase space $P_{\vec x_{t_{k-1}}}(\vec x_{t_{k}}) = (4\pi)^m W_{\Xi_{t_k}}(\vec x_{t_{k-1}} \oplus T \vec x_{t_k})$. It gives us the probability of choosing a phase space value $\vec x_{t_{k}}$, given that we know $\vec x_{t_{k-1}}$. Because the operations are local, (\ref{embeddingLocalOperation}) allows us to keep most of the phase space coordinates constant  from step to step. Furthermore, the first step is simple. Every pair $(x_k, p_k)$ of the initial coordinate $\vec x_{\rm in}$ can be sampled independently because $W_{\rm in}(\vec x_{\rm in})$ factorises. This now gives us the following new algorithm:
\begin{enumerate}
\item Take the initial Wigner function $W_{\rm in}(\vec x^{\, \rm in}) = W^{(1)}_{\rm in}(x_1,p_1)W^{(2)}_{\rm in}(x_2,p_2) \dots W^{(m)}_{\rm in}(x_m,p_m)$ and sample a pair $(x_k,p_k)$ from every single-mode probability distribution $W^{(k)}_{\rm in}(x_k,p_k)$. Put all these pairs together to obtain $\vec x^{\, \rm in} = (x_1,p_1, \dots, x_m, p_m)$.
\item Update the coordinate by sampling new coordinates based on $P_{\vec x_{t_{k-1}}}(\vec x_{t_{k}}) = (4\pi)^m W_{\Xi_{t_k}}(\vec x_{t_{k-1}} \oplus T \vec x_{t_k})$. Because $\Xi_{t_k}$ are local operations, it suffices to only locally update coordinates. Let us make this clear through an example. Say we have $\vec x_{t_{k-1}} = (x^{(k-1)}_1,p^{(k-1)}_1,\dots, x^{(k-1)}_m,p^{(k-1)}_m)^T$ and operation $\Xi_{t_k}$ acts locally on modes with labels $2,5$, and $7$. Take $\vec x^{\, l}_{t_{k-1}} = (x^{(k-1)}_2,p^{(k-1)}_2,x^{(k-1)}_5,p^{(k-1)}_5,x^{(k-1)}_7,p^{(k-1)}_7)$ and use it to evaluate $P_{\vec x_{t_{k-1}}}(\vec x^{\, l}_{t_{k}}) = (4\pi)^l W_{\Xi_{t_k}}(\vec x^{\, l}_{t_{k-1}} \oplus T \vec x^{\, l}_{t_k})$. Now sample a new vector $\vec x^{\, l}_{t_{k}} = (x^{(k)}_2,p^{(k)}_2,x^{(k)}_5,p^{(k)}_5,x^{(k)}_7,p^{(k)}_7)$ from this probability distribution. Then construct the new vector $\vec x_{t_{k}}$ by taking $\vec x_{t_{k-1}}$ and updating the coordinates associated to modes $2, 5,$ and $7$ to the newly sampled coordinates 
\item After the operations $\Xi_1, \dots, \Xi_t$ have been implemented by updating the phase space coordinate, take the final phase space coordinate $\vec x = (x_1,p_1, \dots, x_m, p_m)^T$ and the Wigner function describing the detectors $W_{E_{\bf j}}(\vec x) = W_{E^{(1)}_{j_1}} (x_1,p_1)W_{E^{(2)}_{j_2}} (x_2,p_2) \dots W_{E^{(m)}_{j_m}} (x_m,p_m)$. For each detector $k$, use the phase space coordinate $\vec x$ to generate the probability distribution $P_{(x_k,p_k)}(e^{(k)}_j) = W_{E^{(k)}_{j}} (x_k,p_k)$.
\item Sample an outcome $e^{(k)}_j$ from the distribution $P_{(x_k,p_k)}(e^{(k)}_j)$ for every detector.
\end{enumerate}
Sampling the final phase space coordinate $\vec x\,$ by using a Monte-Carlo-style update rule is time consuming, but if the operations are local it can be done efficiently. This procedure implicitly assumes that we do not just know the state we are sampling from, but that we know the circuit of local operations that is used to create the state from local resources. Ultimately, when one considers the circuit representation of quantum algorithms, this is also how a quantum algorithm works. For example, Subsection \ref{sec:DetnonGaussian} exactly shows that any unitary CV circuit can be built with single- and two-mode gates. The algorithm outlined in this section shows that we can efficiently simulate any protocol where the local input state, the circuit's operations, and the measurements are described by positive Wigner functions.

One may wonder whether any positive Wigner function $W(\vec x)$ can be constructed through such a circuit and, if so, whether there is an efficient way to design such a circuit when we know the Wigner function. If we assume that not only the state $W(\vec x)$ but also all its marginals are known, it is possible to construct a step-wise sampling procedure through the chain rule of probability theory:
\begin{equation}\begin{split}
W&(\vec x) = \\
&W(x_m,p_m \mid x_{m-1}, p_{m-1}, \dots x_1, p_1) \times \dots\\
&W(x_3,p_3 \mid x_2, p_2, x_1,p_1)W(x_2, p_2 \mid x_1,p_1) W(x_1, p_1).
\end{split}
\end{equation}
This process effectively executes a type of random walk with memory. In each step of this walk, we then sample the phase space coordinates for one mode. Nevertheless, this process only works when we have access to all these conditional probabilities, which practically implies having access to all the marginals of the distribution. In practical setups, this will often not be the case. Nevertheless, it is quickly seen that this setup can be efficiently used to sample from Gaussian Wigner functions where these conditional distributions have a particularly simple form.\\

Thus, we have shown that it is impossible to obtain a quantum computational advantage by using only local states, measurements, and operations with positive Wigner functions. This means that Wigner negativity is necessary to reach a quantum advantage in such setups. However, Wigner negativity is certainly not sufficient since there are many setups of quantum systems that involve negative Wigner functions that can be efficiently simulated \cite{garcalvarez2020efficient}. It is thus interesting to take the opposite approach and explore a setup that is known to lead to a quantum advantage. In the spirit of CV setups, the most logical choice for such a discussion is Gaussian boson sampling \cite{PhysRevLett.119.170501}. In literature, this setup has been studied mainly from the point of view of complexity theory \cite{PhysRevA.100.032326, deshpande2021quantum}, but here we rather focus on its physical building blocks. 

Boson sampling \cite{v009a004} is a problem in which one injects a set of $N$ bosons (generally photons) into an $m$-mode interferometer. On the output ports of this interferometer, photodetectors are mounted to count the particles at the output. Simulating this type of quantum Galton board is a computationally hard task, implying that a quantum advantage could be reached by implementing the setup in a quantum optics experiment. On the other hand, it turns out that the required number of photons to implement such an experiment is also hard to come by. This was the motivation for developing a new approach, where the input photons are replaced by squeezed states that are injected in each of the interferometer inputs. Because an interferometer, built out of phase shifters and beam splitters, is a Gaussian transformation the output state will remain Gaussian. We can thus effectively say that we are sampling photons from a state with Wigner function $W_G(\vec x)$. In addition, there is no mean field in the setup such that the entire state is characterised by its covariance matrix $V$.

When we assume that the detectors resolve photon numbers, the probability to detect a string of counts ${\bf n} = (n_1, \dots, n_m)$ can we written as
\begin{equation}\label{eq:WignerGaussianBoson}
P({\bf n}) = (4\pi)^m \int_{\mathbb{R}^{2m}} {\rm d}\vec x\, W_{\bf n}(\vec x) W_G(\vec x).
\end{equation}
We can then use (\ref{eq:WigFock}) to write
\begin{equation}
W_{\bf n}(\vec x) = W_{n_1}(x_1,p_1)\dots W_{n_m}(x_m,p_m).
\end{equation}
Even though the integral (\ref{eq:WignerGaussianBoson}) is hard to compute, it is insightful in the light of (\ref{eq:pj2}) and our discussion regarding the necessity of Wigner negativity. Indeed, we see immediately that the detectors form a crucial element in rendering the setup hard to simulate. The same holds when we replace the number-resolving detectors with their on-off counterparts \cite{PhysRevA.98.062322} such that $n_k = \{0,1\}$ and the Wigner functions are given by $W_{n_k}(x_k,p_k) = (1 - 2\exp[-(x_k^2 + p_k^2)/2])/(4\pi)$. 

When we stick with number-resolving detectors that project on Fock states, it is practical to reformulate the problem in terms of P-functions and Q-functions, such that 
\begin{equation}\label{eq:PFuncGaussianBoson}
P({\bf n}) = \int_{\mathbb{R}^{2m}} {\rm d}\vec x\, P_{\bf n}(\vec x) Q_G(\vec x).
\end{equation}
For the detailed calculation, we refer to Ref.~\cite{PhysRevLett.119.170501}. It turns out that the probabilities $P({\bf n})$ can be expressed in terms of the Hafnian of a matrix \cite{Caianiello}, which establishes a connection to the problem of finding perfect matchings in graph theory. This connection has lead to several suggested applications for Gaussian Boson Sampling \cite{PhysRevA.98.032310,PhysRevA.101.032314,Bromley_2020}. 

In the light of this Tutorial, the most interesting application of Gaussian Boson Sampling is its potential role in quantum state engineering \cite{PhysRevA.100.052301}. When only a subset of modes are measured, we can see Gaussian Boson Sampling as a generalisation of photon subtraction (and even as a generalisation of ``generalised photon subtraction'' \cite{PhysRevA.103.013710}). The idea is reasonably simply explained in the light of Subsection \ref{sec:CondnonGaussian}: when we split the system in two parts $\mathbb{R}^{2m} = \mathbb{R}^{2l} \oplus \mathbb{R}^{2l'}$, such that the Gaussian state that comes out of the interferometer now takes the form $W_{G}(\vec x_{\bf f} \oplus \vec x_{\bf g})$, we can post-select on a measurement outcome ${\bf n} = (n_1, \dots n_{l'})$ for the second subsystem. We thus project on a state $W_{\bf n}(\vec x_{\bf g})$, which is a product of $l'$ Fock states, and from (\ref{eq:CondTwo}), we obtain that the conditional state on the remaining modes is given by
\begin{equation}
W_{{\bf f} \mid {\bf n}}(\vec x_{\bf f}) = \frac{\< [\ket{n_1}\bra{n_1} \otimes \dots \otimes \ket{n_{l'}}\bra{n_{l'}} ] \>_{{\bf g} \mid \vec x_{\bf f}}}{\< [\ket{n_1}\bra{n_1} \otimes \dots \otimes \ket{n_{l'}}\bra{n_{l'}} ] \>} W_{\bf f}(\vec x_{\bf f}) ,
\end{equation} 
with $W_{\bf f}(\vec x_{\bf f})$ defined by (\ref{eq:Wf}). From (\ref{eq:Agf}) we recall the expression
\begin{equation}\label{eq:ConditionalPartGBS}
\begin{split}
&\< [\ket{n_1}\bra{n_1} \otimes \dots \otimes \ket{n_{l'}}\bra{n_{l'}} ] \>_{{\bf g} \mid \vec x_{\bf f}} \\
&= (4\pi)^{l'}\int_{\mathbb{R}^{2l'}}{\rm d}\vec x_{\bf g} \, W_{\bf n}(\vec x_{\bf g}) W_{G}(\vec x_{\bf g} \mid \vec x_{\bf f}),
\end{split}
\end{equation}
and because the initial state $W_G(\vec x_{\bf f} \oplus \vec x_{\bf g})$ is Gaussian, we find that the conditional probability distribution $W_{G}(\vec x_{\bf g} \mid \vec x_{\bf f})$ is given by (\ref{GaussCond}). Ironically, to evaluate $\< [\ket{n_1}\bra{n_1} \otimes \dots \otimes \ket{n_{l'}}\bra{n_{l'}} ] \>_{{\bf g} \mid \vec x_{\bf f}}$ and $\< [\ket{n_1}\bra{n_1} \otimes \dots \otimes \ket{n_{l'}}\bra{n_{l'}} ] \>$ we must essentially solve the same hard problem as for the implementation of Gaussian Boson Sampling itself. Therefore, the exact description of the resulting states is generally complicated.

Nevertheless, in idealised scenarios, even small Gaussian Boson Sampling circuits can be used to prepare interesting non-Gaussian states \cite{PhysRevA.100.052301}. In particular the capacity of Gaussian Boson Sampling to produce GKP states has taken up a prominent place in a recent blueprint for photonic quantum computation \cite{Bourassa2021blueprintscalable}. Furthermore, the results in Subsection \ref{sec:QunatumCorrThroughnonGauss} suggest that states created by performing Gaussian Boson Sampling on a subset of modes can have additional non-Gaussian entanglement. Yet, to be able to use this procedure to produce highly resourceful Wigner negative states, Subsection \ref{SteeringWignerNeg} highlights that the initial Gaussian state needs to be such that the modes in ${\bf f}$ can steer the modes in ${\bf g}$. This condition can be seen as a basic quality requirement for the Gaussian Boson Samplers that are used in \cite{Bourassa2021blueprintscalable}.

Finally, the experimental imperfections are also detrimental for the quantum advantage that is produced in Gaussian Boson Sampling. Clearly, when the Gaussian state $W_G(\vec x)$ can be written as a Gaussian mixture of coherent states (meaning that no mode basis exists in which the quadrature noise is below vacuum noise), the sampling can be simulated efficiently. Because multimode coherent states are always just a tensor product of single-mode coherent states, it suffices to  sample a coherent state from the mixture, calculate all the individual probabilities for the output detectors, and sample independent detector outputs according to these probabilities. The presence of entanglement in the Gaussian state from which we sample is thus crucial. In addition, detector efficiencies must be sufficiently high such that their Wigner functions remain non-positive, otherwise the protocol of \cite{mari_positive_2012} renders the setup easy to simulate (as explained in the first part of this section). A more thorough analysis of how different experimental imperfections render Gaussian Boson Sampling easier to simulate can be found in Ref.~\cite{PhysRevLett.124.100502}.\\

There are clearly still many aspects of the relation between non-Gaussian features of quantum states on the one hand, and the ability to achieve a quantum computational advantage on the other hand, that are not yet fully understood. The Gaussian Boson Sampling setup clearly emphasises the importance of entanglement in combination with Wigner negativity. Furthermore, there is the implicit fact that a simulation scheme such as \cite{mari_positive_2012} requires knowledge of the actual circuit of local operations that was used to create the state. It does make sense to assume that we actually have some ideas of the quantum protocol that we are attempting to simulate, but yet one may wonder whether there could be a reasonable setting (in the sense that we are actually implementing a well-controlled protocol) in which the assumptions of  \cite{mari_positive_2012} do not hold. This clearly shows that many fundamental theoretical aspects of CV quantum computation remain to be uncovered.

\section{Experimental realisations}\label{sec:Exp}
Now that we have provided an overview of some theoretical aspects of non-Gaussian quantum states, we interpret the ``where to find them'' part of the title in a very literal sense. Non-Gaussian states are generally rather fragile, as one should expect from quantum central limit theorem and the fact that thermal states in free bosonic theories are Gaussian. Producing and analysing non-Gaussian states in a laboratory setting is indeed challenging, but nevertheless it has been done numerous times. Our main focus in Subsection \ref{sec:QuantOptExp} will be quantum optics, which is the historical testbed for CV quantum physics. However, in recent years there has been increased attention for CV approaches in other setting such as optomechanics, superconducting circuits, and trapped ions.

\subsection{Quantum optics experiments}\label{sec:QuantOptExp}

{\it This section provides an overview of some of the most important milestones in the generation of non-Gaussian states in optics. For more details, we refer the reader to a specialised review \cite{lvovsky2020production}.}\\

Historically, one might argue that the first experimental realisations of non-Gaussian states in optical setups relied on sufficiently sensitive photon detectors. Initial demonstrations primarily used photoemission of atoms \cite{PhysRevD.9.853,PhysRevLett.39.691} which are prepared in excited states (e.g. by electron bombardment) or via resonance fluorescence in ions \cite{PhysRevLett.58.203}. The development of spontaneous parametric downconversion (SPDC) made it possible to create a single-photon state using only bulk optical elements \cite{PhysRevLett.56.58}. However, all these early non-Gaussian states were characterised through counting statistics, which means that we generally classify them as DV experiments.

It is perhaps intriguing to note that SPDC is also the process that lies at the basis of the creation of squeezed states of light \cite{SCHNABEL20171} which are Gaussian. These states play a key role in the generation of single-photon states, simply because a weakly squeezed vacuum is mainly a superposition of vacuum and a photon pair. By detecting one photon of the pair, the presence of the second photon is heralded. Hence, the approach of \cite{PhysRevLett.56.58} is a basic implementation of a conditional scheme for the generation of non-Gaussian states as presented in Subsection \ref{sec:CondnonGaussian}.

A genuine CV treatment of such non-Gaussian states would only be achieved much later in a work that presents the first tomographic reconstruction of a state with Wigner negativity in optics \cite{PhysRevLett.87.050402}. Due to the developments of an easily implementable maximum-likelihood algorithm for state reconstruction, homodyne tomography became one of the main tools to study non-Gaussian states in CV quantum optics \cite{RevModPhys.81.299}. It did not take long before this also led to the reconstruction of a displaced single-photon Fock state \cite{PhysRevA.66.011801} and a two-photon Fock state \cite{PhysRevLett.96.213601}. The combination of increased squeezing with type-II SPDC and an array of photon detectors to increase the number of heralded photons more recently made it possible to resolve the Wigner function of a three photon Fock state \cite{Cooper:13}. Similar ideas of multiplexed photon detection have also been used the generate superpositions of Fock states \cite{Yukawa:13}.

For non-Gaussian states beyond Fock states, photon subtraction, as described in Subsection \ref{sec:PhotonSubtraction}, is a common experimental tool. Its first experimental implementation successfully showed the capability of generating non-Gaussian statistics in the homodyne measurements, but it failed to demonstrate Wigner negativity \cite{PhysRevLett.92.153601}. Later experiments improved the quality of the generated states, demonstrating Wigner negativity and creating so-called ``Schr\"odinger kittens'' \cite{Ourjoumtsev83,PhysRevLett.97.083604,Wakui:07}. The terminology is chosen because these states resemble cat states $\sim \ket{\alpha} - \ket{-\alpha}$ for small values of the mean field $\alpha$. Even though such Schr\"odinger kittens are ultimately not very different from squeezed single photon states, the nomenclature makes more sense in the context of experiments that ``breed'' cat states \cite{Sychev-breeding}. Here, one mixes two Schr\"odinger kittens on a beamsplitter and performs homodyne detection on one output port. By conditioning on instances where this homodyne detector registers values close to zero, one effectively heralds a larger cat state (the value of $\alpha$ has increased). A variation of photon subtraction has also been used to create a type of CV qubits \cite{PhysRevLett.105.053602}.

As an alternative to photon subtraction, one can also add a photon \cite{Zavatta660}. Even though this operation theoretically equates to applying a creation operator on the state, it is experimentally much harder to implement than photon subtraction as it requires non-linear optics. However, photon subtraction can only produce Wigner negativity when the initial state is squeezed. Photon addition, on the other hand, provides the advantage of always creating a Wigner negative state. A simple way to see this is by applying a creation operator to the state and evaluating the Q-function \eqref{eq:Q1}. When a photon is added to the mode ${g}$, the Q-function after photon addition has the property $Q^+(\vec\alpha)\sim (\vec\alpha^T\vec g)^2 Q_G(\vec \alpha)$, where $Q_G(\vec \alpha)$ is the Q-function of the initial Gaussian state. This relation implies automatically that the Q-function will be exactly zero for $\vec \alpha = \vec 0$, and a zero of the Q-function implies Wigner negativity. This means that one can apply photon addition to highly classical states, such as a coherent state or a thermal state, and still end up creating Wigner negativity. Such photon-added coherent states were also used to experimentally measure \cite{PhysRevA.82.063833} non-Gaussianity $\delta(\hat \rho)$ as defined in (\ref{eq:nonGauss}). Remarkably, combining photon addition and photon subtraction operations in both possible orders provides a way to experimentally verify the canonical commutation relations $[\hat a, \hat a^{\dag}] = 1$, as was shown in \cite{parigi_probing_2007}.

The above methods are all based on Gaussian states as initial resources to generate non-Gaussian states. The non-Gaussian states that are created as such can in turn serve as useful resources to create more intricate non-Gaussian states. Fock states are a commonly used type of input states, for example in the first demonstration of a large Schr\"odinger's cat state \cite{Ourjoumtsev-cats}. Intriguingly, by using non-Gaussian initial states, it suffices to use homodyne detection as the conditional operation. This setup can then be extended to a cat breeding scheme \cite{PhysRevLett.114.193602}. Another method to create large cat states in optics relies on making the light field interact with an atom \cite{Hacker2019}. The presence of entanglement between the ``macroscopic'' coherent state and the ``microscopic'' atomic degrees of freedom make for an experiment that resembles Schr\"odinger's original though experiment \cite{Schrodinger_cat_1935}. Once the atom and the coherent light are entangled, a spin rotation of the atom is followed by a measurement to project the state of the light field in either an even or an odd cat state. This reflects the general idea that atoms still induce much larger nonlinearities than nonlinear crystals. These nonlinearities are the direct source of non-Gaussian effect, but they are also much harder to control. At present, experiments that rely on such higher order nonlinearities to create non-Gaussian states remain rare in the optical regime.  \\


The above methods all focus on the creation of single-mode non-Gaussian states. For multimode systems, much of the experimental progress has concentrated on two-mode systems. As we extensively discussed throughout this tutorial, an important feature in such multimode systems are quantum correlations. Some of the first experimental demonstrations of non-Gaussian quantum correlations were based on the Bell inequality (\ref{eq:CHSHWig}). Homodyne tomography and a single photon, delocalised over two modes by a beam splitter, suffices to violate the inequality \cite{PhysRevLett.92.193601,PhysRevA.74.052114}. However, these works also teach us that extreme high purities are required to do so. 

Motivated by photon subtraction experiments and challenged by the no-go theorem of \cite{PhysRevLett.89.137903,PhysRevLett.89.137904,PhysRevA.66.032316}, entanglement distillation soon became a new focus for non-Gaussian quantum optics experiments. Some of these experiments have focused on adding some form of non-Gaussian noise on the initial state to circumvent the no-go theorem \cite{Dong-2008,Hage-2008}. Entanglement distillation through local photon subtraction from the entangled modes of a Gaussian input state would later be demonstrated in \cite{takahashi_entanglement_2010}. Earlier, it had already been shown that Gaussian entanglement can be increased by photon subtraction in a superposition of the entangled mode \cite{ourjoumtsev_increasing_2007}. Interestingly, in the latter case, the photon is effectively subtracted in a non-entangled mode such that the setup is essentially equivalent to mixing a squeezed vacuum and a photon-subtracted squeezed vacuum on a beam splitter. A similar photon subtraction in a coherent superposition of modes was later carried out to entangle two Schr\"odinger kittens \cite{Ourjoumtsev-entagled-cats}. This can probably be seen as the first realisations of purely non-Gaussian entanglement in CV. 

Photon addition has also been considered as a tool for creating entanglement between pairs of previously uncorrelated modes \cite{Hybrid-entanglement-addition}. The resulting state can be seen as a hybrid entangled state $\sim \ket{0}\ket{\alpha} + \ket{1}\ket{-\alpha}$, such states have also been produced using techniques similar to photon subtraction \cite{Morin:2014aa}. For two modes, photon addition can be implemented in a mode-selective way \cite{PhysRevLett.124.033604}. This setup is particularly useful to create entanglement between coherent states by adding a photon in a superposition of displaced modes.\\

Going beyond two modes has always remained a challenging task. For mode-selective photon subtraction from a multimode field, one must abandon the typical implementation based on a beam splitter. For two modes, such an alternative photon subtraction scheme was for example realised in the time-frequency domain, by subtracting a photon from a side-band \cite{PhysRevLett.121.143602}. Yet, going to a genuine multimode scenario required the design of a whole new photon subtractor based on sum-frequency generation \cite{PhysRevA.89.063808,ra_tomography_2017}. This finally permitted the first demonstration of multimode non-Gaussian state in a CV setting, demonstrating non-Gaussian features in up to four entangled modes \cite{Ra2019}.

Such highly multimode states of more than two modes are confronted with a considerable problem: the exponential scaling of the required number of measurements for a full state tomography. This makes it highly challenging to demonstrate non-Gaussian features such as Wigner negativity in multimode non-Gaussian states. For single-photon-subtracted states, it has been pointed out that good analytical models can be used to train machine learning algorithms to recognise Wigner negativity based on single-mode measurements \cite{PhysRevLett.125.160504}. Furthermore, the techniques of \cite{chabaud2021witnessing} combined with \cite{chabaud2021efficient} should also make it possible to use multiplexed double homodyne detection to witness Wigner negativity in certain classes of multimode states.\\

In multimode systems, we are confronted with the limitations of homodyne tomography. Recently, it has been shown that machine learning techniques can be used to implement an improved form of CV tomography based on homodyne measurements \cite{Tiunov:20}. Even though this setup is computationally heavy in single mode setups, it uses a smaller set of states as a basis for state reconstruction which might make multimode versions of the protocol more scalable. Alternatively, one can also bypass homodyne measurements all together. Photon-number-resolving detectors such as transition edge sensors \cite{Lita:08} make it possible to use the identity (\ref{eq:wigexpvalueparity}) to directly measure the Wigner function \cite{Nehra:19}. Intriguingly, this implies that a photon-number-resolving detector and a setup to generate displacements of the state in arbitrary modes makes it possible to directly measure the full multimode Wigner function. Nevertheless, such a multimode protocol has so far not been realised in any experiment.

\subsection{Other experimental setups}\label{sec:OtherExp}

Given all the experimental work in CV quantum optics, it is perhaps surprising that the first experimental demonstrations of quantum states with Wigner negativity happened in different fields. The very first realisation of such a state was achieved with trapped ions. Even though one often uses the atomic transitions in these systems to isolate qubits for potential quantum computers, trapped ions also have interesting motional degrees of freedom. By exploiting  a Jaynes-Cummings type interaction between the atom and the trapping field, it is possible to use the ions' internal atomic degrees of freedom to create well-controlled non-Gaussian states such as a Fock state \cite{PhysRevLett.77.4281} and a Schr\"odinger's cat state \cite{Monroe1131} in the motional degrees of freedom.

Mathematically, this setup is equivalent to cavity QED, where it was shown that photons in a cavity can be manipulated through interactions with atoms \cite{PhysRevA.45.5193} and the Rabi-oscillations of the injected Rydberg atoms can in turn be used to probe the field within the cavity \cite{PhysRevLett.76.1800}. These methods would then be combined to experimentally generate a single-photon Fock state of the microwave field in a cavity \cite{Nogues-99}, confirm its Wigner negativity \cite{PhysRevA.62.054101}, and probe its full Wigner function \cite{PhysRevLett.89.200402}. A few years later, similar techniques were used to finally generate Schr\"odinger cat states and higher order Fock states \cite{Deleglise08}. 

A third setup with very similar physics is found in circuit QED. In this field, the macroscopic microwave cavities are replaced by superconducting circuits, and nonlinearities are induced by Josephson junctions rather than atoms \cite{Vion886}. Even though these setups are often used in a DV approach, the microwave fields involved can equally be treated in a CV approach. The large nonlinearities rather naturally create non-Gaussian states, but getting a good sense of control over them can be challenging. Nevertheless, a wide range of non-Gaussian states such as Fock state \cite{Hofheinz:2008} and large Schr\"odinger's cat states \cite{Vlastakis607} have been experimentally realised. The latter have furthermore been stabilised by engineering the decoherence processes in the system \cite{Leghtas853}. Very recently these systems have also been used to demonstrate the deterministic generation of photon triplets \cite{PhysRevX.10.011011}.

In recent years, both, trapped ions \cite{Fluhmann:2019} and superconducting circuits \cite{Campagne-Ibarcq} were used to achieve another important milestone in CV quantum computing: the experimental generation of a GKP state. These highly non-Gaussian states are useful for encoding a fault-tolerant qubit in a CV degree of freedom. By exploiting the redundancy that is offered by the infinite dimension Hilbert space of a CV system, one can create a qubit with a certain degree of robustness. This effectively makes it possible to implement error correction routines, as shown in Ref.~\cite{Campagne-Ibarcq}. In other words, these systems have managed to generate CV states that are so non-Gaussian that they can be effectively used as fault-tolerant DV states. 

A final field that has shown much potential over the last decades is cavity optomechanics. Here, an optical field is injected into a cavity with one moving mirror (more generally also other types of ``dynamic cavities'' can be used). The goal is to cool this mirror to its ground state to observe its quantum features. This way, one hopes to create non-classical states of motion in reasonably large objects. A wide variety of such optomechanical devices exist \cite{RevModPhys.86.1391}. Several theoretical schemes have been proposed to generate non-Gaussian states in such optomechanical setup \cite{PhysRevLett.117.143601,PhysRevA.98.063801}. Even though quantum features such photon-phonon entanglement have been demonstrated in such systems \cite{Palomaki710,Riedinger:2016}, it remains highly challenging to obtain good experimental control over the motional quantum state. Nevertheless, some CV non-Gaussian states states in the form of superpositions between vacuum and a single-phonon Fock state have been experimentally realised \cite{Reed2017}.\\

A common problem in these setups is the creation of entanglement between the CV degrees of freedom in different modes. Some degree of such CV entanglement has been experimentally achieved in trapped ion \cite{Jost:2009} and circuit QED setups \cite{PhysRevLett.109.183901}. However, the number of entangled modes is much lower than what has been achieved in optics \cite{Asavanant:2019aa,Larsen:2019aa,cai-2017,gerke_full_2015,PhysRevLett.112.120505}, where even non-Gaussian entangled states of more than two modes have been created \cite{Ra2019}. This shows clearly how different experimental setups have different strengths and weaknesses. Optics comes with the advantage of spatial, temporal, and spectral mode manipulations, which allows to create large entangled states. However, the resilience of optical setups to decoherence is due to limited interaction with the environment. The latter implies that it is also difficult to find controlled ways to make these systems strongly non-Gaussian. On the other hand, the other setups which we discussed require much more significant shielding from environmental degrees of freedom. When this coupling to other degrees of freedom can be controlled, it provides the means to create non-Gaussian quantum states. In this context, it is appealing to combine the advantages of different regimes. Optomechanics offers a potential pathway to achieve this by converting between microwave and optical degrees of freedom \cite{Forsch:2020,PhysRevX.10.021038}.\\

As a last remark, it is interesting to mention that phase space descriptions and non-Gaussian states also appear in atomic ensembles. This framework relies on the fact that an ensemble of a large number of atoms can be described by collective observables that behave very similar to bosonic systems. The associated phase space behaves differently from the optical phase space, in the sense that it is compact. More specifically, the phase space will cover a sphere and the radius of this sphere will depend on the number of atoms. Effectively, we would recover a bosonic system in the limit of an infinite number of atoms. However, the compactness of phase space for a finite ensemble comes with interesting side-effects: a sufficiently high amount of spin squeezing can create non-Gaussian states. We will not go into details for these systems, but it should nevertheless be highlighted that non-Gaussian spin states have received considerable attention in literature \cite{RevModPhys.90.035005} and have been produced in a range of experiments \cite{PhysRevLett.108.183602, Haas180,Strobel424, SpinNeg}.

\section{Conclusions and outlook}\label{sec:conclusions}

In this Tutorial, we have presented a framework based on phase-space representations to study continuous-variable quantum systems. We then focused on the various aspects of non-Gaussian states, where we first represented different ways to structure the space of continuous-variable states in a single mode in Fig.~\ref{fig:nonGaussianOnion}. Whenever possible, we generalised results from literature to a multimode setting. However, for certain properties such as the stellar rank, these generalisations become insufficient to classify all possible quantum states. 

We introduced two paradigms to create non-Gaussian states, where one is a deterministic approach based on unitary transformations, reminiscent of the circuit approach for quantum information processing. The second approach is conditional, in the sense that it relies on conditioning one part of a state on measurement outcomes for another part of a state, which is more narrowly related to a measurement-based approach to quantum protocols. Throughout the remainder of the Tutorial, we have largely focused on conditional operations, since it is the most commonly used approach in experiments. It also provides a natural avenue to start studying the relation between quantum correlations and non-Gaussian features. We show how the conditional approach requires certain correlations in the initial Gaussian state to be able to induce certain type of non-Gaussianity in the conditional state, as summarised in Fig.~\ref{fig:nonGaussianCorr}. 

On the other hand, non-Gaussian operations can also create a type of non-Gaussian entanglement as introduced in Subsection \ref{sec:QunatumCorrThroughnonGauss}. This kind of entanglement is particular as it can not be identified with typical techniques that rely on the state's covariance matrix. Nevertheless, we use R\'enyi-2 entanglement as a measure to illustrate the existence of such purely non-Gaussian quantum correlations in photon-subtracted states. Even though its existence is known from pure state examples, it has only received limited attention in both theoretical and experimental work. One possible reason is the difficulty of studying this type of entanglement for mixed states, since convex roof constructions tend to become highly intractable for non-Gaussian states. 

As a final theoretical aspect of the Tutorial, we highlight the need of Wigner negativity to achieve some of the most striking features in quantum technologies. On the one hand, we show that Wigner negativity in either the state or the measurement is necessary to violate a Bell inequality. This observation can be understood in the broader context of non-locality and contextuality: Wigner negativity is often seen as a manifestation of the contextual behaviour of quantum systems, and non-locality can be understood as a type of contextuality of measurements on different subsystems. On the other hand, we also present results that show how Wigner negativity is a requirement to achieve a quantum computational advantage. Intuitively, it is perhaps not surprising that states, operations, and measurements that can all be described by probability distributions on phase space can be efficiently simulated on a classical computer. However, as we showed in Section \ref{sec:Advantages}, the actual simulations protocol contains many subtle points. Here, too, we conclude that there are still many open questions surrounding the physics of quantum computational advantages in continuous-variable setups.

As a last step of this Tutorial, we provided an overview of the experimental realisations of non-Gaussian states with continuous-variables. Quantum states of light are indeed the usual suspects for continuous-variable quantum information processing, but it turns out to be remarkably challenging to engineer highly non-Gaussian states in such setups. We highlighted how trapped ions, cavity QED, and circuit QED have proven to be better equipped for this task, but in return they are confronted with other problems. Optomechanics presents itself as an ideal translator between these two regimes, which may soon make it possible to combine the scalability of optical setups with the high nonlinearities of the microwave domain.\\

In a more general sense, there are definitely many open question to be resolved in the domain of continuous-variable quantum physics. In this Tutorial, we have focused extensively on questions related to non-Gaussian features, notably in multimode systems. In the greater scheme of things, this is only one of the many challenges in the field. The recent demonstration of a quantum computational advantage with Gaussian Boson Sampling has set an important milestone for continuous-variable quantum technologies \cite{Zhong1460}, but we are still far away from useful computational protocols as set out in the roadmap of Ref.~\cite{Bourassa2021blueprintscalable}. Even though the quest for Gottesman-Kitaev-Preskill state \cite{PhysRevA.64.012310} is one of the main experimental priorities, there are still many open challenges in designing the Gaussian operations that form the basis of such a setup \cite{larsen2020deterministic,Arrazola:2021}.

Beyond universal fault-tolerant quantum computers, there are many other potential applications for continuous-variable systems. They are widely used in quantum communications for quantum key distribution \cite{Jouguet:2013} and and secret sharing \cite{Bell:2014}. These protocols are largely based on Gaussian states and measurements, such that also the best possible attacks to these systems are Gaussian \cite{PhysRevLett.97.190502}. Nevertheless, non-Gaussian protocols for quantum key distribution, based on photon subtraction, have been proposed \cite{PhysRevA.95.032304}. Such non-Gaussian quantum computation protocols and their security still involve many open questions.

Continuous-variable systems also provide a natural link to other bosonic systems, which is why they have been suggested as a platform to simulate molecular vibronic spectra \cite{Huh:2015}. The continuous-variable approach also plays an important role in quantum algorithms for other chemistry-related problems such as drug discovery through molecular docking \cite{Banchieaax1950} and the simulation of electron transport \cite{Jahangiri:2021}.

Furthermore, the continuous-variable setting is also suitable to implement certain elements for quantum machine learning such as quantum neural networks \cite{PhysRevResearch.1.033063}. Even though this is a promising platform for tackling a wide range of problems, the proposal is highly ambitious on several points. In the context of this Tutorial, we emphasise the need of non-Gaussian unitary transformations. In principle, neural networks require linear couplings between different ``neurons'' which each implement some form of nonlinear operation. Non-Gaussian operations play the role of this nonlinear element, making them a crucial step in the scheme. To implement such continuous-variable neural networks we thus require either new developments on the implementation of non-Gaussian operations, or theoretical modifications in the protocol to make it fit for implementable conditional non-Gaussian operations. It should be highlighted that other machine learning approaches exist, such as reservoir computing, which can be entirely based on Gaussian states \cite{Nokkala:2021}.

A final quantum technology that may benefit from the use of non-Gaussian states is quantum metrology, as was recently demonstrated with motional Fock states of trapped ions \cite{Wolf2019Ions}. Even though early work has shown that there is no clear benefit in using non-Gaussian operations such as photon subtraction for parameter-estimation \cite{PhysRevA.90.013821}, there may still be other settings where such states are beneficial. Non-Gaussian entanglement could for example have a formal metrological advantage that is reflected in the quantum Fisher information \cite{Gessner2017entanglement}. On the other hand, ideas from quantum metrology also provide a possible approach for measuring non-Gaussian quantum steering \cite{YadinGessner}. The effects of non-Gaussian features on the sensitivity of the state can in principle be captured by higher moments of the quadrature operators \cite{PhysRevLett.122.090503}. It was recently shown that post-selected measurements could, indeed, offer a quantum advantage for metrology \cite{Arvidsson-Shukur2020}. This result is narrowly connected to the field of weak measurements and makes a connection to yet another phase-space representation: the Kirkwood-Dirac distribution \cite{PhysRev.44.31,RevModPhys.17.195}. Hence, we circle back to the fundamental physics of continuous-variable systems and conclude that there are still many connections to be made.\\

Beyond the technological applications that continuous-variable systems may have to offer, there is an important down-to-earth perspective that must be emphasised. With the improvement of detectors throughout the years, we have reached a point where theory and experiment can be considered mature to tackle single-mode problems. In multimode systems, the same cannot be said. With the exponential scaling of standard homodyne tomography, experimental tools for studying large multimode states beyond the Gaussian regime are limited. We may have to accept that the full quantum state is out of reach for experimental measurements. Even theoretically, highly multimode Wigner functions quickly become cumbersome to handle. Treating them with numerical integration techniques becomes a near-impossible task, once the number of modes is drastically increased. This makes even numerical simulations challenging. How then can we understand and even detect the non-Gaussian features of these systems?

One clear and important future research goal in this field is to provide an answer to this question. For quantum technologies, this may provide us with new ways to benchmark our systems, but more fundamentally it might teach us something new about the physics of these systems. One place where one might look for inspiration is the field of statistical mechanics, where statistical methods show that even highly complex systems can produce clear emergent signatures. We recently took a first step in exploring such ideas by looking at emergent network structures for continuous-variable non-Gaussian states \cite{walschaers2021emergent}. The most exciting lesson from such preliminary work is that there is still much to be learned about non-Gaussian quantum states.

\begin{acknowledgments}
First, I thank M. Genoni for several useful suggestions for this Tutorial. More generally speaking, the content of this text was influenced by stimulating discussions throughout the years with many colleagues, notably F. Grosshans, R. Filip, Q. He, M. Gessner, and G. Ferrini. Furthermore, I am very grateful to M. Fannes for teaching me the mathematical foundations that lie at the basis bosonic quantum systems. This point of view was complemented by colleagues in the multimode quantum optics group of the Laboratoire Kastler Brossel, V. Parigi, N. Treps, and C. Fabre, who have introduced me to the wonderful world of experimental quantum optics. 

Still, the main source of inspiration for this Tutorial are the many excellent students and postdocs that I have worked with in the last few years. Their questions and struggles have been essential to highlight the barriers that I try to overcome in this Tutorial. Among these students and postdocs, I want to express explicit gratitude to U. Chabaud, G. Sorelli, and D. Barral for their careful and detailed reading of the manuscript and for their useful comments. I also acknowledge the many useful discussions with K. Zhang, who notably made me aware of the possibility of reducing entanglement through photon subtraction (here shown in Fig.~\ref{fig:EntCreation}). Last but not least, I want to express special thanks to C. Lopetegui, who started reading this Tutorial as a newcomer to the field of CV quantum optics and thus had the perfect point of view to help fine-tune the content. 
\end{acknowledgments}

\appendix
\section{Mathematical remarks}

Here we present some important well-known mathematical concepts that are regularly used in the Tutorial to make the text more self-contained. The comments and definitions given here are not very rigorous and mainly aim at giving the reader an intuitive understanding, for a more formal introduction one should consult a standard textbook \cite{Schaefer1971,Conway2007,bratteli_operator_1987}.

\subsection{Topological vector spaces}\label{sec:TopologicalVectorSpace}

Throughout this article, we often deal implicitly with topological vectors spaces. Vectors spaces are well-known from linear algebra and can be thought of as sets of mathematical objects called vectors, which can be added together in a commutative way and multiplied by scalars. When we consider a vectors space ${\cal V}$ on a field ${\cal F}$, this means that for any $\vec v_1, \vec v_2 \in {\cal V}$ and any $\alpha_1, \alpha_2 \in {\cal F}$ the object $\alpha_1 \vec v_1 + \alpha_2 \vec v_2 \in {\cal V}$. This means that the vectors space is closed under addition and scalar multiplication. In this Tutorial, the field ${\cal F}$ is either identified as $\mathbb{R}$ (for phase space) or $\mathbb{C}$ (for Hilbert spaces).

The spaces that are considered it the Tutorial have much more structure than what is given by the vector space. First of all, we generally deal with normed spaces, which means that our vector spaces are topological vector spaces in the sense that there is a notion of distance defined upon them. Generally speaking, topological vector spaces can be equipped with exotic topologies, but here we simply deal with norms. On top, we again add an additional structure when we assume that these norms are generated by inner products (depending on exact properties, these inner products go by different names such as ``positive-definite sesquilinear form'', which is what we typically consider in quantum mechanics). 

As we deal with infinite dimensional spaces to describe bosonic quantum states and Fock space, it is important to set some terminology straight. When we talk about a Hilbert space, there is the assumption that the space is complete. In an infinite-dimensional inner-product space, we can define sequences of elements in ${\cal V}$. If we consider a sequence $(\vec v_j)_{j \in {\mathbb{N}}}$ such that for any $\epsilon$ we can find a value $N>0$ such that $\norm{\vec v_j - \vec v_k} < \epsilon$ for all $j,k > N$, we call the sequence a Cauchy sequence. In other words, the distance between elements in the Cauchy sequence shrinks as we proceed further into the sequence. The fact that a Hilbert space is closed means that all Cauchy sequences converge in the sense that $\lim_{j \rightarrow \infty} \vec v_j = \vec v \in {\cal V}$. Finite-dimensional inner product spaces automatically have this property, but for infinite-dimensional spaces it must be imposed explicitly.

Another class of structured vector space, that is often encountered in the Tutorial, is a real symplectic space. These spaces appear when we consider phase space, and the are given by a real vector space with an additional symplectic form $\sigma$ instead of the usual inner product. In mathematical literature, one often encounters the notation $({\cal V}, \sigma)$ for a symplectic space, where the symplectic for has the following properties: we consider $\vec{v}_1.\vec v_2 \in {\cal V}$ and find that $\sigma(\vec v_1, \vec v_2) \in \mathbb{R}$, $\sigma$ is bilinear, and $\sigma(\vec v_1, \vec v_2) = -\sigma(\vec v_2, \vec v_1).$ In all cases in this Tutorial, we also consider that $\sigma$ is non-degenerate, which means that $\sigma(\vec v_1, \vec v_2) = 0$ for all $\vec v_1 \in {\cal V}$ if and only if $\vec v_2 = \vec 0$. When the symplectic space is finite-dimensional, it is often practical to represent the symplectic form in terms of a matrix. In the Tutorial this is done by associating $\sigma(\vec v_1, \vec v_2) = \vec v_1^T \Omega \vec v_2$.

In principle, a real symplectic space is all that is needed to develop the mathematical framework of the CCR algebra. However, it is often natural when dealing with bosonic systems to include an additional structure in the form of an inner product. In the Tutorial, this is done implicitly by also using the standard inner product $\vec v_1^T \vec v_2$ on phase space. This allows us to ultimately get the isomorphism \eqref{eq:PhaseHilb}. In the quantum statistical mechanics literature, it is common to see references to a ``pre-Hilbert space'', rather than a phase space or a symplectic space. When we refer to a pre-Hilbert space, we consider an inner-product space which is not necessarily complete and one must consider the closure to be guaranteed to obtain a full Hilbert space. The reason is that phase space, as a real vector space ${\cal V}$ with an inner product, given by a bilinear form $s(.,.)$, and a symplectic form $\sigma(.,.)$ is equivalent to a complex pre-Hilbert space ${\cal H}$. For finite dimensional spaces, the equivalence between the vector spaces is obtained via isomorphism \eqref{eq:PhaseHilb}:
\begin{equation}
\vec f \in {\cal V} \mapsto \ket{\psi_f} = \sum_{j} (f_{2j-1} +i f_{2j})\ket{\phi_j} \in \mathcal{H},
\end{equation}
where $\ket{\phi_j}$ for a basis of ${\cal H}$.
As we are talking about an isomorphism between structured vectors spaces, we also need an identity between additional structures, which is given by
\begin{equation}
\<\psi_{f_1} \mid \psi_{f_2}\> = s(\vec f_1, \vec f_2) - i \sigma( \vec f_1, \vec f_2).
\end{equation}
This isomorphism holds very generally and can be extended to infinite-dimensional spaces. It provides a very formal connection between the single-particle Hilbert space for a many-boson system and its phase space associated with the modes of the bosonic field. Technically, we note that the phase space is equivalent to a pre-Hilbert space, and the closure of this space is the single-particle Hilbert space. Whenever the phase space (and thus the single-particle Hilbert space) is finite-dimensional, the pre-Hilbert space is closed such that phase space and single-particle Hilbert space really are equivalent. For a very rigorous treatment on all these points, we refer to \cite{bratteli_operator_1997}.

\subsection{Span}\label{sec:Span}
Throughout the Tutorial, we often refer to the ``span'' of a certain set of vectors. These vectors can be members of a vector space, symplectic space, topological vectors space, pre-Hilbert space, or Hilbert space, the definition of the span is always the same. Let us here assume that ${\cal V}$ denotes any type of vector space over a field ${\cal F}$ and consider a set $\vec v_1, \dots \vec v_n \in {\cal V}$. We can now define the span of this set of vectors as the set of all linear combinations that can be made with these vectors
\begin{equation}
{\rm span} \{\vec v_1, \dots \vec v_n\} \coloneqq \{\alpha_1 \vec v_1 + \dots + \alpha_n \vec v_n \mid \alpha_1, \dots, \alpha_n \in {\cal F} \}.
\end{equation}
We emphasise that there is no need for the set $\vec v_1, \dots \vec v_n$ to form a basis, not for the vectors to be linearly independent, nor for the vectors to be normalised, nor for the vectors to be orthogonal to one another. 

Throughout the Tutorial, the vector spaces we will encounter are either real (in the case of phase space) such that ${\cal F} = \mathbb{R}$ or complex (in the case of Hilbert spaces for quantum systems) such that ${\cal F} = \mathbb{C}$. In the case where the vector spaces have some topological structure (which we can colloquially understand as a mathematical sense of distance that allows to define limits), it can make sense to consider the closure of a span, denoted by
\begin{equation}
\overline{{\rm span} \{\vec v_1, \dots \vec v_n\} },
\end{equation}
such that any convergent sequence built with elements of the span has its limit also included in the closure.

\bibliography{notes_steering}

\end{document}